\documentclass[10pt,journal,compsoc]{IEEEtran}
\ifCLASSOPTIONcompsoc
  \usepackage[nocompress]{cite}
\else
  \usepackage{cite}
\fi
\ifCLASSINFOpdf
  \usepackage[pdftex]{graphicx}
   \graphicspath{{Images/}}
 
\else
\fi

\usepackage{diagbox}
\usepackage{rotating}
\usepackage{url}
\usepackage{xcolor}

\begin{document}
%
% paper title
% Titles are generally capitalized except for words such as a, an, and, as,
% at, but, by, for, in, nor, of, on, or, the, to and up, which are usually
% not capitalized unless they are the first or last word of the title.
% Linebreaks \\ can be used within to get better formatting as desired.
% Do not put math or special symbols in the title.
\title{Finding Critical Scenarios for Automated Driving Systems: A Systematic Literature Review}
%
%
% author names and IEEE memberships
% note positions of commas and nonbreaking spaces ( ~ ) LaTeX will not break
% a structure at a ~ so this keeps an author's name from being broken across
% two lines.
% use \thanks{} to gain access to the first footnote area
% a separate \thanks must be used for each paragraph as LaTeX2e's \thanks
% was not built to handle multiple paragraphs
%
%
%\IEEEcompsocitemizethanks is a special \thanks that produces the bulleted
% lists the Computer Society journals use for "first footnote" author
% affiliations. Use \IEEEcompsocthanksitem which works much like \item
% for each affiliation group. When not in compsoc mode,
% \IEEEcompsocitemizethanks becomes like \thanks and
% \IEEEcompsocthanksitem becomes a line break with idention. This
% facilitates dual compilation, although admittedly the differences in the
% desired content of \author between the different types of papers makes a
% one-size-fits-all approach a daunting prospect. For instance, compsoc 
% journal papers have the author affiliations above the "Manuscript
% received ..."  text while in non-compsoc journals this is reversed. Sigh.

\author{Xinhai Zhang,
        Jianbo Tao,
        Kaige Tan,~\IEEEmembership{Member,~IEEE,}
        Martin Törngren,~\IEEEmembership{Senior Member,~IEEE,}
        José Manuel Gaspar Sánchez,~\IEEEmembership{Member,~IEEE,}
        Muhammad Rusyadi Ramli,~\IEEEmembership{Member,~IEEE,}
        Xin Tao,~\IEEEmembership{Member,~IEEE,}
        Magnus Gyllenhammar,
        Franz Wotawa, ~\IEEEmembership{Member,~IEEE,}
        Naveen Mohan, ~\IEEEmembership{Member,~IEEE,}
        Mihai Nica, ~\IEEEmembership{Member,~IEEE,}
        and~Hermann Felbinger,~\IEEEmembership{Member,~IEEE,}
         
        % and~Jane~Doe,~\IEEEmembership{Life~Fellow,~IEEE}% <-this % stops a space
\IEEEcompsocitemizethanks{
\IEEEcompsocthanksitem X. Zhang is with Sigma Technology Consulting AB and the autonomous group of Scania CV AB in Sweden. He was with the Mechatronics division of KTH for most of the time when the paper was writing. \protect\\
% note need leading \protect in front of \\ to get a newline within \thanks as
% \\ is fragile and will error, could use \hfil\break instead.
E-mail: xinhai@kth.se
\IEEEcompsocthanksitem J. Tao, M. Nica and H. Felbinger are with AVL.
\IEEEcompsocthanksitem K. Tan, M. Törngren, J. Gaspar, M. R. Ramli, M. Gyllenhammar and N. Mohan are with the Mechatronics division at KTH, Stockholm, Sweden
\IEEEcompsocthanksitem X. Tao is with the Integrated Transport Research Lab (ITRL) of KTH.
\IEEEcompsocthanksitem M. Gyllenhammar is with Zenseact AB, Gothenburg, Sweden
\IEEEcompsocthanksitem F. Wotawa is with the CD Laboratory for Quality Assurance Methodology for Autonomous Safety Critical Systems (QAMCAS), Inst. for Software Technology, TU Graz, Graz, Austria
}% <-this % stops a space
\thanks{
%Manuscript received April 19, 2005; revised August 26, 2015.
}}

\IEEEtitleabstractindextext{%

\begin{abstract}
%The abstract should briefly summarize the contents of the paper in 15--250 words.

Scenario-based approaches have been receiving a huge amount of attention in research and engineering of 
%, from design to assurance, of 
%regarding the development, verification and validation (V\&V) of 
automated driving systems. Due to the complexity and uncertainty of the driving environment, and the complexity of the driving task itself, the number of possible driving scenarios that an ADS or ADAS may encounter is virtually infinite. Therefore it is essential to be able to reason about the identification of scenarios and in particular critical ones that may impose unacceptable risk if not considered. Critical scenarios are particularly important to support design, verification and validation efforts, and as a basis for a safety case.
In this paper, we present the results of a systematic literature review in the context of autonomous driving. The main contributions are: (i) introducing a comprehensive taxonomy for critical scenario identification methods; (ii) giving an overview of the state-of-the-art research based on the taxonomy encompassing 86 papers between 2017 and 2020; and (iii) identifying open issues and directions for further research.
%\textcolor{red}{}
% PREVIOUS TEXT HERE!! , to analyze (1) how critical Scenarios are used in the development and V\&V processes; (2) how scenarios are defined and modeled for different purposes; (3) how critical scenarios are identified or generated; and (4) how real traffic data are used in the identification and generation processes. 
The provided taxonomy comprises three main perspectives encompassing the problem definition (the why), the solution (the methods to derive scenarios), and the assessment of the established scenarios. %The identified critical scenarios are classified accordingly to if they are safety-critical or functional-critical, and if they are implementation-specific. The adopted surrogate criticality measures are analyzed in each class. CSI methods are grouped into five clusters according to the similarity of their solutions. Detailed classification of the CSI solutions is conducted in each cluster.
% As a result of this process, we derive present a Thereafter, this paper also discusses how the finding from this work will contribute to a systematic literature survey on this topic.
In addition, we discuss open research issues considering the perspectives of coverage, practicability, and scenario space explosion.
\end{abstract}

% Note that keywords are not normally used for peerreview papers.
%\keywords{  Scenario-based Approach     \and 
%            Critical Scenarios          \and
%           Scenario Modeling           \and
%          Automated Driving           \and
%         Literature Survey           
%    }

}

% make the title area
\maketitle

% To allow for easy dual compilation without having to reenter the
% abstract/keywords data, the \IEEEtitleabstractindextext text will
% not be used in maketitle, but will appear (i.e., to be "transported")
% here as \IEEEdisplaynontitleabstractindextext when compsoc mode
% is not selected <OR> if conference mode is selected - because compsoc
% conference papers position the abstract like regular (non-compsoc)
% papers do!
\IEEEdisplaynontitleabstractindextext
% \IEEEdisplaynontitleabstractindextext has no effect when using
% compsoc under a non-conference mode.

% For peer review papers, you can put extra information on the cover
% page as needed:
% \ifCLASSOPTIONpeerreview
% \begin{center} \bfseries EDICS Category: 3-BBND \end{center}
% \fi
%
% For peerreview papers, this IEEEtran command inserts a page break and
% creates the second title. It will be ignored for other modes.
\IEEEpeerreviewmaketitle

\ifCLASSOPTIONcompsoc
\IEEEraisesectionheading{\section{Introduction}\label{sec:intro}}
\else
\section{Introduction}
\label{sec:intro}
\fi

% Computer Society journal (but not conference!) papers do something unusual
% with the very first section heading (almost always called "Introduction").
% They place it ABOVE the main text! IEEEtran.cls does not automatically do
% this for you, but you can achieve this effect with the provided
% \IEEEraisesectionheading{} command. Note the need to keep any \label that
% is to refer to the section immediately after \section in the above as
% \IEEEraisesectionheading puts \section within a raised box.

% The very first letter is a 2 line initial drop letter followed
% by the rest of the first word in caps (small caps for compsoc).
% 
% form to use if the first word consists of a single letter:
% \IEEEPARstart{A}{demo} file is ....
% 
% form to use if you need the single drop letter followed by
% normal text (unknown if ever used by the IEEE):
% \IEEEPARstart{A}{}demo file is ....
% 
% Some journals put the first two words in caps:
% \IEEEPARstart{T}{his demo} file is ....
% 
% Here we have the typical use of a "T" for an initial drop letter
% and "HIS" in caps to complete the first word.
%\IEEEPARstart{T}{his} demo file is intended to serve as a ``starter file''
%for IEEE Computer Society journal papers produced under \LaTeX\ using IEEEtran.cls version 1.8b and later.
% You must have at least 2 lines in the paragraph with the drop letter
% (should never be an issue)
%I wish you the best of success.
%\hfill mds
%\hfill August 26, 2015

%%%%% set the scene
%Once upon a time there was a system that found itself immersed into a complex environment with more than one scenario ...

\IEEEPARstart{T}{he} long and winding road towards high levels of automated driving is fascinating and highly representative of an ongoing technological paradigm shift, see e.g., \cite{torngren_martin_2021_4710500} and further references therein. The past years have seen enormous amounts of funding going into a new automated driving ecosystem of startups, new players and the existing automotive industry, \cite{Frazzoli}, attracting in excess of 80 billion US dollars during the period 2014 to 2017 \cite{KerryKarsten}, with likely more than 100 billion US dollars invested the past decade, \cite{SHETTY2021103133}. 
As a result, impressive advances have been made and technology has matured. However, the resulting complexity still poses formidable socio-technical challenges for introducing automated driving on a larger scale. 

One key challenge refers to reasoning about what can happen during automated driving with relevance for safety engineering activities, from design to assurance. 
The complexity and uncertainty of the driving environment, and the complexity of the driving task itself, imply that the number of possible scenarios that an Automated Driving Systems (ADS) or Advanced Driving-Assistance Systems (ADAS) may encounter is virtually infinite. 
It is clear that traditional mileage validation is infeasible for the safety validation of ADS and ADAS, and moreover that "driven miles" and disengagement reports are far from sufficient for reasoning about risk, see e.g., \cite{KALRA2016182,SHETTY2021103133}. 
This has led to a strong interest in virtual evaluation and verification, allowing for "safe testing" and exploration, and potentially for cost-efficient assurance based on the created evidence. 
However, this still begs the following questions: Assuming you have a faithful simulation environment (a challenge on its own), what tests are you to select (i.e. which are critical), how do you derive them, and how do you reason about coverage and the completeness of the safety analysis? 
%However, simulation still has many challenges: Scenarios; Abstraction levels; methodologies

A typical approach to limit the potentially infinite scope of what an ADS may encounter, including unknowns and uncertainty, is to limit the Operational Design Domain (ODD), i.e., the operating conditions under which a given ADS or one or more of its features are designed to function, \cite{J3016}. This is indeed the most common approach taken for current roll-outs of ADS at high levels of automation.
A complementary proposed approach is to attempt to define and limit the behaviors to be encountered on the roads, see e.g., \cite{RSS}. While this may be very promising in the long term, it is hard to accomplish in the short term given mixed-mode traffic (e.g., manually driven cars together with vehicles at level 4 of automated driving), the complexity of automated driving, and the need to agree on what represents a suitable risk level (compromise between performance and safety). A third avenue would be to leverage collaborating driving and smart infrastructure, however this requires multi-stakeholder coordination,  infrastructure investments and dealing with new security related threats, thus complicating and delaying such an approach, see e.g., \cite{SHETTY2021103133}. 

In any case, it becomes essential to be able to reason about the identification of scenarios - as "the temporal development between several scenes in a sequence of scenes", \cite{Ulbrich2015}. Scenarios could be seen as "test cases" within an ODD. As such, scenarios are tangible and concrete and support ADS design and evaluation, with the key challenge that the number of possible driving scenarios that an AD or ADAS may encounter is virtually infinite, and correspondingly for covering a given ODD.  
%"infinitely thin lines"  [Ref. quote by Jan Reich, The Autonomous seminar]. 
In particular, for the purpose of supporting safety engineering, it becomes imperative how "critical" scenarios are identified, motivated and used. With "critical" scenarios we refer to situations which cause potential risk of harm (safety risks), that need explicit consideration for risk investigation and potential risk reduction measures (note: we summarize the key terms used in this paper in Section \ref{sec:Background}).

%%%%% scenario based methods
It is thus not surprising that scenario-based approaches are receiving a huge amount of attention in research and engineering regarding the development, verification and validation (V\&V) of ADAS and ADS.

The area is very active in terms of research, standardization, and engineering, encompassing, for example, scenario and ODD modeling, exchange formats, methods for scenario generation, and scenario space exploration, scenario catalogs and standards/guidelines, see e.g.,
\cite{ASAM, SOTIF, UL4600, P2846, ISO-AWI-TS-5083, ISO-SAE-21434, Amersbach2019}. 
%\textcolor{red}{[INCLUDE REPRESENTATIVE REFERENCES HERE PAS1883, ISO3450} Pegasus \textcolor{red}{add pegasus} project, scenario modeling languages such as Scenic, Foretellix, Traffic sequence charts, ASAM - Open scenario, Open ODD, etc.

During our research and work, we also discovered that many different approaches are taken for how to reason about scenarios and critical scenarios, ranging from different abstraction levels, assumptions, methods for scenario generation, exploration, etc. 
% THE FOLLOWING SENTENCE MIGHT BE USEFUL IN THE DISCUSSION!:
%these different approaches represent different viewpoints (concerns and stakeholders) regarding scenarios and how/what you perceive as critical. ... Use of scenarios: to support design, as an attempt to reduce the V\&V effort, and as a basis for reasoning about safety (as in a safety case).
%lack of systematic surveys
%lack of a framework and taxonomy to provide overview and structure all the research in this area
Moreover, we were missing a way to classify the different approaches we found, as a way to organize and compare them, and we
were not, to the best of our ability, able to find existing comprehensive surveys on this topic.

%%%%% A survey on Critical scenario identification
This led us to formulate a systematic literature review focused on Critical Scenario Identification (CSI) methods for ADS and ADAS.
Our literature review focuses on approaches to identify critical scenarios to support the development of ADS and ADAS. This, for example, means that scenario modeling languages like OpenScenario are out of the scope of the survey, other than when used as part of a CSI method.
The detailed scope of the literature review is described in Section \ref{sec:planning}. Other directions of scenario-based methods are briefly introduced and exemplified in Section \ref{sec:Other_Dir}.

%%%%% Contributions
%Critical Scenario Identification (CSI) methods for Automated Driving Systems (ADS) or Advanced Driving-Assistance Systems (ADAS)
The main contributions of the paper are as follows:
\begin{itemize}
\item A comprehensive taxonomy of CSI methods for the development of ADS and ADAS based on a systematic literature review.
\item An overview of the reviewed CSI methods based on the taxonomy.
\item Identified open issues and directions for further research.
\end{itemize}

\begin{figure}[!t]
\centering
\includegraphics[width=0.42\textwidth]{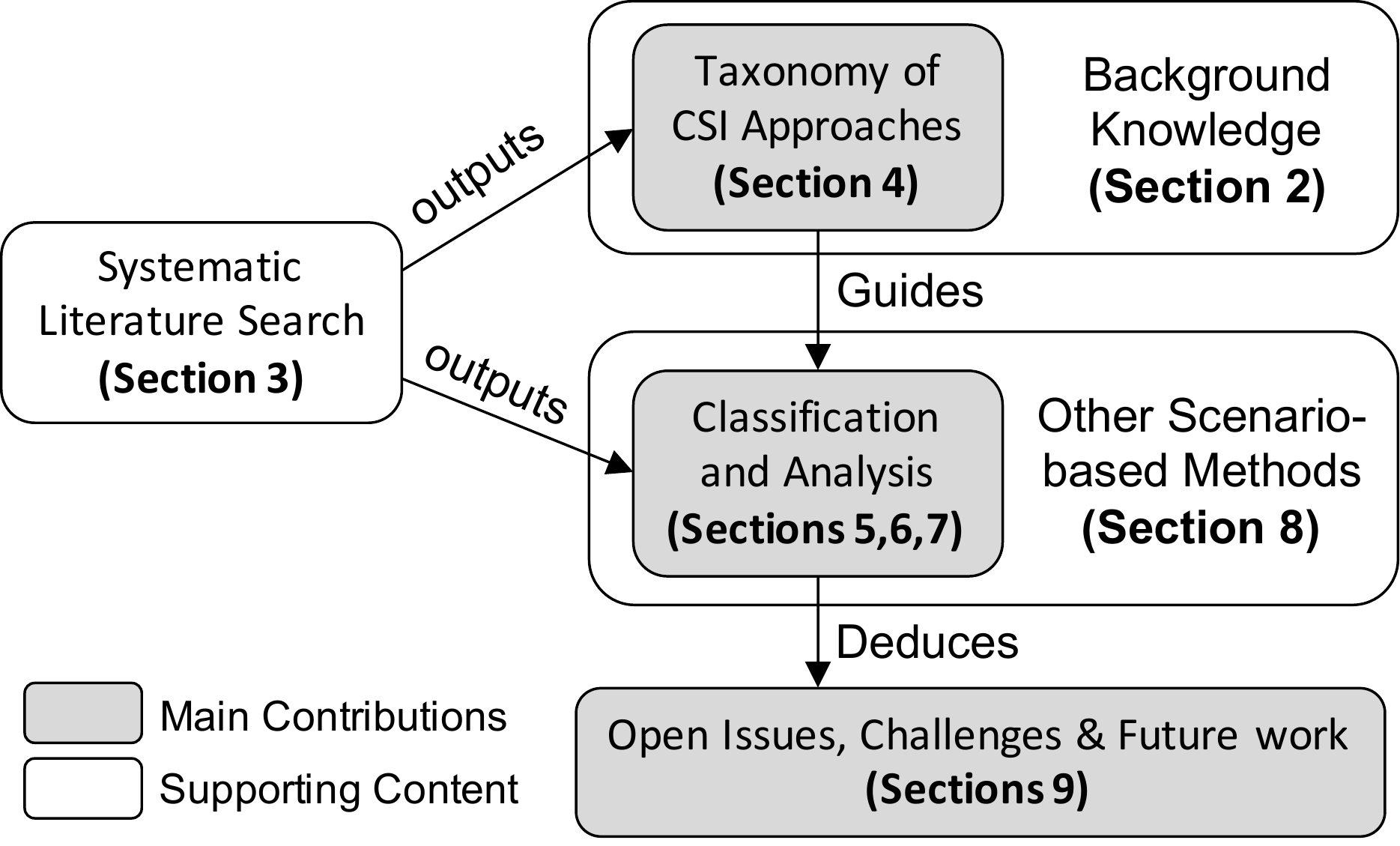}
\caption{Graphical outline of the rest of the paper}
\label{fig:outline}
\end{figure}

The rest of the paper is organized as shown in Fig.~\ref{fig:outline}. Section \ref{sec:Background} introduces key concepts, related standards and surveys of relevance for CSI methods for ADS/ADAS systems. Section \ref{sec:methodology} describes the employed literature review methodology, including the detailed scope, the research questions and the processes to search for relevant primary studies, to analyze the selected studies and to document the findings. The main result of this literature review includes a taxonomy for the studied CSI methods (Section \ref{sec:Tax}) and the corresponding classification of the primary studies (Sections \ref{sec:Problem}, \ref{sec:Solution} and \ref{sec:Validation}). Other related topics (but outside the scope of this literature review) are identified and briefly introduced in Section \ref{sec:Other_Dir}. Before concluding the paper in Section \ref{sec:Conc}, Section \ref{sec:Diss} summarizes and discusses all the findings from this systematic literature review, and accordingly suggests future work.

% MOVE TO COVERAGE!! “A notion of sufficient coverage must be found and metrics have to be defined to support this (i.e. when target is reached, the risk of overlooked hazardous scenarios is acceptably low)”
% and in particular critical ones that may impose unacceptable risk if not considered, to support design, as an attempt to reduce the V\&V effort, and as a basis for reasoning about safety (as in a safety case).

\section{Key Terminology and Related Surveys}
\label{sec:Background}
This section introduces the background knowledge and key concepts/terminology to facilitate the explanation of the taxonomy in Section \ref{sec:Tax}. The definitions of the terms are given in TABLE~\ref{tab:all_def} based on the following standards: SAE J3016 \cite{J3016}, ISO 26262 \cite{ISO26262}, ISO/PAS 21448 \cite{SOTIF}, BSI/PAS 1883 \cite{PAS1883}, UL4600 \cite{UL4600} and ISO 3450 series under development~\cite{ISO34501,ISO34502,ISO34503}. In TABLE~\ref{tab:all_def}, definitions without a citation refer to definitions proposed by the authors of this survey to serve the explanation of the taxonomy. If a definition has multiple citations, it means that the term has been defined in all those citations, and the definition in this paper considers all those definitions.

\begin{table}[!t]
\centering
\caption{Summary of Relevant Definitions}
\label{tab:all_def}
\begin{tabular}{|l|l|} 
\hline
Terminology                                                                      & Definition                                                                                                                                                \\ 
\hline
ADAS  &  %%%%%%%%%%%%%%%%%%%%%%%%%
\begin{tabular}[c]{@{}l@{}}
The term ADAS (Advanced Driving-Assistance \\
System) is used to describe assistant \\
systems based on Level 3 or active safety \\
systems. 
\end{tabular}  \\
\hline
ADS  &  %%%%%%%%%%%%%%%%%%%%%%%%%
\begin{tabular}[c]{@{}l@{}}
The term ADS (Automated Driving System) is\\ 
used specifically to describe a Level 3, 4, or 5\\
driving automation system. 
\end{tabular}  \\
\hline
Active Safety &     %%%%%%%%%%%%%%%%%%%%%%%%%
\begin{tabular}[c]{@{}l@{}}
Vehicle systems that sense and monitor \\
conditions inside and outside the vehicle\\
for the purpose of identifying perceived \\
present and potential dangers to the vehicle.~\cite{J3016} 
\end{tabular}  \\

\hline
Safety &       %%%%%%%%%%%%%%%%%%%%%%%%%                                                                            
The absence of unreasonable risk of harm.  \\ 
\hline
FuSa &      %%%%%%%%%%%%%%%%%%%%%%%%%
\begin{tabular}[c]{@{}l@{}}
Functional Safety: The absence of unreasonable \\ 
risk due to hazards caused by malfunctioning \\
behavior of E/E systems \cite{ISO26262} 
\end{tabular}  \\ 
\hline
SOTIF &     %%%%%%%%%%%%%%%%%%%%%%%%%
\begin{tabular}[c]{@{}l@{}}
Safety of the intended functionality: The absence\\
of unreasonable risk due to hazards caused by\\
the functional limitations of the intended\\
functionality \cite{SOTIF}
\end{tabular}  \\
\hline
Harm &      %%%%%%%%%%%%%%%%%%%%%%%%%
\begin{tabular}[c]{@{}l@{}}
Physical injury or damage to the health of\\
persons \cite{ISO26262}
\end{tabular}  \\
\hline
\begin{tabular}[c]{@{}l@{}}Risk of \\Harm \end{tabular} &   %%%%%%%%%%%%%%%%%%%%%%%%%
\begin{tabular}[c]{@{}l@{}}
Combination of the probability of occurrence of \\
harm and the severity of that harm \cite{ISO26262} 
\end{tabular}  \\
\hline
\begin{tabular}[c]{@{}l@{}}Hazardous \\Event\end{tabular} &     %%%%%%%%%%%%%%%%%%%%%%%%%
\begin{tabular}[c]{@{}l@{}}
Combination of a hazard and an operational\\
situation \cite{ISO26262}
\end{tabular}  \\
\hline
Hazard  &   %%%%%%%%%%%%%%%%%%%%%%%%%
\begin{tabular}[c]{@{}l@{}}
Potential source of harm caused by malfunction-\\
ing behavior of the implementation of a vehicle-\\
level function \cite{ISO26262}
\end{tabular}  \\
\hline
Failure  &  %%%%%%%%%%%%%%%%%%%%%%%%%
\begin{tabular}[c]{@{}l@{}}
Termination of an intended behavior of a system\\
due to a fault manifestation \cite{ISO26262}
\end{tabular}  \\
\hline
Fault  &    %%%%%%%%%%%%%%%%%%%%%%%%%
\begin{tabular}[c]{@{}l@{}}
Abnormal condition that can cause a system to \\
fail \cite{ISO26262}
\end{tabular}  \\
\hline
\begin{tabular}[c]{@{}l@{}}Unintended \\Behavior\end{tabular} &     %%%%%%%%%%%%%%%%%%%%%%%%%
\begin{tabular}[c]{@{}l@{}}
Behavior going beyond the intended behavior of \\
a system due to functional insufficiencies
\end{tabular}  \\
\hline
\begin{tabular}[c]{@{}l@{}}Functional\\Insufficiency\end{tabular} & %%%%%%%%%%%%%%%%%%%%%%%%%
\begin{tabular}[c]{@{}l@{}}
Incomplete specification or insufficient imple-\\
mentation of the intended functionality with an \\
unreasonable level of risk \cite{SOTIF}
\end{tabular}  \\
\hline
\begin{tabular}[c]{@{}l@{}}Triggering \\Condition\end{tabular}  &   %%%%%%%%%%%%%%%%%%%%%%%%%
\begin{tabular}[c]{@{}l@{}}
Specific conditions of a scenario that serve as an \\
initiator for a subsequent system reaction, \\
possibly leading to a hazardous behavior \cite{SOTIF}
\end{tabular}  \\
\hline
\begin{tabular}[c]{@{}l@{}}Safety-Critical\\Operational~\\Situation\end{tabular} &    %%%%%%%%%%%%%%%%%%%%%%%%%
\begin{tabular}[c]{@{}l@{}}
Traffic conditions (within the ODD), where a \\ 
hazard is very likely to propagate to a harm\\
\cite{ISO26262,SOTIF,DriveFI}
\end{tabular}  \\
\hline
\begin{tabular}[c]{@{}l@{}}Influential~\\(Scenario) \\ Factors\end{tabular}  &    %%%%%%%%%%%%%%%%%%%%%%%%%
\begin{tabular}[c]{@{}l@{}}
A parameterized scenario factor (e.g., sun \\
angle or road friction), which may affect \\
the performance of at least one AD function.
\end{tabular}  \\
\hline
Scenario &  %%%%%%%%%%%%%%%%%%%%%%%%%
\begin{tabular}[c]{@{}l@{}}
The temporal development between several \\
scenes in a sequence of scenes. ~\cite{Ulbrich2015}
\end{tabular}  \\
\hline
Scene   &   %%%%%%%%%%%%%%%%%%%%%%%%%
\begin{tabular}[c]{@{}l@{}}
A snapshot of the environment including the\\
scenery and movable objects, as well as all \\
actors’ and observers’ self-representations, and\\
the relationships among those entities.~\cite{Ulbrich2015}
\end{tabular}  \\                                                                                                                                                
\hline
\begin{tabular}[c]{@{}l@{}}Critical \\Scenario\end{tabular} &   %%%%%%%%%%%%%%%%%%%%%%%%%
\begin{tabular}[c]{@{}l@{}}
Scenarios that cause potential risks of harm,\\
which need explicit consideration for risk inves-\\
tigation and potential mitigation measures
\end{tabular}  \\                                                                 
\hline
CSI Method  &   %%%%%%%%%%%%%%%%%%%%%%%%%
\begin{tabular}[c]{@{}l@{}}
Methods to find triggering conditions, safety-\\
critical operational situations, or combinations\\
of the two that will lead to harm
\end{tabular}  \\
\hline
ODD  &      %%%%%%%%%%%%%%%%%%%%%%%%%
\begin{tabular}[c]{@{}l@{}}
Operational Design Domain: Operating condi-\\
tions under which a given ADS or AD feature\\
thereof is specifically designed to function \cite{J3016}. \\
It contains the set of all the influential factors \\
and the possible combinations of these factors.
%including, but not limited to, environmental, geo-\\
%graphical, and time-of-day restrictions, and/or \\
%the requisite presence or absence of certain traffic\\
%or roadway characteristics.\cite{SAE2002}
\end{tabular}  \\ 
\hline
\begin{tabular}[c]{@{}l@{}}Functional\\Scenario\end{tabular}&   %%%%%%%%%%%%%%%%%%%%%%%%%
\begin{tabular}[c]{@{}l@{}}
Scenario space representation on a semantic or \\
a high level of abstraction via linguistic \\
notations \cite{menzel2018scenarios}
\end{tabular}  \\                                                                                     
\hline
\begin{tabular}[c]{@{}l@{}}Logical \\Scenario\end{tabular}  &   %%%%%%%%%%%%%%%%%%%%%%%%%
\begin{tabular}[c]{@{}l@{}}
Scenario space representation on a state-space \\
level with parameter ranges \cite{menzel2018scenarios}
\end{tabular}  \\   
\hline
\begin{tabular}[c]{@{}l@{}}Concrete\\Scenario\end{tabular}  &   %%%%%%%%%%%%%%%%%%%%%%%%%
\begin{tabular}[c]{@{}l@{}}
A concretization of a logical scenario with\\
concrete parameter values \cite{menzel2018scenarios}
\end{tabular}  \\ 
\hline
\end{tabular}
\end{table}

%considered standards (SAE J3016, PAS1883, ISO21448)
%%%%%%%%%%%%%%%%%%%%%%%%%%%%%%%%%%%%%%%%%%%%%%%%%%
\subsection{ADS and ADAS}
\label{sec:ADS}
We adopt relevant ADAS and ADS terminology from the SAE J3016(TM) "Standard Road Motor Vehicle Driving Automation System Classification and Definition" \cite{J3016}.

\begin{figure}[!t]
\centering
\includegraphics[width=0.48\textwidth]{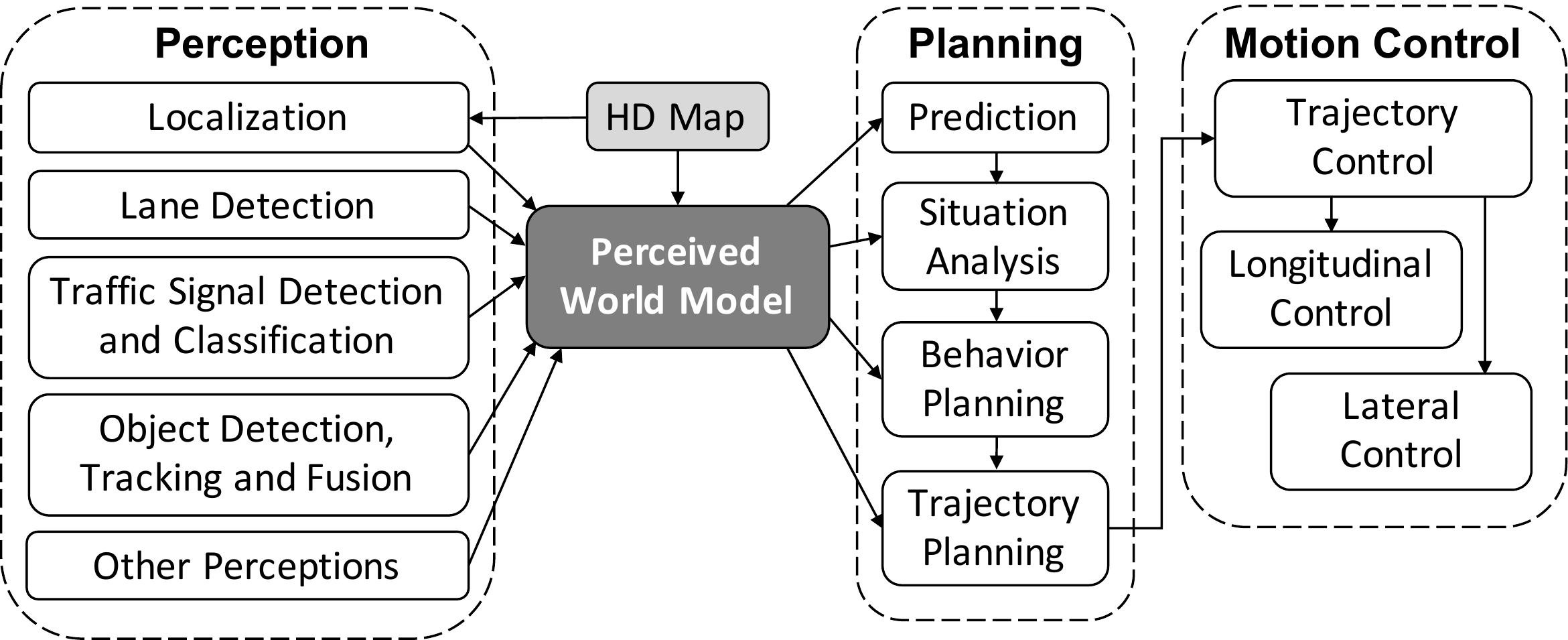}
\caption{Nominal AD functions according to Autoware\cite{AutoWare2}, Apollo\cite{Apollo} and Elektrobit open robinos\cite{EBOpenRob}}
\label{fig:AD_functions}
\end{figure}

\textbf{\textit{ADS}:} In J3016, the term ADS (Automated Driving System) is used specifically to describe a highly-automated driving system, which can perform the entire Dynamic Driving Task (DDT) on a sustained basis without human supervision.~\cite{J3016}.
Figure~\ref{fig:AD_functions} shows the main functions of ADS according Autoware\footnote{Autoware: \url{https://www.autoware.auto/}} \cite{AutoWare,AutoWare2}, Apollo\footnote{Apollo: \url{https://apollo.auto/developer.html}} \cite{Apollo} and Elektrobit open robinos\footnote{Elektrobit open robinos: \url{https://www.elektrobit.com/products/automated-driving/eb-robinos/}} \cite{EBOpenRob}.

\textbf{\textit{ADAS}:} The term “Advanced Driver Assistance Systems” (ADAS) is defined in J3016~\cite{J3016} to describe a broad range of features. ADAS system also includes the \textit{active safety systems}, which are excluded from the scope of this driving automation taxonomy because they do not perform part or all of the DDT on a sustained basis, but rather provide momentary intervention during potentially hazardous situations, such as lane keeping assistance (LKA) systems and automatic emergency braking (AEB) systems.

In our paper, we used the definition of levels of automation from the SAE J3016~\cite{J3016} to classified the systems in this survey into L3+ (with a human driver), L3- (without a human driver), and active safety system (no continues control of the vehicle). Detailed definitions of these three classes are given in Section \ref{sec:SoI}.

%Brief introduction of ADS and ADAS.

%\textbf{\textit{Level of automation}:} 

%skip the definition of ADAS, only keep ADS and active safety systems
%Levels of automation

%ADS: L3+ (Driver behavior is not considered and the system has both longitudinal and lateral controls) L3- (driver behavior is considered or the system has only longitudinal control), continuous control --> ADS, not continuous control --> active safety. 
%Levels of Driving Automation
 %Based on the SAE's classification, autonomous driving technology is divided into six levels, L0-L5. L0 represents traditional human driving without the addition of autonomous driving, while L1-L5 is graded with the technical configuration of autonomous driving. L1-L5 are assisted driving, partially autonomous driving, conditional autonomous driving, highly autonomous driving, and fully autonomous driving, respectively. In short, L0 to L2 is ADAS, and L3 and above levels are ADS.

%Challenges for design, verification and validation of AD and ADAS

%%%%%%%%%%%%%%%%%%%%%%%%%%%%%%%%%%%%%%%%%%%%%%%%%%
\subsection{Safety and the Sources of Harm}
\label{sec:source_harm}
\begin{figure*}[!t]
\centering
\includegraphics[width=0.85\textwidth]{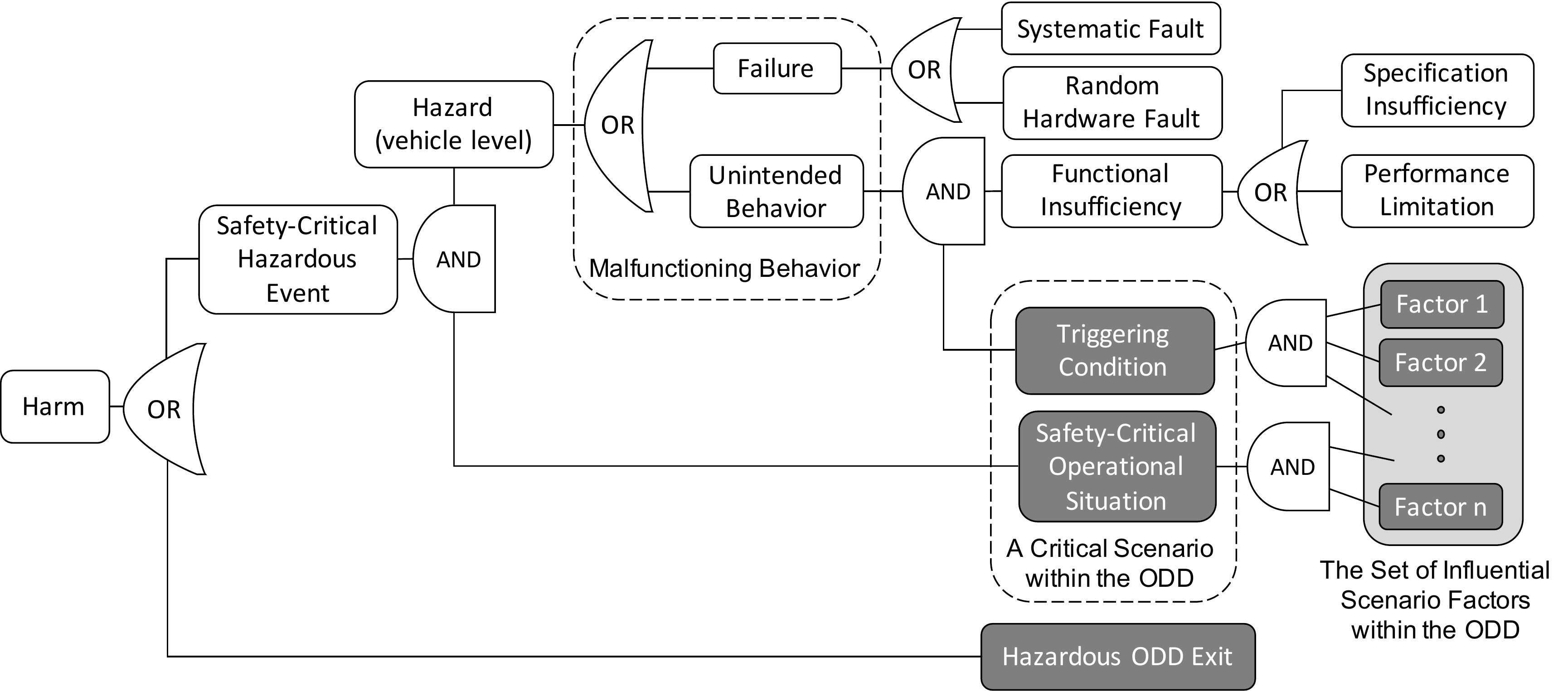}
\caption{Sources of harm according to ISO 26262, ISO/PAS 21448 and SAE J3016} 
\label{fig:SOTIF}
\end{figure*}

In this paper, safety is considered as a combination of Functional Safety (FuSa) \cite{ISO26262} and the Safety Of The Intended Functionality (SOTIF) \cite{SOTIF} within a specified operational design domain (ODD) \cite{J3016}. According to ISO 26262 \cite{ISO26262} and ISO/PAS 21448 \cite{SOTIF}, FuSa and SOTIF have the goal to mitigate residual risk associated with different sources of harm. Extending the discussion in \cite{Torngren2018a}, Fig. \ref{fig:SOTIF} illustrates the potential sources of harm considered in this paper.

FuSa considers the malfunctioning behavior caused by systematic faults and random hardware faults. It assumes that the design intent of the system is safe. In other words, it assumes that if the vehicle implementation perfectly follows its specification, the vehicle is safe. However, this assumption is not always valid for ADS for the following two reasons: 

\textbf{Reason 1:} Due to the openness of the driving environment, the ADS may be exposed to a nearly infinite set of environmental conditions. Therefore the specifications of AD functions may not be sufficient (i.e., not all the conditions are completely considered, e.g., an incomplete traffic sign beside the road). Consequently, some unidentified conditions may lead to safety problems.

\textbf{Reason 2:} Deviation from the intended functionality can also stem from inevitable performance limitations of some advanced functions due to the complexity or the inductive nature (e.g., learning-based methods) of the employed algorithm.

To this end, SOTIF complements FuSa by considering the malfunctioning behaviors stemming from functional inefficiencies, which include the aforementioned insufficient specifications and performance limitations. As shown in Fig. \ref{fig:SOTIF}, a functional insufficiency will lead to an unintended behavior of that function (e.g., mis-detection of the front vehicle) under a certain triggering condition (e.g., glare). If this unintended behavior is not resolved by the resilience of the downstream functions (e.g., object tracking and sensor fusion) \cite{DriveFI}, it may propagate to a vehicle level hazard (e.g., a failure to detect a pedestrian implying that braking is not initiated in time). 

A hazard may propagate to harm if it occurs in a safety-critical operational situation (e.g., the ego vehicle is running on a highway with a small distance to the front vehicle). For ADS with lower levels of automation, this propagation also depends on the reaction of the involved person (e.g., the driver of an L2 ADS).

Note that the propagation from an unintended behavior to a vehicle-level hazard is not necessarily deterministic owing to the resilience of the system. For example, if the camera-based vehicle detection function fails to detect the front vehicle at one frame because of a sudden glare, this unintended behavior will very likely be fixed by the downstream object tracking function. In addition, hazards stemming from the triggering conditions may not lead to a harm without a safety-critical operational situation. 

Another source of harm considered in \cite{Torngren2018a} is the misbehavior of other traffic participants. In our framework, an ADS should be able to tolerate a certain level of misbehavior of other traffic participants. The set of all tolerable behaviors should be covered in the (functional) ODD. If an ADS could not survive a tolerable misbehavior of another traffic participant, it will be considered as a functional insufficiency. \footnote{If this misbehavior is explicitly specified in the logical scenario, it will be considered as a performance limitation, otherwise it will be considered as a specification insufficiency.} If the other traffic participants behave beyond the ODD and cause a safety problem, it will be considered as a hazardous ODD exit. However, as far as the authors know, there is no method to completely define an ODD, therefore the boundary between functional insufficiency and ODD exit might be vague.

%%%%%%%%%%%%%%%%%%%%%%%%%%%%%%%%%%%%%%%%%%%%%%%%%%
\subsection{Scenarios}

The term "scenario-based testing" was first applied to the development of software systems. ~\cite{carroll1995scenario} A standardized definition of the term scenario in the context of verification and validation of automated vehicles is introduced in ISO/PAS 21448 \cite{SOTIF}.

\subsubsection{Scenario, Scene and ODD}
\label{sec:scenario_def}

In this paper, the definitions of scenario and scene are directly reused from ~\cite{Ulbrich2015} as follows.

\textbf{\textit{Scenario}:} “A scenario describes the temporal development between several scenes in a sequence of scenes. Every scenario starts with an initial scene. Actions \& events as well as goals and values may be specified to characterize this temporal development in a scenario. Other than a scene, a scenario spans a certain amount of time.” ~\cite{Ulbrich2015} As assumed in \cite{SOTIF} and illustrated in Fig. \ref{fig:SOTIF}, a scenario can be described by a set of influential factors.

\textbf{\textit{Scene}:} “A scene describes a snapshot of the environment including the scenery and movable objects, as well as all actors’ and observers’ self-representations, and the relationships among those entities. Only a scene representation in a simulated world can be all-encompassing (objective scene, ground truth). In the real world it is incomplete, incorrect, uncertain, and from one or several observers’ points of view (subjective scene).” ~\cite{Ulbrich2015}

All the relevant scenarios that conform to the same scenario description compose a scenario space. An important scenario space defined in \cite{J3016} is the Operational Design Domain (ODD), within which, the ego vehicle is supposed to drive safely. The definition of ODD is as follows:

\textbf{\textit{ODD}:} “Operating conditions under which a given driving automation system or feature thereof is specifically designed to function, including, but not limited to, environmental, geographical, and time-of-day restrictions, and/or the requisite presence or absence of certain traffic or roadway characteristics”.\cite{J3016}
ODD essentially defines the operating environment, for which an ADS is designed. 

The Pegasus project~\cite{winner2019pegasus} presented a 6-layer \textit{scenario description model} to categorize the influential factors for scenario and ODD description. The 6 layers are introduced in TABLE \ref{tab:6_layers}. To facilitate the discussion in later sections, an additional layer L0 (Internal Ego-vehicle) is added to describe the internal states of the ego-vehicle and the behavior of the driver. Detailed description of these layers can be found in TABLE \ref{tab:6_layers}.
%content:
%6-layer model
\begin{table*}
\centering
\caption{The 6-layer scenario description model \cite{weber2018}}
\label{tab:6_layers}
\begin{tabular}{|l|l|l|} 
\hline
\multicolumn{1}{|c|}{Layer} & \multicolumn{1}{c|}{Name}           & \multicolumn{1}{c|}{Description and Examples~ ~}                                                                                                                    \\ 
\hline
L0                          & Internal ego-vehicle                & \begin{tabular}[c]{@{}l@{}}Internal states of the other systems of the ego-vehicle \\other than the SOI + the states or behavior of the driver\end{tabular}         \\ 
\hline
L1                          & Road-level                          & Road topology, road surface, road elevation (slope, waviness)~ ~                                                                                                    \\ 
\hline
L2                          & Traffic infrastructure              & Including traffic rules                                                                                                                                             \\ 
\hline
L3                          & Temporary manipulation of L1 and L2 & Geometry, topology (overlaid)~ ~                                                                                                                                    \\ 
\hline
L4                          & Objects                             & \begin{tabular}[c]{@{}l@{}}Moving objects (traffic-related objects, e.g. vehicles, pedestrians, \\movement relative to the vehicle being measured)~ ~\end{tabular}  \\ 
\hline
L5                          & Environment                         & Weather, wind (may affect control), temperature (may affect camera)                                                                                                 \\ 
\hline
L6                          & Communication~ ~                    & V2X information, digital map                                                                                                                                        \\
\hline
\end{tabular}
\end{table*}

%Layer 4:
%introduce the terms:
%Ego vehicle: the vehicle equipped with the SOI.
%Other traffic participants:
   % Dynamic: e.g., vehicles, pedestrians, bikes,
  %  Static: parked vehicle, 

\subsubsection{Scenario Representation}
\label{sec:scenario_rep}

Menzel et al.~\cite{menzel2018scenarios} classified scenario representations into three levels of abstraction, namely functional scenario, logical scenario and concrete scenario. Functional scenario and logical scenario describe scenario spaces on two different levels of abstraction, while concrete scenario describes a particular scenario.

\textbf{Functional scenario:} A scenario space representation on a semantic level with linguistic scenario notations.~\cite{menzel2018scenarios} "The vocabulary used for the description of functional scenarios is specific for the use case and the domain and can feature different levels of detail.”~\cite{menzel2018scenarios}

\textbf{Logical scenario:} A scenario space representation on a state-space level with parameter ranges in the state space. Each parameter correlates to one influential factor. The parameter ranges can optionally be specified with probability distributions. A logical scenario includes a formal notation of the scenario space.~\cite{menzel2018scenarios} 
Additionally, "the relations of the parameter ranges can optionally be specified with the help of correlations or numeric conditions.”~\cite{menzel2018scenarios} 
In this paper, it is assumed that a logical scenario cannot fully reflect its corresponding functional scenario since relevant parameters cannot be completely listed.

\textbf{Concrete scenario:} A parameterized representation of a particular scenario. Each concrete scenario is an instantiation of a logical scenario, with a concrete value for each parameter.~\cite{menzel2018scenarios}

According to a concrete scenario, an \textbf{executable scenario} can be constructed, which can be either a simulation model or a real test. An executable scene refers to an image from a camera or a point cloud from a LiDAR.

In the rest of this paper, functional scenario, logical scenario and concrete scenario are also used to denote a scenario space or a scenario represented with the corresponding levels of abstraction. Fig. \ref{fig:LoA} depicts the transitions between the three levels of abstraction, which are defined as follows:

%Explain reasoning, formalization, instantiation, improvisation and refinement.

\begin{figure}[!t]
\centering
\includegraphics[width=0.45\textwidth]{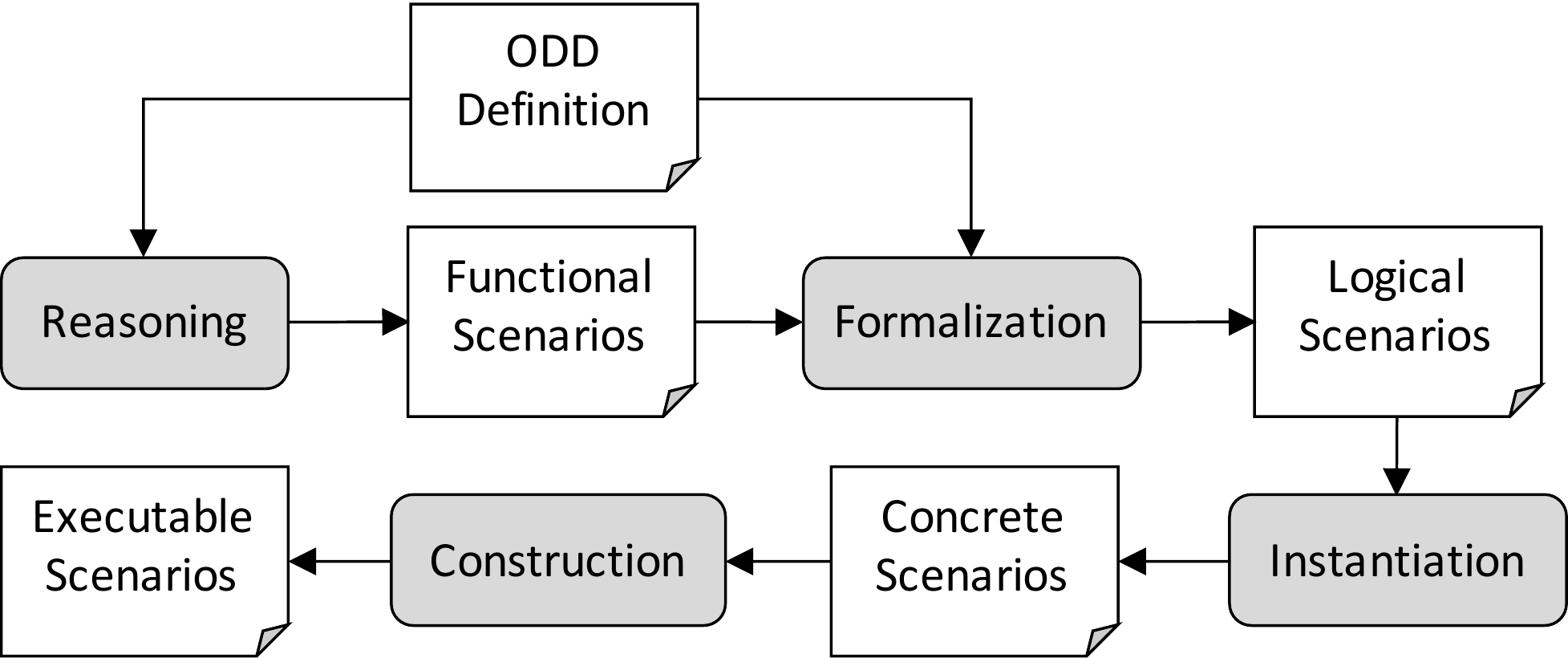}
\caption{Relationships between scenario description at different levels of abstraction} \label{fig:LoA}
\end{figure}

%\begin{itemize}
\textbf{Reasoning:}
This refers to the methods that reason (inductively or deductively) about critical functional scenarios based on knowledge, experience, and information described in the ODD.

\textbf{Formalization:} 
Using input from the ODD definition, a functional scenario will be formalized and parameterized to a logical scenario with all the parameter definition and their value ranges. In this paper, even though the formalized logical scenario contains more information than its functional scenario, it represents a smaller scenario space since not all the influential factors may have been identified and considered. 

\textbf{Instantiation:}
In this phase sampling or optimization approaches are used to instantiate the concrete scenario by naive search or guided search methods.

\textbf{Construction:}
The concrete scenarios will be converted into executable scenarios with the help of formats like OpenX (OpenDRIVE, OpenSCENARIO)\cite{ASAM}, for use in simulators.
%which fcan be directly executed in simulator.
%\end{itemize} 

Since an ODD is also a scenario space, ODD definitions can also be classified as functional ODD or logical ODD, explained as follows:
\begin{itemize}
\item \textbf{Functional ODD:} It describes the entire intended ODD on a high level of abstraction. e.g. a particular highway in Stockholm in sunny weather.
\item \textbf{Logical ODD:} It refers to a parameterized ODD description. It can support the design of the ODD exit detection algorithm. It can also support the formalization of functional scenarios.
\end{itemize} 

Similar as the relation between a functional scenario and its logical scenario, functional ODD represents a bigger scenario space. The misalignment between the functional ODD and the logical ODD constitutes a major source of specification insufficiency. 
In the rest of this paper, ODD refers to the functional ODD.
In Fig. \ref{fig:LoA}, it is the logical ODD that supports the formalization of a functional scenario.

%%%%%%%%%%%%%%%%%%%%%%%%%%%%%%%%%%%%%%%%%%%%%%%%%%
\subsection{Critical Scenarios}
\label{sec:critical_sce}
%introduce our definition of critical scenarios
%definitions: corner case, edge case, critical scenario (from UL and other standards)

The concept of a critical scenario is defined in a number of (different) ways in the research literature. Some papers also use related terms (sometimes used as direct synonyms), such as edge case or corner case. This section lists the standard definitions of these synonyms and provides our definition of critical scenario.
%introduction of some definition of edge corner cases critical. and 26262 probability (our definition incl. both edge and corner cases ). why to cite these two definition from the references. propose our definition of critical scenario from figure 3.

\textbf{Corner case and Edge case:}
In many cases, corner cases and edge cases are related terms and are often used as synonyms. The probability of occurrence is the most significant difference between them. Corner cases are combinations of normal operational parameters and a rare or unusual condition~\cite{SOTIF}. \cite{Koopman2019b} states that not all edge cases are corner cases, and vice versa. Only corner cases with a special combination of conditions that are both uncommon and novel are considered edge cases. %A rare situation that will occur only occasionally, but still needs specific design attention to be dealt with in a reasonable and safe way\cite{UL4600}. 
According to \cite{Koopman2019b} the definition of rare is relative, and it generally refers to situations or conditions that occur frequently enough in a fully deployed fleet to be a problem, but are not documented during the design or requirements process.

%\textbf{Corner case:}

\textbf{Critical scenario:}
In this survey, critical scenarios are defined as relevant scenarios for system design, safety analysis, verification or validation, with a potential risk of harm. In ISO 26262 \cite{ISO26262}, risk of harm is defined based on the likelihood of the scenario and the severity of the consequential harm. In this survey, we consider critical scenarios as all the scenarios that may lead to harm. As shown in Fig.~\ref{fig:SOTIF}, a critical scenario may contain a triggering condition and a safety-critical operational situation.
%“We define critical scenarios as relevant scenarios for system design, safety analysis, verification and validation, with critical referring to a potential risk of harm if not considered appropriately”.(potential here means that it is possible that a set of scenarios will not pose unacceptable risk, but that this is not obvious, and requires an in depth analysis and potentially appropriate design measures)

%Our definition of critical scenario: relevant scenarios for safety analysis, verification or validation.

\subsection{Critical Scenario Identification Methods}

One major goal of ISO/PAS 21448 \cite{SOTIF} is to identify unknown critical scenarios, and thereafter make them safe. Triggering conditions and safety-critical operational situations are two major components of a critical scenario within an ODD. Therefore, an unknown critical scenario can stem from either an unknown triggering condition or an unknown safety-critical operational situation. In addition, Annex B2 of ISO/PAS 21448 \cite{SOTIF} further assumes that a scenario condition/situation can be modeled as a combination of several influential scenario factors (e.g., heavy rain, glare, slippery road surface, a sudden cut-in of another vehicle, etc.). Under this assumption, an unknown critical scenario can be attributed to either an unknown scenario factor or an unknown combination of known scenario factors. 

\begin{figure}[!t]
\centering
\includegraphics[width=0.48\textwidth]{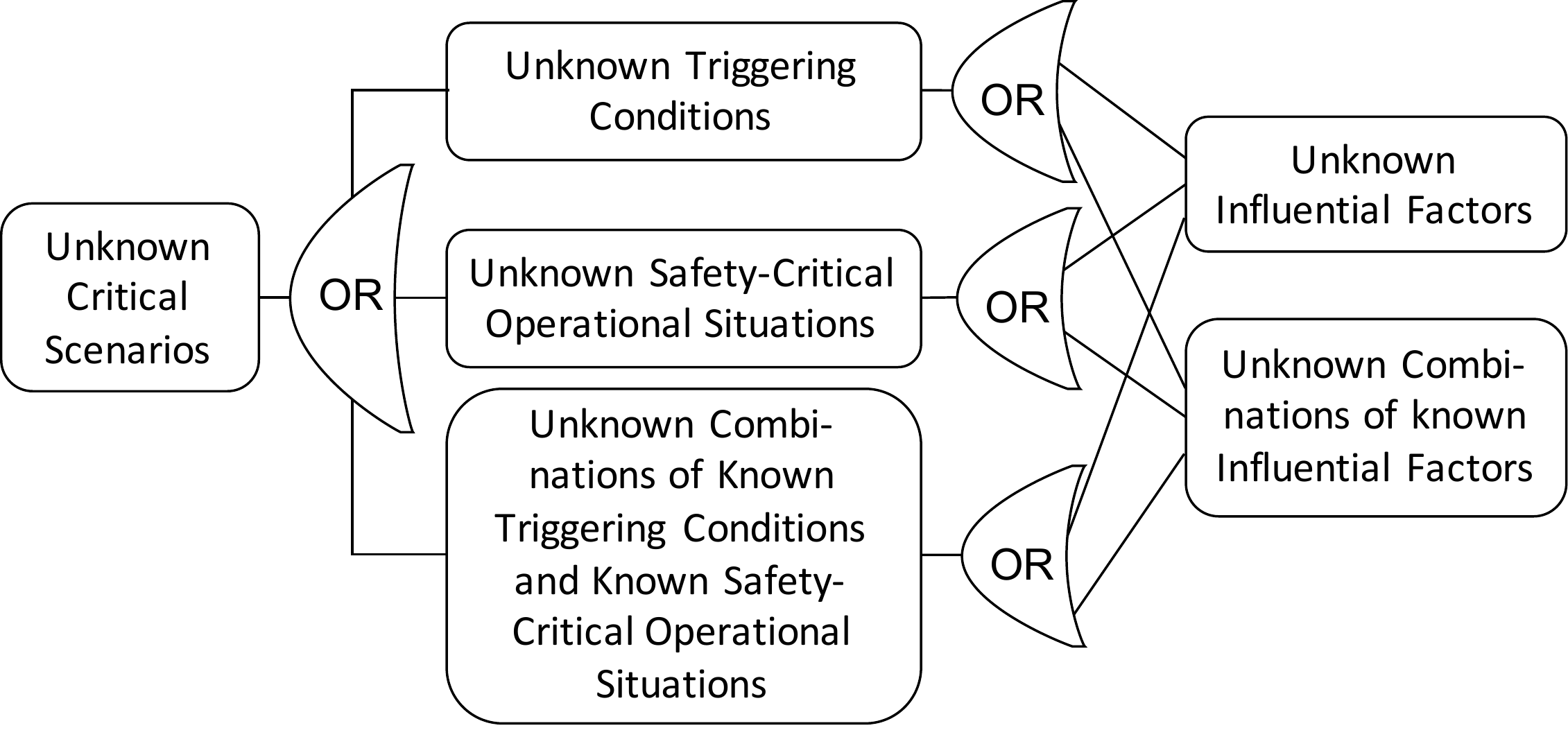}
\caption{Sources of unknown critical scenarios} 
\label{fig:source_unknown_CS}
\end{figure}

To this end, Critical Scenario Identification (CSI) Methods are defined as the methods to find triggering conditions, safety-critical operational situations, or combinations of the two that will lead to harm. An ODD definition is considered as the input of a CSI method to delimitate the scenario space.

\begin{figure}[!t]
\centering
\includegraphics[width=0.4\textwidth]{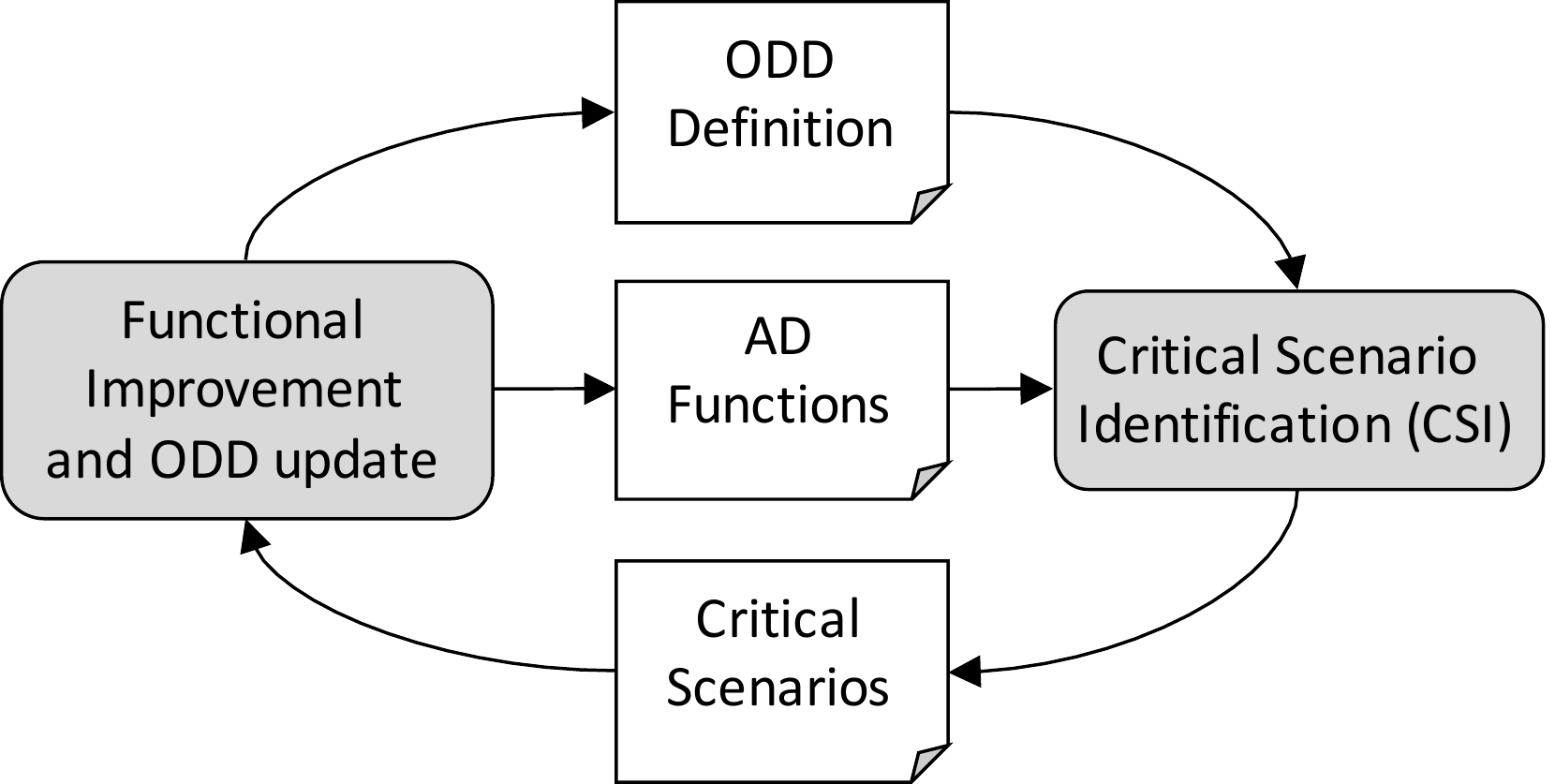}
\caption{An iterative process to improve the safety of the intended functionality}
\label{fig:ODD_CS}
\end{figure}

As shown in Fig.~\ref{fig:ODD_CS}, the identified critical scenarios will support the refinement of the automated driving functions to make the ADS safer. They may also help to complete the ODD definition, especially when an identified critical scenario points to an unconsidered aspect of the ODD definition. Meanwhile, functional refinement may also lead to an ODD change, which will initiate a new CSI process in the next iteration.

\subsection{Related Survey Papers}
\label{sec:related_survey}

While there is a vast amount of literature on various aspects of CSI methods, there are much fewer related survey papers.
Other related topics, although outside the scope of this review, are briefly introduced in Section \ref{sec:Other_Dir}. 

According to our literature search, to the best of our knowledge, related relevant survey papers are \cite{Neurohr2020} and \cite{Riedmaier2020}. 
Neurohr et al.~\cite{Neurohr2020} reviewed and analyzed the literature about the scenario-based testing method for automated vehicles. The authors presented fundamental arguments, principles and assumptions of the scenario-based approach. They also proposed a generic framework (from scenario elicitation to test evaluation) for scenario-based testing and analyzed in detail the individual steps based on the reviewed articles. As a result, they presented various considerations for using and implementing scenario-based testing to support automated vehicle homologation.

Riedmaier et al.~\cite{Riedmaier2020} performed a survey of scenario-based approaches for safety assessment of automated vehicles. The authors provided an overview of various approaches. They also developed a taxonomy for the scenario-based approach and compared the summarized methods with each other. In the end, this paper integrated the formal verification with the scenario-based approach as an alternative concept.

The differences between this paper and those two survey papers are:
\begin{itemize}
  \item This paper is dedicated to CSI methods, which is a subset of the focus of the other papers' scenario-based methods.
  \item This paper provides a systematic literature review.
   \item This paper provides a taxonomy for critical scenarios identification approaches.
\end{itemize}

% Relevant ODD factors summarized by P. Koopman et al. \cite{koopman2019} 
% - Topography, and associated location-related features.

% - Environmental and weather conditions.

% - Operational infrastructure, such as operational road surfaces, aids to navigation, traffic management equipment, avoidance zones, special road use rules and vehicle availability to infrastructure.

% - Rules and expected behaviour interrelated with other aspects of the environment and operational state space, including traffic regulations, social norms and customary signalling and negotiation procedures with other agents (both autonomous and human, including explicit signals as well as implicit signals controlled through vehicle movements).

% - Considerations for deployment in different regions/countries (e.g. blue stop signs, right turn keep moving stop sign modifiers, horizontal vs. vertical traffic signal directions, kerb changes).

% - Communication mode, bandwidth, latency, stability.

% - Availability and freshness of infrastructure characterisation data, e.g. mapping level of detail and identification of temporary deviations from baseline data (e.g. construction zones, traffic jams, hurricane evacuation, etc. temporary traffic rules).

% - Expected distribution of operational state space elements, including which elements are considered rare but within range (e.g. toll booths, police traffic stops), and which elements are considered outside the state space area in which the system is intended to operate.

\section{Literature Review Methodology}
\label{sec:methodology}
\begin{figure}[!t]
\centering
\includegraphics[width=0.42\textwidth]{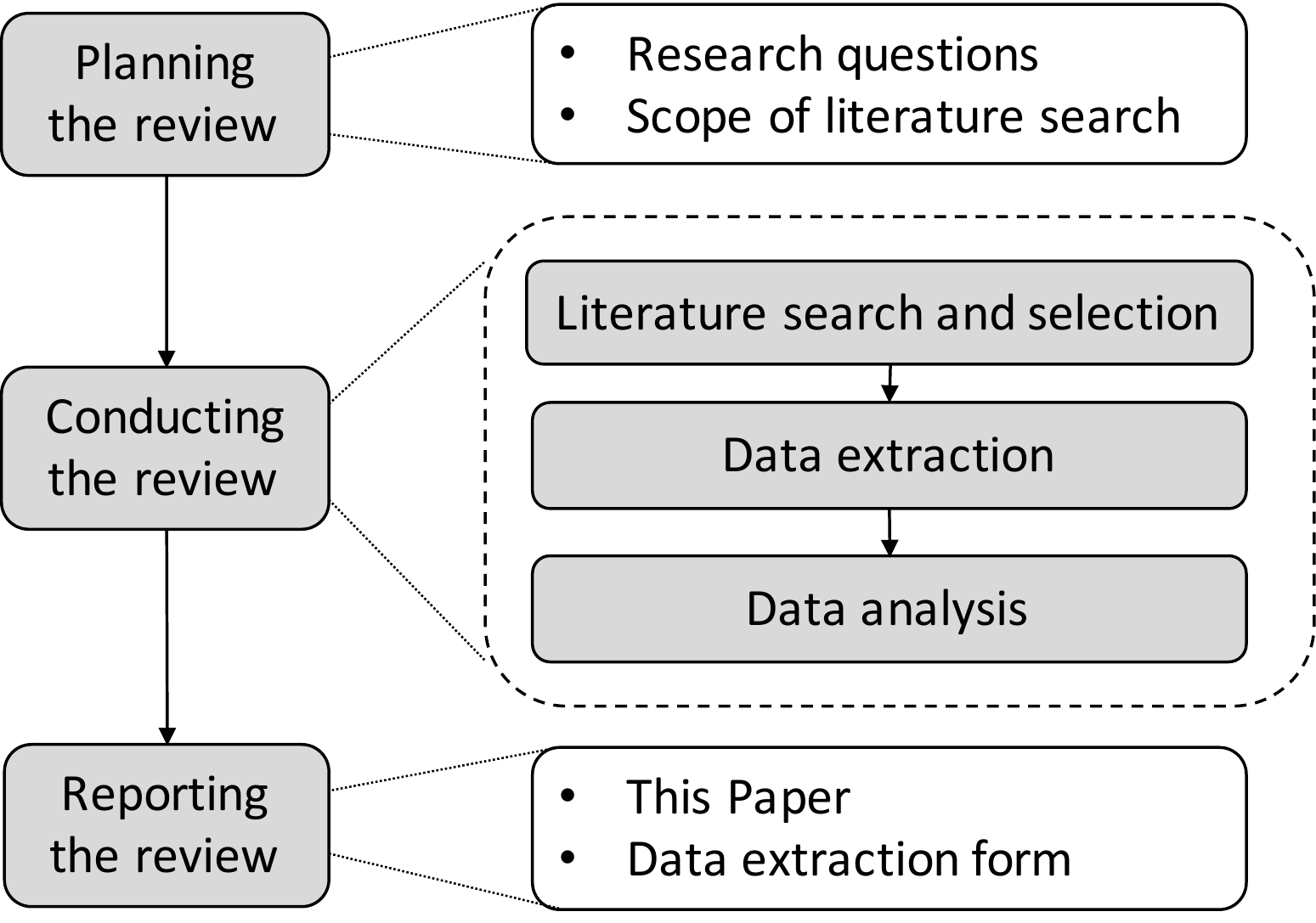}
\caption{Overview of the literature review process} \label{fig:LR_protocol}
\end{figure}

This literature review follows the guidelines proposed by Keele \cite{Kitchenham2007}, which divides the whole literature review process into three stages: planning, conducting and reporting. Based on this guideline, our literature review protocol is illustrated in Fig. \ref{fig:LR_protocol} and detailed in the rest of this section.

%%%%%%%%%%%%%%%%%%%%%%%%%%%%%%%%%%%%%%%%%%%%%%%%%%%%%%%%%%%%%%
\subsection{Planning the Review}
\label{sec:planning}
This section describes the research questions and the planned scope of this literature review.

As discussed in the previous sections, the goal of this paper is to propose a framework to categorize and analyze the CSI methods for ADS and ADAS. Based on this objective, the following research questions are derived.
\begin{enumerate}
%\item[\textbf{RQ1:}] How can the state-of-the-art CSI methods for ADS and ADAS be classified?
\item[\textbf{RQ1:}] What would be a taxonomy that allows to systematically categorize and compare state-of-the-art CSI methods for ADS and ADAS?
\item[\textbf{RQ2:}] What is the current status of CSI methods research with respect to this taxonomy?
\item[\textbf{RQ3:}] What are the remaining problems and challenges for further investigation?
\end{enumerate}

The taxonomy in RQ1 can provide a systematic structure of common characteristics to enable the classification and comparison of different CSI methods. With this taxonomy, further researchers or engineers can easily pinpoint a CSI method on the big picture, and thereby understand its strength and limitations. The answer to RQ2 presents the state of the art of CSI methods. It can also help new researchers or engineers in this field have a quick and comprehensive understanding of these methods. RQ3 tries to identify the further directions of this research topic.

To clarify the scope of this literature review, as suggested in \cite{Kitchenham2007}, the research questions are broken down into individual facets (Population, Intervention, Comparison, Outcomes and Context - PICOC \cite{Brereton2007}) as:
\begin{itemize}
\item \textbf{Population:} Software-intensive ADS or ADAS;
\item \textbf{Intervention:} Approaches applied during development time to identify Critical scenarios;
\item \textbf{Comparison:} Not applicable;
\item \textbf{Outcomes:} Improved safety;
\item \textbf{Context:} Peer-reviewed publications from academia and industry.
\end{itemize} 

The scope of this literature review can be further narrowed down by clarifying the definitions of scenario and criticality. As discussed in Section \ref{sec:scenario_def}, the definition of a scenario should follow the one given in \cite{Ulbrich2015}, and cover at least one layer of the 6-layer model\cite{weber2018}. The included studies must distinguish critical scenarios from other scenarios. General-purpose scenario modeling methods, such as \cite{Bach2016}, are excluded if they do not consider the identification or generation of critical scenarios. For a similar reason, we also excluded general-purpose data augmentation approaches for training machine learning models (e.g., ~\cite{AbuAlhaija2018,Johnson-Roberson2017}). 

%The identified critical scenarios should be be used in the development time to improve the design, verification or validation of a software-intensive system. 

In addition, since CSI methods play an important role for SOTIF \cite{SOTIF}, this literature review only considers the studies published after the initiation of the SOTIF standard, ISO/PAS 21448 (i.e. 2017).

To this end, the following inclusion and exclusion criteria are formulated. The included papers should satisfy all the inclusion criteria, and should not be covered by any of the exclusion criteria.  
\begin{enumerate}
\item[\textbf{I1:}] Studies describing approaches to identify critical scenarios for ADS or ADAS during development time;
\item[\textbf{I2:}] The scenario considered in the study should contain environmental aspects, i.e. content covered by the 6-layer model;
\item[\textbf{I3:}] The identified critical scenarios should serve the development of a software intensive system;
\item[\textbf{I4:}] Studies published between January 2017 to August 2020;
\item[\textbf{I5:}] Peer-reviewed studies written in English, and available in full text;
\item[\textbf{E1:}] Papers with main focus on cyber security;
\item[\textbf{E2:}] Approaches to identify misuse scenarios;
\item[\textbf{E3:}] On-line methods, e.g., online risk assessment;
\item[\textbf{E4:}] Approaches to identify critical scenarios for a hardware component, e.g., radar or LiDAR;
\item[\textbf{E5:}] Papers focusing on how to model scenarios and how to identify influential factors;
\item[\textbf{E6:}] Papers introducing a framework for scenario-based methodology instead of a particular method to identify critical scenarios;
\end{enumerate}

%%%%%%%%%%%%%%%%%%%%%%%%%%%%%%%%%%%%%%%%%%%%%%%%%%%%%%%%%%%%%%
\subsection{Conducting the review}
\label{sec:conducting}

As illustrated in Fig. \ref{fig:LR_protocol}, to answer the research questions, three research tasks were conducted.
This section provides the details of these three research tasks.

%%%%%%%%%%%%%%%%%%%%%%%%%%%%%%%%%%%%%
\subsubsection{Literature search and selection}
\label{sec:search_selection}

\begin{figure*}[!t]
\centering
\includegraphics[width=0.8\textwidth]{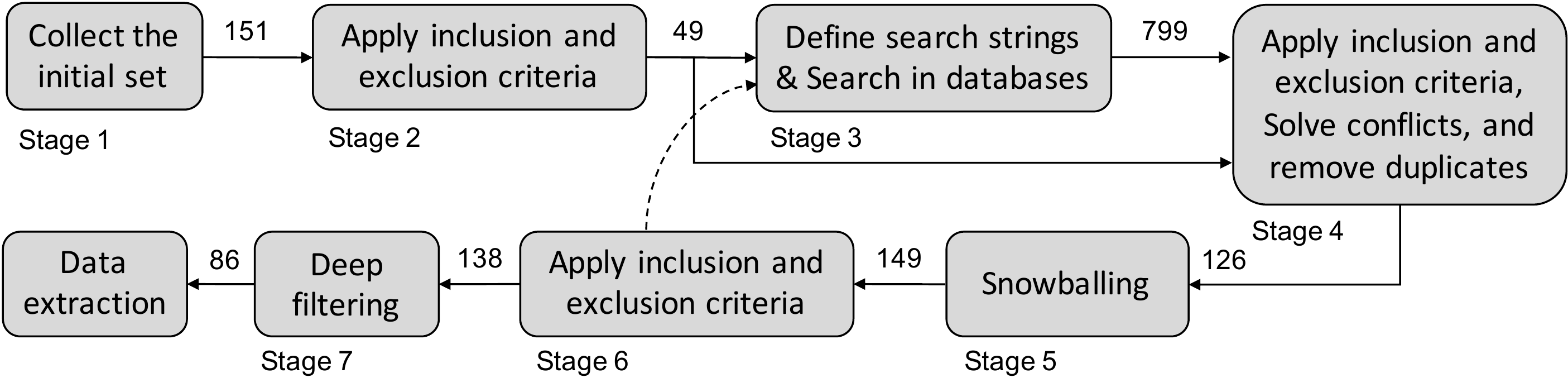}
\caption{The stages to collect the primary studies. The numbers on the arrows indicate the numbers of studies given to the next stage} 
\label{fig:search_selection}
\end{figure*}

The primary studies for this literature review are collected through an iterative process with automatic search and snowballing as shown in Fig. \ref{fig:search_selection}. This section describes the details of how we conducted each stage.

\textbf{Stage 1:}
A comprehensive initial search string can reduce the number of iterations to determine the final search string as described in \textbf{Stage 3}. To better define the initial search strings, an initial set of relevant studies was gathered from two sources. The first source was publications from recent relevant research projects including AutoDrive, Prystine, Pegasus, Enable-S3 and AdaptIVe. The second source are the relevant papers included in the two relevant survey papers \cite{Riedmaier2020,Neurohr2020} introduced in Section \ref{sec:related_survey}. The initial set contains 151 potentially relevant studies. After being filtered by the inclusion and exclusion criteria, 49 studies remained.

\textbf{Stages 2 and 6:}
To filter a given set of potentially relevant studies, we reviewed the title abstract and author keywords of each paper with respect to the inclusion and exclusion criteria defined in Section \ref{sec:planning}. After reviewing, each paper was labelled as either \textit{"included"}, \textit{"excluded"} (together with the violated criterion) or \textit{"unclear"}. Unclear studies were further checked by going through their introduction, conclusion and some other parts. When necessary, a discussion among multiple researchers would be conducted to determine the inclusion of an unclear study.

\textbf{Stage 3:}
We performed an automatic literature search with the same search string on different databases.

To cover as many relevant studies as possible, we conducted an automatic search on the five major electronic databases on computer science, electronic engineering and automotive technology, namely: IEEE Xplore Digital Library, ACM Digital Library, Scopus, ScienceDirect and SAE Mobilus.

According to the PICOC analysis presented in Section \ref{sec:planning}, the search string is designed as:
\[
    \textnormal{Search String} = \$AD \textbf{ AND } \$CS \textbf{ AND } \textnormal{"safe*"}
\]
where $\$AD$ denotes the identified synonyms of automated driving, and $\$CS$ represents the synonyms of critical scenario or CSI methods. These synonyms are listed in Table \ref{tab:synonyms}. $\$AD$ and "safe*"are searched within the whole text, and $\$CS$ is searched only in title, abstract and author keywords. After stage 7, new synonyms of $\$AD$ and $\$CS$ could be identified to extend the search string. This process was iterated three times until no more synonyms could be found. Table \ref{tab:synonyms} lists the final set of all the identified synonyms. The search result from each database is illustrated in the second column (Found) of Table \ref{tab:search_rst}.

\begin{table}[!t]
\caption{Synonyms in $\$AD$ and $\$CS$}
\label{tab:synonyms}
\centering
\begin{tabular}{|l|l|}
\hline
     & \multicolumn{1}{c|}{Synonyms} \\ 
\hline
$\$AD$ & \begin{tabular}[c]{@{}l@{}}\big((automated \textbf{OR} autonomous \textbf{OR} intelligent) \textbf{AND} \\ 
                                (vehicle \textbf{OR} driving)\big) \textbf{OR} "self-driving" \textbf{OR} ADAS\end{tabular}\\
\hline
$\$CS$ & \begin{tabular}[c]{@{}l@{}}\big((critical \textbf{OR} challenging \textbf{OR} risky \textbf{OR} “high-risk” \textbf{OR}\\ 
                                    hazardous \textbf{OR} complex \textbf{OR} "safety-relevant" \textbf{OR}\\ 
                                    accident) \textbf{AND} scenario\big) \textbf{OR} "corner case" \textbf{OR}\\ 
                                    "falsification" \textbf{OR} \big((stress \textbf{OR} adversarial) \textbf{AND} test*\big) \\ 
                                    \textbf{OR} "test* scenario" \textbf{OR} "test case"\end{tabular}       \\ 
\hline
\end{tabular}
\end{table}

\begin{table}[!t]
\centering
\caption{Literature search and selection result before snowballing}
\label{tab:search_rst}
\begin{tabular}{|l|c|c|c|c|} 
\hline
\textbf{Database} & \multicolumn{1}{l|}{\textbf{Found}} & \multicolumn{1}{l|}{\textbf{Selected}} & \multicolumn{1}{l|}{\textbf{Unclear}} & \multicolumn{1}{l|}{\textbf{Included}}  \\ 
\hline
Initial set       & 151                                 & 49                                     & 0                                     & 49                                        \\ 
\hline
ScienceDirect     & 29                                  & 3                                      & 2                                     & 4                                         \\ 
\hline
Scopus            & 261                                 & 61                                     & 27                                    & 67                                        \\ 
\hline
IEEE              & 311                                 & 35                                     & 10                                    & 32                                        \\ 
\hline
ACM               & 112                                 & 1                                      & 6                                     & 3                                         \\ 
\hline
SAE               & 152                                 & 14                                     & 19                                    & 18                                        \\ 
\hline
\textbf{Total}    & \multicolumn{2}{l|}{\textbf{Found: 929}}                                     & \multicolumn{2}{l|}{\textbf{Included: 126}}                                       \\
\hline
\end{tabular}
\end{table}

\textbf{Stage 4:}
Compared to stages 2 and 6, stage 4 had a much larger input set. To share the workload, and also to guarantee the correctness of the filtering, this stage was conducted by two researchers. Since Scopus has overlaps with all the other databases, we assigned one researcher for Scopus and the other for the rest of the databases. The percentage of conflicts (studies that were included by one researcher but excluded by the other) on the overlap was used as a metric to evaluate the explicitness of the criteria and the quality of the filtering. The result of this evaluation is analyzed in Section \ref{sec:TTV}.

Each conflict was resolved by a discussion among the two researchers conducting stage 4 and an additional senior researcher.

The result of this stage is listed in Table \ref{tab:search_rst}. Studies selected by this stage might also be excluded in \textbf{Stage 7}.

\textbf{Stage 5:}
Snowballing was conducted in parallel with deep filtering and data extraction. When reading through each study, relevant references were collected. The relevance of each reference was judged by its title and how it is described in the study under review. 

\textbf{Stage 7:}
The exclusion criteria \textbf{E5} and \textbf{E6} are sometimes difficult to evaluate by only reading the title and abstract. This stage examines the papers (selected by the previous stages) with a special focus on \textbf{E5} and \textbf{E6}.

%%%%%%%%%%%%%%%%%%%%%%%%%%%%%%%%%%%%%
\subsubsection{Data Extraction}
\label{sec:data_extraction}

\begin{figure}[!t]
\centering
\includegraphics[width=0.4\textwidth]{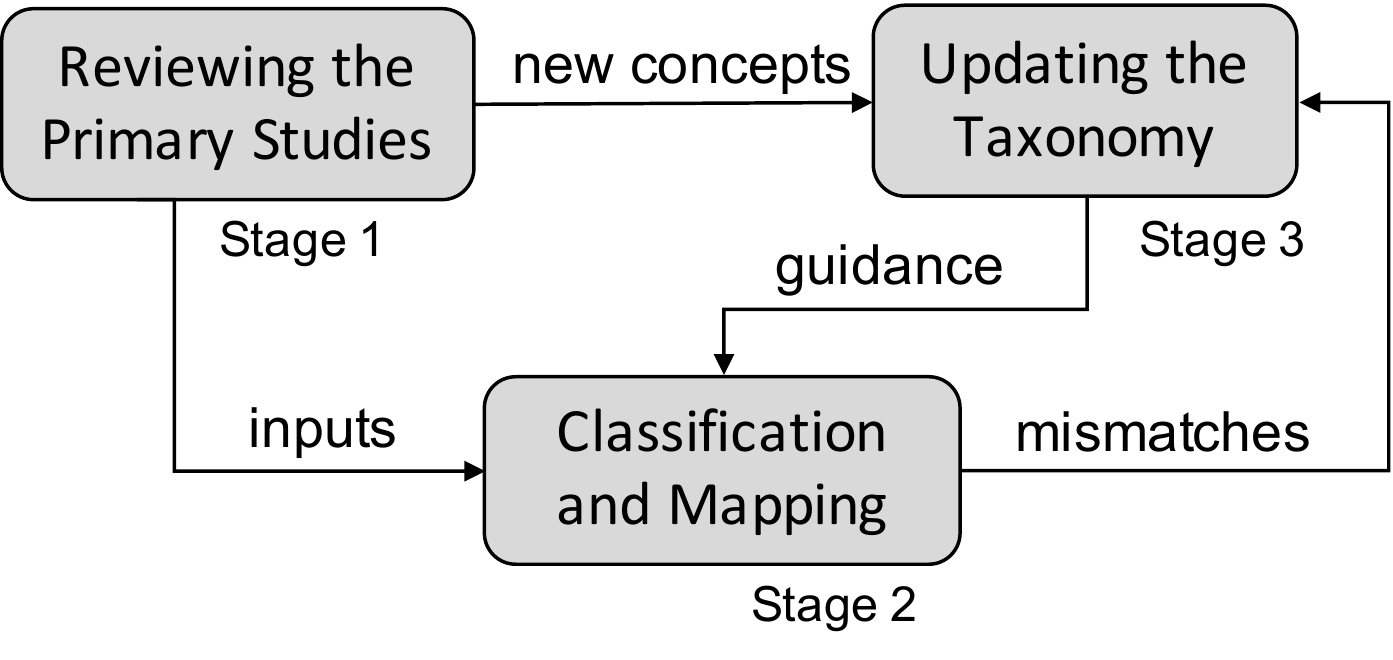}
\caption{The stages to extract data from primary studies} \label{fig:data_extraction}
\end{figure}

This task is to answer RQ1. Relevant information needs to be extracted from the primary studies according to a taxonomy. Meanwhile, the taxonomy needs to be updated during the extraction. To start, an initial taxonomy was proposed based on (1) the relevant industrial standards introduced in Section \ref{sec:Background}, (2) the concepts identified from the initial set of the primary studies, and (3) the previous project experience of the authors. The structure of the taxonomy was inspired by \cite{Aleti2013}. The initial taxonomy was, thereafter, iteratively updated when reviewing the primary studies, following the process illustrated in Fig. \ref{fig:data_extraction}. This section elaborates how each stage in this process was conducted.

\textbf{Stage 1:}
All the primary studies were thoroughly read by at least one researcher to extract information according to the taxonomy, and to identify new concepts to update the taxonomy. 

\textbf{Stage 2:}
As suggested by Keele \cite{Kitchenham2007}, the extracted information was coded according to the taxonomy and  documented in a data extraction form, which can be found in \cite{review_table}. The result for each study was reviewed by another two researchers to guarantee correctness. For the information that could not be explicitly found in the study, it was marked as "not given" in the data extraction form and if possible, a reasonable assumption was given according to our understanding.

\textbf{Stage 3:}
The taxonomy was iteratively updated when (1) a new concept, which had not been included in the taxonomy, was identified from stage 1; or (2) a mismatch (i.e. a study that cannot be reasonably classified by the taxonomy) was detected from stage 2. When the taxonomy was updated, the extracted data would also be updated according to the new taxonomy. 

Section \ref{sec:Tax} presents the final taxonomy, and the data extraction form can be found online as \cite{review_table}. This task is also the basis of the next task to answer RQ2 and RQ3.

%%%%%%%%%%%%%%%%%%%%%%%%%%%%%%%%%%%%%
\subsubsection{Data Analysis}
\label{sec:clustering_analysis}

The goal of this task is to answer RQ2 and RQ3. In the previous task, information was extracted from each primary study according to the taxonomy. This task analyzes the extracted information with systematic and statistical approaches. The analysis results answer RQ2. The findings from the analysis answer RQ3. The analysis results and findings are presented in Sections \ref{sec:Problem}, \ref{sec:Solution} and \ref{sec:Validation} and discussed in Section \ref{sec:Diss}. 
%%%%%%%%%%%%%%%%%%%%%%%%%%%%%%%%%%%%%%%%%%%%%%%%%%%%%%%%%%%%%%
\subsection{Reporting the review}
\label{sec:Reporting}
The reporting contains two parts. The first part is this paper, which presents the relevant terminology, the literature review methodology, and the answers to the research questions. The second part is the data extraction form \cite{review_table}, which summarizes each primary study according to the taxonomy.

%%%%%%%%%%%%%%%%%%%%%%%%%%%%%%%%%%%%%%%%%%%%%%%%%%%%%%%%%%%%%%
\subsection{Threats to Validity}
\label{sec:TTV}

This section discusses the potential threats to the completeness of the literature search and our mitigation approaches. 
One of the main threats to the validity of this systematic literature review is incompleteness. The risk of this threat highly depends on the selected search string, the limitation of the employed search engine, and the quality of the literature filtering (stages 2,4 and 6 in Fig. \ref{fig:search_selection}). To reduce the risk of an incomplete search string, the search string was constructed iteratively until no new synonyms of $\$AD$ and $\$CS$ could be added to Table \ref{tab:search_rst}. In order to omit the limitations of the employed search engines, multiple search engines were used on different databases with overlaps. To avoid the exclusion of relevant papers during literature filtering, the label \textit{unclear} was introduced for the case where the inclusion of a paper could not be determined with high confidence. This label guaranteed that papers were excluded with high confidence. In addition, the risk of incorrect exclusion could be implicitly evaluated by the number of conflicting papers (i.e. papers that were included by one researcher but excluded by the other) within the overlap between Scopus and other databases. Among the total 77 overlapping papers, there were only 3 conflicts. After further confirmation, only 1 paper (out of 3) was included.

Another important threat is the robustness of the taxonomy to describe any CSI method. To guarantee the taxonomy has sufficient concepts to cover all the selected papers, an iterative approach was conducted to update the taxonomy and the extracted data until the taxonomy converged. Furthermore, the capability of the taxonomy to describe a new paper was verified by the 9 papers (after deep filtering) found in the snowballing phase. All these papers were successfully classified by the taxonomy without the need to add any new concepts.

%%%%%%%%%%%%%%%%%%%%%%%%%%%%%%%%%%%%%%%%%%%%%%%%%%%%%%%%%%%%%%

\section{Overview of the Taxonomy}
\label{sec:Tax}

\begin{figure*}
\centering
\includegraphics[width=0.8\textwidth]{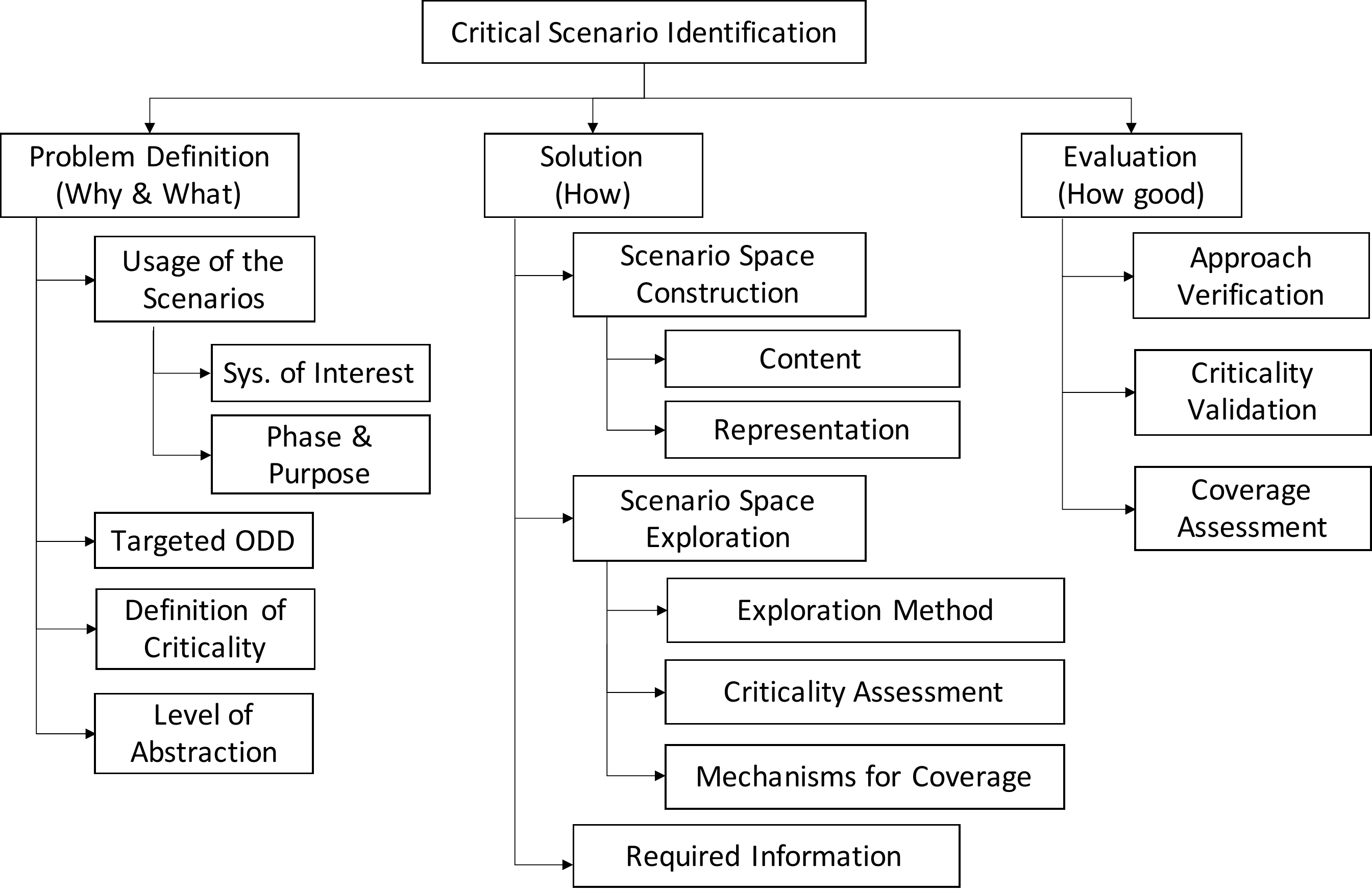}
\caption{The taxonomy for critical scenarios identification approaches, achieved from the reviewed literature} \label{fig:Taxonomy}
\end{figure*}

% \begin{figure*}
% \centering
% \includegraphics[width=0.6\textwidth]{Information_flow}
% \caption{A generic information flow to find critical scenarios} \label{fig:Info_flow}
% \end{figure*}
Employing the methodology introduced in Section \ref{sec:data_extraction}, a hierarchical taxonomy for CSI methods was identified as depicted in Fig. \ref{fig:Taxonomy}.

Inspired by \cite{Aleti2013}, the first level of the taxonomy hierarchy structures the studied CSI approaches with the following three fundamental categories, which reflect the common logic of a complete research.
\begin{itemize}
\item \textbf{\textit{Problem Definition}:} What kind of scenario is being identified? Why are these identified scenarios important? (Section \ref{sec:Problem})
\item \textbf{\textit{Solution}:} What techniques are applied to identify the critical scenarios? What external information / data is needed? (Section \ref{sec:Solution})
\item \textbf{\textit{Evaluation}:} How are the validity of the approach and the identified critical scenarios assessed? (Section \ref{sec:Validation})
\end{itemize} 
Each of these three top-level categories is decomposed into multiple subcategories. Each leaf category in the taxonomy has a number of possible values to categorize CSI methods. The rest of this section introduces all the subcategories.

%%%%%%%%%%%%%%%%%%%%%%%%%%%%%%%%%%%%%%%%%%%%%%%%%%
\subsection{Subcategories of \textit{Problem Definition}}
\label{sec:tax_prob_def}
The \textit{Problem Definition} category specifies the problem that the approaches aim to solve. It is decomposed into the following subcategories.

\textbf{\textit{Usage of the Scenarios}:} 
This category classifies the CSI methods according to how the identified critical scenarios are supposed to be used. Subcategories include 1) the System of Interest (SoI), i.e. the AD system/function whose development is supported by the identified scenarios; and 2) how are the identified critical scenarios used in different development phases. For example, Li et al. \cite{Li2020} proposed a method to find critical scenarios as test cases (\textit{purpose}) for the verification (\textit{phase}) of a whole ADS (\textit{SoI}).

\textbf{\textit{Target ODD}:} 
As depicted in Fig. \ref{fig:LoA}, a clear and sufficient ODD definition is necessary for the reasoning of critical functional scenarios and the formalization of a functional scenario to a logical scenario. This category analyzes how ODD is defined and used in each primary study.

\textbf{\textit{Definition of Criticality}:} 
The studied CSI methods aim to select or generate critical scenarios, which are distinguished from other scenarios from a specific perspective. Even though the definitions of criticality are not explicitly given  in most of the primary studies, this category explicitly classifies the criticality according to the characteristics of the identified scenarios.

\textbf{\textit{Level of Abstraction}:} 
This category classifies the studied approaches according to their inputs and outputs in terms of the levels of abstraction of their scenario representation, as described in Section \ref{sec:scenario_rep}. For example, the approach proposed by Li et al. \cite{Li2020} identifies critical concrete scenarios within a given logical scenario. Therefore this approach is classified as "\textit{logical $\rightarrow$ concrete}". Similar approaches to identify critical scenarios according to a scenario with a higher level of abstraction are called deductive approaches. In contrast, inductive approaches find critical scenarios based on a set of lower-level scenarios.

%%%%%%%%%%%%%%%%%%%%%%%%%%%%%%%%%%%%%%%%%%%%%%%%%%
\subsection{Subcategories of \textit{Solution}}
The category \textit{Solution} classifies the primary studies according to how they find critical scenarios within the constructed scenario space. This section explains all its subcategories.

\textbf{\textit{Scenario Space Construction}:}
To systematically identify critical scenarios, scenarios need to be consistently represented to construct a scenario space. This category classifies the CSI methods in terms of 1) \textbf{\textit{Content}:} what aspects are included in the scenario model regarding the 6-layer model introduced in Table \ref{tab:6_layers}; and 2) \textbf{\textit{Representation}:} how they are represented (e.g., formally, semi-formally or qualitatively). %\textit{Representation} also summarizes how OpenX standards (as introduced in Section \ref{sec:OpenX}) are considered in the primary studies.

\begin{figure}
\centering
\includegraphics[width=0.42\textwidth]{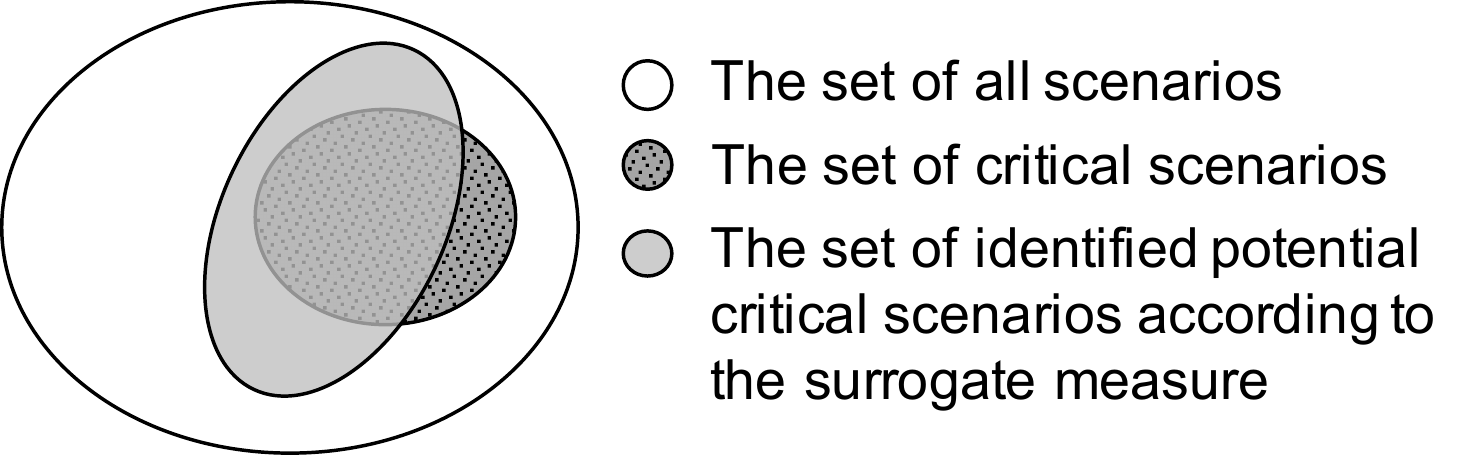}
\caption{Relationship between critical scenarios and potential critical scenarios identified according to a surrogate measure} \label{fig:surrogate_measure}
\end{figure}

\textbf{\textit{Scenario Space Exploration}:}
Under this category, the studied CSI approaches are categorized in terms of 1) which algorithm or technique is adopted to explore the scenario space; 2) how the criticality of a scenario is assessed during the exploration; and 3) what mechanisms are used to guarantee or improve the coverage of the exploration. Since the criticality of a scenario can hardly be directly measured, a surrogate measure is commonly used for \textit{criticality assessment}. One common example of a surrogate measure is the time-to-collision (TTC) with the front vehicle in simulation. However, since a surrogate measure cannot completely reflect the criticality of a scenario, as shown in Fig.~\ref{fig:surrogate_measure}, the set of potential critical scenarios (identified according to a surrogate measure) may slightly mismatch with the set of all the critical scenarios in reality. The \textit{criticality assessment} category collects all the surrogate measures used in the primary studies, and maps them to different criticality definitions.

\textbf{\textit{Required Information}:} This category summarizes the necessary information each CSI method needs to identify critical scenarios, such as relevant databases, assumed vehicle dynamic/kinematic models, functional models or implementations of the SoI, pre-defined surrogate measures for criticality assessment and environmental information provided by a simulator (e.g., CARLA or PreScan, etc.).

In Section \ref{sec:Solution}, CSI approaches are clustered into several groups according to the similarity between the approaches based on all the subcategories under the \textit{Solution} category. 

%%%%%%%%%%%%%%%%%%%%%%%%%%%%%%%%%%%%%%%%%%%%%%%%%%
\subsection{Subcategories of \textit{Evaluation}}
For the \textit{Evaluation} category of the taxonomy, three subcategories are considered. They are explained as follows:

\textbf{\textit{Approach Verification}:} This category describes the techniques used to assess the availability and efficiency of the approaches proposed in the primary studies. 

\textbf{\textit{Criticality Validation}:} In contrast to the \textit{Approach Verification}, this category focuses on the evaluation of the identified scenarios. As shown in Fig. \ref{fig:surrogate_measure}, identified potential critical scenarios based on surrogate measures may not fully overlap with the set of all the critical scenarios. Reasons for this mismatch include, but are not limited to, the low fidelity of the employed simulator, the assumptions made to simplify the exploration and the limitations of the selected surrogate measure. This category summarizes the approaches to validate the criticality of the identified critical scenarios.

\textbf{\textit{Coverage Assessment}:} This category enumerates the approaches to assess the coverage of the identified scenarios with respect to the whole scenario space.

\section{The Problem Definition Category}
\label{sec:Problem}
This section presents the classification result of the primary studies according to the \textit{problem definition} category of the taxonomy. 

%%%%%%%%%%%%%%%%%%%%%%%%%%%%%%%%%%%%%%%%%%%%%%%%%%%%%%%%%%%%%%%%%%%%%%%%%%%
\subsection{Usage of the Identified Scenarios}

In the primary studies, the identified critical scenarios can be used for different Systems of Interest (SoI) under different development phases.

%%%%%%%%%%%%%%%%%%%%%%%%%%%%%%%%
\subsubsection{Systems of Interest}
\label{sec:SoI}
Section \ref{sec:ADS} introduces the common functions of ADS or ADAS. As discussed in Section \ref{sec:planning}, the studied CSI approaches are used to support the safety analysis, verification and validation of the whole ADS (or ADAS) or a particular function of the ADS (or ADAS). 

Many of the primary studies do not explicitly specify their Systems of Interest (SoIs). In this section, SoIs of the primary studies are classified according to their targeted levels of automation and their functionality. 

\begin{figure}[!t]
\centering
\includegraphics[width=0.45\textwidth]{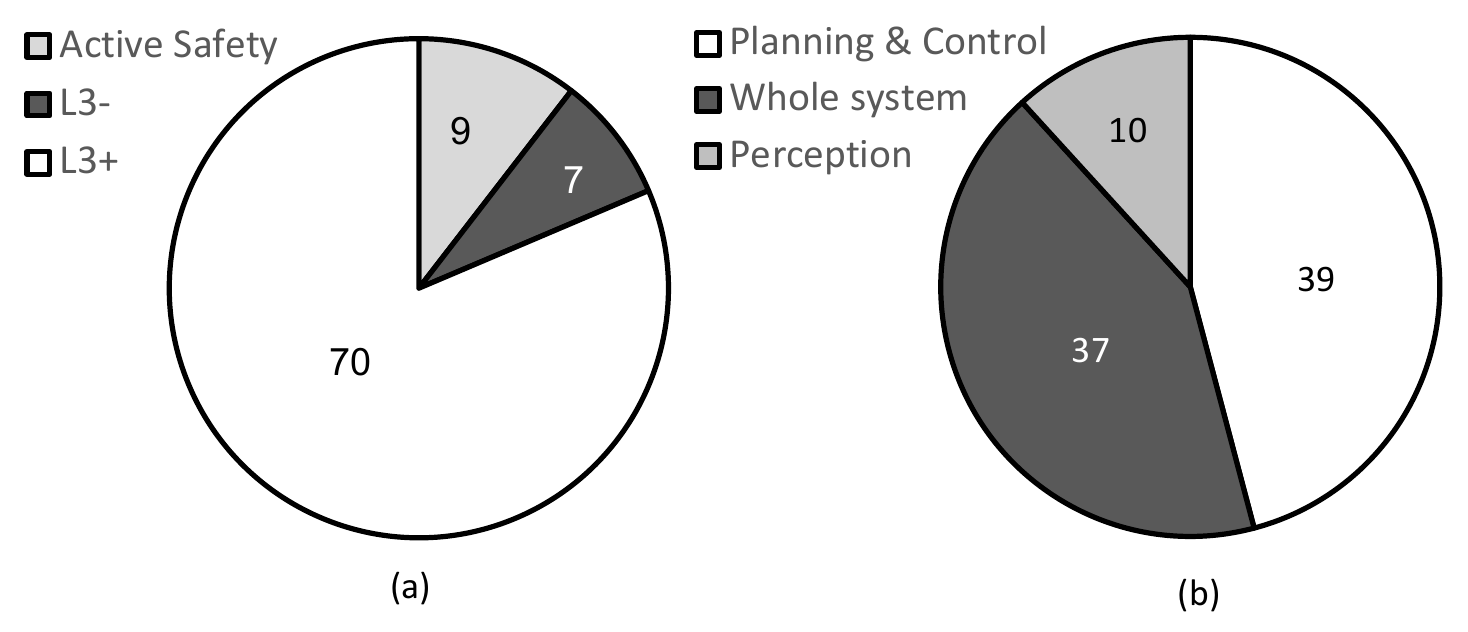}
\caption{Number of primary studies in each SoI category}
\label{fig:SoI_Pie}
\end{figure}

Section \ref{sec:ADS} introduces how ADS and ADAS can be classified according to the level of automation. Accordingly, the SoIs of the primary studies are classified into the following categories based on the corresponding criteria. 
\begin{itemize}
\item \textbf{\textit{L3+}:} CSI methods will be classified into this category if their SoIs have both longitudinal and lateral control, and they do not require a driver.\footnote{Since most of the primary studies are classified in this category, references to the primary studies are only given for the other two categories. The rest of the primary studies belong to this category.}
\item \textbf{\textit{L3-}:}  CSI methods will be classified into this category if their SoIs have only longitudinal control \cite{DeGelder2017a,reiterer2019beyond,Huang20173078,zhao2016accelerated,Koschi2019,7963024} or a driver's operations \cite{Khastgir2017} are required.
\item \textbf{\textit{Active Safety}:} CSI methods will be classified into this category if their SoIs cannot provide continuous control of the vehicle \cite{Abdessalem2018,pop00010 ,pop00045,Beglerovic2018,Khastgir2017,pop00220,Hu2020,Kim20181,SUI2019}.
\end{itemize}
It is assumed that CSI methods for higher levels of automation can also work for lower levels. Therefore, if a CSI method is designed for multiple levels (e.g., the methods to find safety critical operational situations can be used for all the levels), it is classed into the highest level. The statistical result of this classification is shown in Fig. \ref{fig:SoI_Pie} (a). 

An SoI can be the entire ADS (or ADAS) or a particular functionality of it. From this perspective, the primary studies are classified into the following categories based on the corresponding criteria derived from Fig. \ref{fig:source_unknown_CS} and Fig. \ref{fig:AD_functions}. The statistical result of this classification is shown in Fig. \ref{fig:SoI_Pie} (b). Detailed classification of all the primary studies can be found in the data extraction form \cite{review_table}.
\begin{itemize}
\item \textbf{\textit{Perception}:} CSI methods in this category try to find triggering conditions of unintended behaviors (e.g., wrong object detection and wrong object classification) of the perception functions. A CSI method will be classified into this category if the inputs of its SoI are the driving environment, and the outputs are the information in the Perceived World Model.  
\item \textbf{\textit{Planning and Motion Control}:} CSI methods in this category try to find triggering conditions of unintended behaviors (e.g., collision or other risky behaviors) of the planning and motion control functions. The input of these functions is the Perceived World Model, and the outputs are the control commands to the actuators (e.g., the steering wheel, the throttle and the brake.) Therefore a CSI method will be classified in this category if it assumes the Perceived World Model is given as ground truth, or ground truth with added noises \cite{Tuncali2020,Koren2018,Corso2019,Koren2019}.
\item \textbf{\textit{The Whole System}:} CSI methods in this category try to find 1) potential safety-critical operational situations; or 2) critical scenarios as combinations of both triggering conditions and safety-critical operational situation. The input of the whole system is the driving environment, and the outputs are the control commands. A CSI method will be classified into this category if it does not assume the Perceived World Model is given and tries to find scenarios that may lead to vehicle level hazards (e.g., unintended brake) or accidents (e.g., crash).
\end{itemize}

Unintended behaviors of different functionalities may have different triggering conditions attributed to scenario factors on different layers introduced in TABLE \ref{tab:6_layers} \cite{Amersbach2019}. TABLE \ref{tab:SoI_vs_Layers} illustrates the numbers of primary studies that consider influential factors on a particular layer for a particular SoI. Detailed models of these influential factors are analyzed in Section \ref{sec:Solution}.

\begin{table}[!t]
\centering
\caption{Relations between SoIs and influential factors on different layers}
\label{tab:SoI_vs_Layers}
\begin{tabular}{|r|c|c|c|c|c|c|c|} 
\hline
                                & \textbf{L0}   & \textbf{L1}   & \textbf{L2}   & \textbf{L3}   & \textbf{L4}   & \textbf{L5}   & \textbf{L6}  \\ 
\hline
\textbf{Perception}             & 0             & 0             & 0             & 0             & 0             & 0             & 0   \\ 
\hline
\textbf{Planning and Control}   & 0             & 5             & 2             & 0             & 37            & 2             & 0   \\ 
\hline
\textbf{The Whole System}       & 3             & 31            & 5             & 0             & 44            & 28             & 0   \\ 
\hline
\textbf{Total}                  & 3             & 36            & 7             & 0             & 81            & 30            & 0   \\
\hline
\end{tabular}
\end{table}

%%%%%%%%%%%%%%%%%%%%%%%%%%%%%%%%
\subsubsection{Phase and Purpose}
\label{sec:problem_purpose}

\begin{figure*}[!t]
\centering
\includegraphics[width=0.82\textwidth]{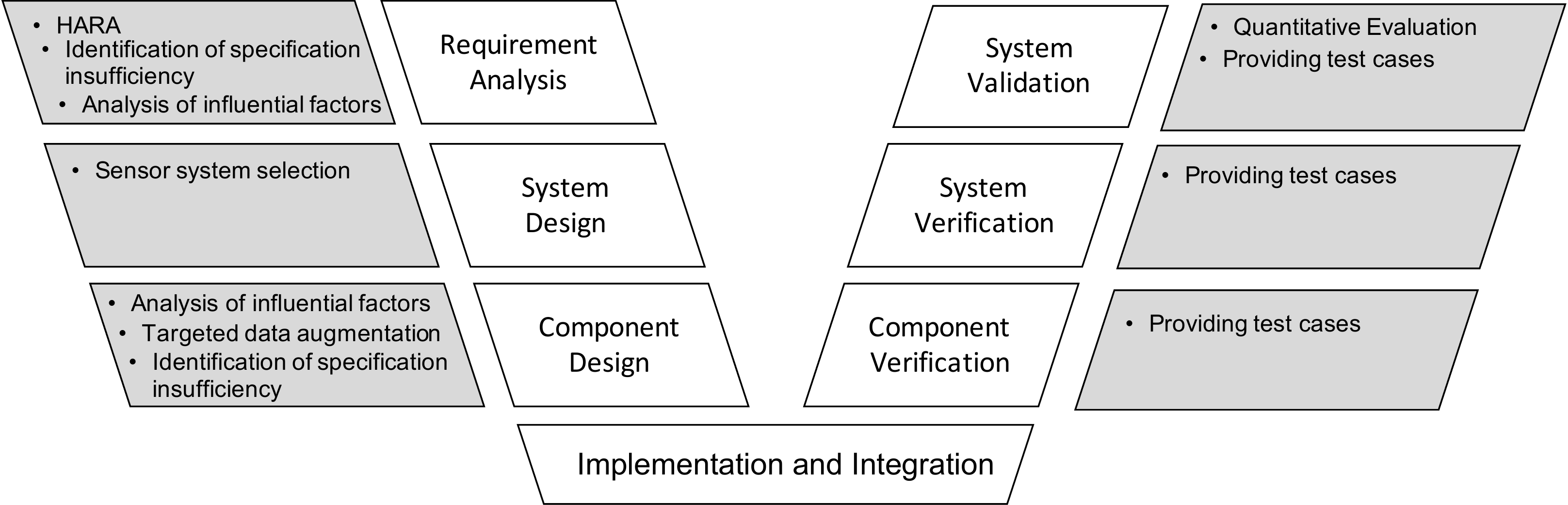}
\caption{How CSI methods support different development phases}
\label{fig:V}
\end{figure*}

As illustrated in Fig.~\ref{fig:V}, in the primary studies, the identified critical scenarios are used in every phase of the development V model. The gray boxes in Fig. \ref{fig:V} list what the identified critical scenarios can support in the corresponding development phase. 

\textbf{Requirement Analysis:}

This phase analyzes functional requirements or safety requirements at the vehicle level.
In ISO 26262, Hazard Analysis and Risk Assessment (HARA) is an essential step to identify all the potential hazardous events. As depicted in Fig. \ref{fig:SOTIF}, each hazardous event is a combination of a hazard and an operational condition. With the methods proposed in \cite{Zhou2017a,weber2018,Huang2018,Xie2018}, the identified pre-crash functional scenarios can be used as the set of all the operational conditions for HARA as described in \cite{Neurohr2020b}. 

Most of the studied CSI approaches treat the identified critical scenarios as test cases used in the verification and validation phases. The failed test cases or the identified critical scenarios through simulation can also support the identification of specification insufficiency.

Some of the studied CSI approaches can also support the analysis of influential factors by determining the critical regions (i.e. particular values or value ranges of a set of parameters) where critical scenarios are significantly more probable \cite{Tuncali2018,Tuncali2020,Koren2018,Corso2019,Koren2019,Tuncali2019}.
The identification of unknown influential factors is out of the scope of this survey but is briefly discussed in Section \ref{sec:Ontology_Design}.

\textbf{System Design:}

This phase decides the system configuration and the decomposition of vehicle level requirements to component level. However, in this survey, no CSI approach was found to support requirement decomposition. System configuration includes the selection of sensors \cite{Kim20181} and sensor ranges \cite{batsch2019performance}.

\textbf{Component Design:}

In this phase, the SoIs are certain AD functions, whose requirements are decomposed from the vehicle level. Since different AD functions may be sensitive to different environmental factors, influential factors are commonly analyzed on the component level rather than the vehicle level \cite{Amersbach2019}. One type of approach is to evaluate whether the variance of a particular parameter will significantly affect the criticality of the scenario \cite{Scenic}. For example, if changing the color of a vehicle in front will significantly affect the success rate to detect it, vehicle color will be considered as an influential factor. Critical regions can also be determined for a certain AD function such as the perception \cite{Tuncali2020}, and the planning and control functions \cite{Tuncali2020,Klischat2019,Althoff2018,Tuncali2017,7963024}. In addition, for machine learning based algorithms, identified critical scenes can also be added to the training set to support targeted data augmentation \cite{Scenic}. If the critical scenarios are used as test cases, the failed test cases can help with the identification of functional insufficiency, which includes specification insufficiency and performance limitations.

\textbf{Component and System Verification:}
 
Nearly 55\% of the CSI methods proposed in the primary studies focus on the identification of critical concrete scenarios, which are used in the verification phase as test cases. The generated test cases can be used to verify the whole AD system (e.g., \cite{Li2020} ) or a particular AD function (e.g., \cite{DeGelder2017a} ). Detailed analysis of these methods can be found in Sections \ref{sec:Ins_no_mp} and \ref{sec:Ins_mp}.

\textbf{System Validation:} 

A common validation method is to estimate the accident rate or failure rate through the Monte-Carlo simulation. Since critical scenarios are relatively rare, Monte-Carlo simulation with a small sample size may lead to a bad coverage of the critical scenarios and hence increase the estimation error. As a variant of Monte-Carlo simulation, importance sampling reduces the estimation error by assigning more samples to the critical but relatively rare scenarios. Therefore importance sampling methods entail the identified critical scenarios or critical regions as input. 

In addition, as discussed in \cite{koopman2019,UL4600}, test cases that are potentially critical for most of the ADS implementations can also serve validation purposes, e.g., scenarios with heavy fog or a sudden cut-in in front of the ego vehicle. This type of critical scenario is defined in Section \ref{sec:criticality_def} as a non-implementation-specific critical scenario. More analysis on how to find these scenarios can be found in Section \ref{sec:Solution}.
%Sections \ref{sec:C1_Critical_assessment_method}, \ref{sec:C2_Critical_assessment_method}, \ref{sec:C3_Critical_assessment_method}, \ref{sec:C4_Critical_assessment_method} and \ref{sec:C5_Critical_assessment_method}.

%%%%%%%%%%%%%%%%%%%%%%%%%%%%%%%%%%%%%%%%%%%%%%%%%%%%%%%%%%%%%%%%%%%%%%%%%%%
\subsection{Targeted ODD}
As shown in Fig. \ref{fig:LoA}, a clear ODD definition is necessary for the reasoning of critical functional scenarios, and the formulation (together with improvisation) of a functional scenario to a logical scenario. As discussed in Section \ref{sec:planning}, improvisation and formulation are out of the scope of this literature review. Therefore this section only focuses on how ODD supports the reasoning of critical functional scenarios. 
A clear ODD definition is essential to derive a complete set of critical concrete scenarios. However, none of the primary studies provide an explicit ODD definition.
Some of the primary studies explicitly or implicitly provide the scope of the ODD, e.g., driving on a highway \cite{Zhou2017a,weber2018} and driving on a structured road \cite{Huang2018,Xie2018,Stark2019a,Yue9129787}. Methods in these studies make assumptions about the environment (e.g., the behavior of other vehicles) within the ODD, and propose systematics to reason about critical functional scenarios based on these assumptions.

%%%%%%%%%%%%%%%%%%%%%%%%%%%%%%%%%%%%%%%%%%%%%%%%%%%%%%%%%%%%%%%%%%%%%%%%%%%
\subsection{Definitions of Criticality}
\label{sec:criticality_def}

As discussed in Section \ref{sec:critical_sce}, in this paper, critical scenarios are considered to be more important than non-critical scenarios in terms of safety analysis or verification. The definitions of criticality are tightly connected with the usages of the identified critical scenarios. However, most of the primary studies do not provide an explicit definition of criticality. Instead, they explicitly define the employed surrogate measures for criticality. In this survey, these measures of criticality are classified according to two dimensions, as illustrated in Table \ref{tab:criticality_def}. This classification is also used to specify the definition of criticality in each primary study. The two dimensions are explained in the rest of this section. The surrogate measures are detailed in Section \ref{sec:Solution}.

% solution space -> drivable area
\begin{table*}
\centering
\caption{Surrogate measures of each criticality definition}
\label{tab:criticality_def}
\begin{tabular}{|m{1em}|l|l|} 
\hline
\multicolumn{1}{|l|}{}   & 
\multicolumn{1}{c|}{Implementation-specific}  & \multicolumn{1}{c|}{Non-implementation-specific}  \\ 
\hline
\rotatebox{90}{Safety-Critical}   & 
\begin{tabular}[c]{@{}l@{}}
\textbf{\textit{KPIs in Simulation:}}
    \textbf{C1} \cite{Li2020} \cite{DeGelder2017a} \cite{reiterer2019beyond} \cite{Huang20173078} \cite{zhao2016accelerated} \cite{pop00010}  \cite{pop00045} \cite{pop00220} \\ 
                \cite{pop00003} \cite{pop00005} \cite{Mullins2018} \cite{Nabhan2019a} \cite{akagi2019risk} \cite{Feng2019b} \cite{Vehicles2020} \cite{Feng2020} \cite{Feng2020c} \cite{Gambi2019273} \cite{pop00022}  \cite{stumper2018towards}  \cite{pop00261} \cite{wagner2019virtual} \\
                \cite{Gladisch2019} \cite{Junietz201860} \cite{Chance2020} \cite{Wagner2018} 
    ~ \textbf{C2} \cite{Corso2019} \cite{Du18}
    ~ \textbf{C3} \cite{Kim20181} \cite{Stark2019a} \cite{VAAFO}\\
\textbf{\textit{Collision:}}
    ~ \textbf{C1} \cite{Abdessalem2018} \cite{Beglerovic2018} \cite{Khastgir2017} \cite{gangopadhyay2019identification}
    ~ \textbf{C2} \cite{Koschi2019} \cite{Koren2018} \cite{Koren2019} \cite{Kuutti2020} \cite{Abeysirigoonawardena2019}\cite{OKelly2018} \\
    ~ \textbf{C3} \cite{Gambi2019a} \cite{pop2019} \cite{pop2019} \cite{Qi2019} \cite{VAAFO}
    ~ \textbf{C4} \cite{Neurohr2020}\\
\textbf{\textit{Formal Specification:}}
    ~ \textbf{C1} \cite{Tuncali2018} \cite{Tuncali2020} 
    ~ \textbf{C2} \cite{Qin2019} \cite{pop00009}\\
\textbf{\textit{Performance Boundary:}}
    ~ \textbf{C1} \cite{Tuncali2018} \cite{batsch2019performance}
    ~ \textbf{C2} \cite{Tuncali2019}\\
\textbf{\textit{Hazardous Event (ISO 26262):}}
    ~ \textbf{C4} \cite{Neurohr2020}\\
%\textbf{\textit{Harm (ISO 26262):}}
    \\ 
\end{tabular} & 
\begin{tabular}[c]{@{}l@{}}
\textbf{\textit{KPIs in Simulation:}}
    ~ \textbf{C1} \cite{8431291} \cite{cutrone2018framework} \cite{stumper2018towards} \cite{ma2019driver} \cite{Bithar2019} \cite{8490321}\\
\textbf{\textit{Driveable Area:}}
    ~ \textbf{C1} \cite{Klischat2019} \cite{Althoff2018} \\
\textbf{\textit{Collision:}}
    ~ \textbf{C2} \cite{Chance2020} \cite{Jenkins20183340}
    ~ \textbf{C3} \cite{So2019} \cite{Hu2020} \cite{SUI2019} \cite{Yue9129787} \cite{Jenkins20183340}   \\
    ~ \textbf{C4}  \cite{Zhou2017a} \cite{weber2018} \cite{Huang2018} \cite{Xie2018} \cite{Neurohr2020} \cite{Ponn} 
%\textbf{\textit{Hazardous Event (ISO 26262):}}
 %   ~ \textbf{C4} \cite{Neurohr2020} 
\end{tabular}  \\ 
\hline
\rotatebox{90}{Function-Critical} & 
\begin{tabular}[c]{@{}l@{}}
~~~~~~~~~~~~~~~~~~~~~ Consequence Aware:\\
\textbf{\textit{KPIs in simulation:}}
    ~ \textbf{C1} \cite{pop00003}
    ~ \textbf{C2} \cite{7963024}\\
\textbf{\textit{Formal Specification:}} 
    ~ \textbf{C1} \cite{Tuncali2020}
    ~ \textbf{C2} \cite{Qin2019} \\
\textbf{\textit{Failures:}}
    ~ \textbf{C3} \cite{Qi2019}
    ~ \textbf{C5} \cite{Scenic} \cite{S.Yang.2017} \cite{CycleGANs} \cite{Dreossi.2019}\\
~~~~~~~~~~~~~~~~~~~~ Consequence Unaware:\\
\textbf{\textit{Performance Boundary:}}
    ~ \textbf{C1} \cite{Mullins2018}\\
\textbf{\textit{Differential Behaviors:}}
    ~ \textbf{C5} \cite{DeepTest} \cite{DeepXplore}
\end{tabular}  & 
\begin{tabular}[c]{@{}l@{}}
~~~~~~~~~~~~~~~~~~~~~ Consequence Unaware:\\
\textbf{\textit{Complexity:}}
    ~ \textbf{C1} \cite{Gao2019} \cite{Xia2017} \cite{xia2018test} \cite{8985542}
    ~ \textbf{C5} \cite{Scenic} \cite{DeepRoad} \cite{J.Wang.2018} \cite{C.Zhang.2018}\\
\textbf{\textit{Predictability:}}
    ~ \textbf{C5} \cite{Bolte.2019}
\end{tabular}  \\ 
\hline
\rotatebox{90}{Other}   & 
\begin{tabular}[c]{@{}l@{}}
\textbf{\textit{Fuel consumption:}}
    ~ \textbf{C1} \cite{Mullins2018}\\
\textbf{\textit{Comfort:}}
    ~ \textbf{C1} \cite{Nabhan2019a}
    ~ \textbf{C2} \cite{Tuncali2017} (KPI)\\
\textbf{\textit{Overall Traffic Quality:}}
    ~ \textbf{C1} \cite{Hallerbach2018b}
\end{tabular}   &    \\
\hline
\end{tabular}
\end{table*}

The first dimension specifies whether the identified scenarios are implementation-specific. An implementation-specific critical scenario is only determined to be critical for a particular implementation of an AD function or system (set of functions). It may or may not be critical for other implementations of the same function or system. In contrast, non-implementation-specific critical scenarios refer to the ones that are critical for most of the implementations of the same function or system. For example, scenarios with a heavy fog are critical for most of the camera-based object detection/classification functions. 

The second dimension classifies criticality according to consequence. Scenarios that are highly likely to cause a hazardous event are defined as safety-critical, while scenarios that may lead to a malfunctioning behavior are classified as function-critical. 
According to Fig. \ref{fig:SOTIF}, both safety-critical and function-critical scenarios can support the identification of unknown functional insufficiency and the corresponding triggering conditions. There are also primary studies whose criticality definition includes other perspectives such as comfort and traffic impact. These perspectives are out of the scope of this survey. 

Function-critical scenarios can be further classified according to the awareness of the consequential malfunctioning behavior. If the consequential malfunctioning behavior is pre-defined, (e.g., a collision or a misclassification of a certain object), the criticality of the identified scenarios is consequence-aware. Approaches to find consequence-unaware critical scenarios tend to find scenarios where malfunctioning behaviors are tend to be triggered. These scenarios may help to find unknown influential scenario factors \cite{Bolte.2019,DeepTest,DeepXplore,DeepRoad}.

Different primary studies may implicitly adopt different definitions of criticality. TABLE~\ref{tab:criticality_def} explicitly classifies the definitions of criticality according to the aforementioned three dimensions. As shown in TABLE~\ref{tab:criticality_def}, most of the primary studies focus on the identification of safety-critical scenarios. The number of primary studies that find implementation-specific critical scenarios is larger than those finding non-implementation-specific critical scenarios.

Different criticality definition entails different surrogate measures, which are further summarized in TABLE \ref{tab:criticality_def} and discussed in the next section. 

% Safety-critical scenarios can be further classified into
% Scenarios that may lead to a harm
% Scenarios that may lead to a harm under a certain hazard/hazardous behavior

% Function-critical
% Scenarios that may lead to a pre-defined hazard/hazardous behavior
% Scenarios that may lead to a hazard/hazardous behavior

%%%%%%%%%%%%%%%%%%%%%%%%%%%%%%%%%%%%%%%%%%%%%%%%%%%%%%%%%%%%%%%%%%%%%%%%%%%
\subsection{Level of Abstraction}
\label{sec:level_of_abstraction}

Critical scenarios can be identified on different levels of abstraction. As introduced in Section \ref{sec:tax_prob_def}, this section classifies the CSI methods according to their inputs and outputs in terms of the level of abstraction of scenario description. The classification result is given in Table \ref{tab:input_output}.

\begin{table}[!t]
\caption{Number of studies classified according to their inputs and outputs}
\label{tab:input_output}
\centering
\begin{tabular}{|l|c|c|c|c|} 
\hline
\diagbox{In.}{Out.}     & \multicolumn{1}{l|}{\textbf{Functional}} & \multicolumn{1}{l|}{\textbf{Logical}} & \multicolumn{1}{l|}{\textbf{Concrete}} & \multicolumn{1}{l|}{\textbf{Criticality}}  \\ 
\hline
\textbf{Others}                    & 2                               & 2                            & 0                             & 0                                 \\ 
\hline
\textbf{Functional}              & 2                               & 0                            & 4                             & 0                                 \\ 
\hline
\textbf{Logical}                 & 0                               & 2                            & 52                            & 6                                 \\ 
\hline
\textbf{Concrete}                & 3                               & 0                            & 12                            & 6                                 \\
\hline
\end{tabular}
\end{table}

Most of the studied CSI methods find concrete critical scenarios within a given logical scenario (i.e., \textit{logical $\rightarrow$ concrete}). An important step within this process is to evaluate the criticality of a given concrete scenario (i.e. \textit{concrete $\rightarrow$ criticality}). These two classes cover most of the CSI methods introduced in Sections \ref{sec:Ins_no_mp}, \ref{sec:Ins_mp} and \ref{sec:CV}. 

As shown in Fig. \ref{fig:LoA}, critical functional scenarios can be reasoned from the ODD definition and/or functional specifications and/or previous project experience, etc. (i.e. \textit{others $\rightarrow$ functional}) \cite{Zhou2017a,weber2018,Neurohr2020}. In some papers, these reasoned functional scenarios are also formulated into logical scenarios (i.e. \textit{others $\rightarrow$ logical}) \cite{Huang2018,Xie2018}. These methods are analyzed in Section \ref{sec:DR}.

Critical functional scenarios can also be induced from accident databases. Depending on whether the accidents in the database are described qualitatively or quantitatively, these methods can be classified as \textit{functional $\rightarrow$ functional} \cite{So2019} or \textit{concrete $\rightarrow$ functional} \cite{Hu2020}. These methods are introduced in Section \ref{sec:IR}. Another type of \textit{functional $\rightarrow$ functional} method is to combine triggering conditions with hazardous operational situations \cite{Neurohr2020}.

% \textit{logical $\rightarrow$ functional}

Studies evaluating the criticality of a logical scenario (i.e. \textit{logical $\rightarrow$ criticality}) actually evaluate the failure rate of an SOI under the given logical scenario. These studies estimate the failure rate through importance sampling \cite{Feng2019b, Vehicles2020, Feng2020, Feng2020c, pop00022, DeGelder2017a}, which is a variant of Monte-Carlo simulation, considering the distribution of critical scenarios in the space of the given logical scenario. It should be noticed that the focus of this survey does not include how the failure rate is evaluated. We only care about how the critical region can be discovered at the first step of the importance sampling.

There are two types of \textit{concrete $\rightarrow$ concrete} methods. The first type refines a concrete scenario to make it more critical by tuning its parameters \cite{Tuncali2018,Tuncali2020,wagner2019virtual,Tuncali2017,7963024,DeepRoad,DeepTest,DeepXplore}. These methods can be found in Sections \ref{sec:Ins_no_mp}, \ref{sec:Ins_mp} and \ref{sec:CV}. The second type requires multiple critical concrete scenarios to synthesize new critical scenarios ~\cite{Stark2019a,Yue9129787,VAAFO,Jenkins20183340}. This type of method is introduced in Section \ref{sec:IR}.

\section{The Solution Category}
\label{sec:Solution}
\begin{figure}[!t]
\centering
\includegraphics[width=0.3\textwidth]{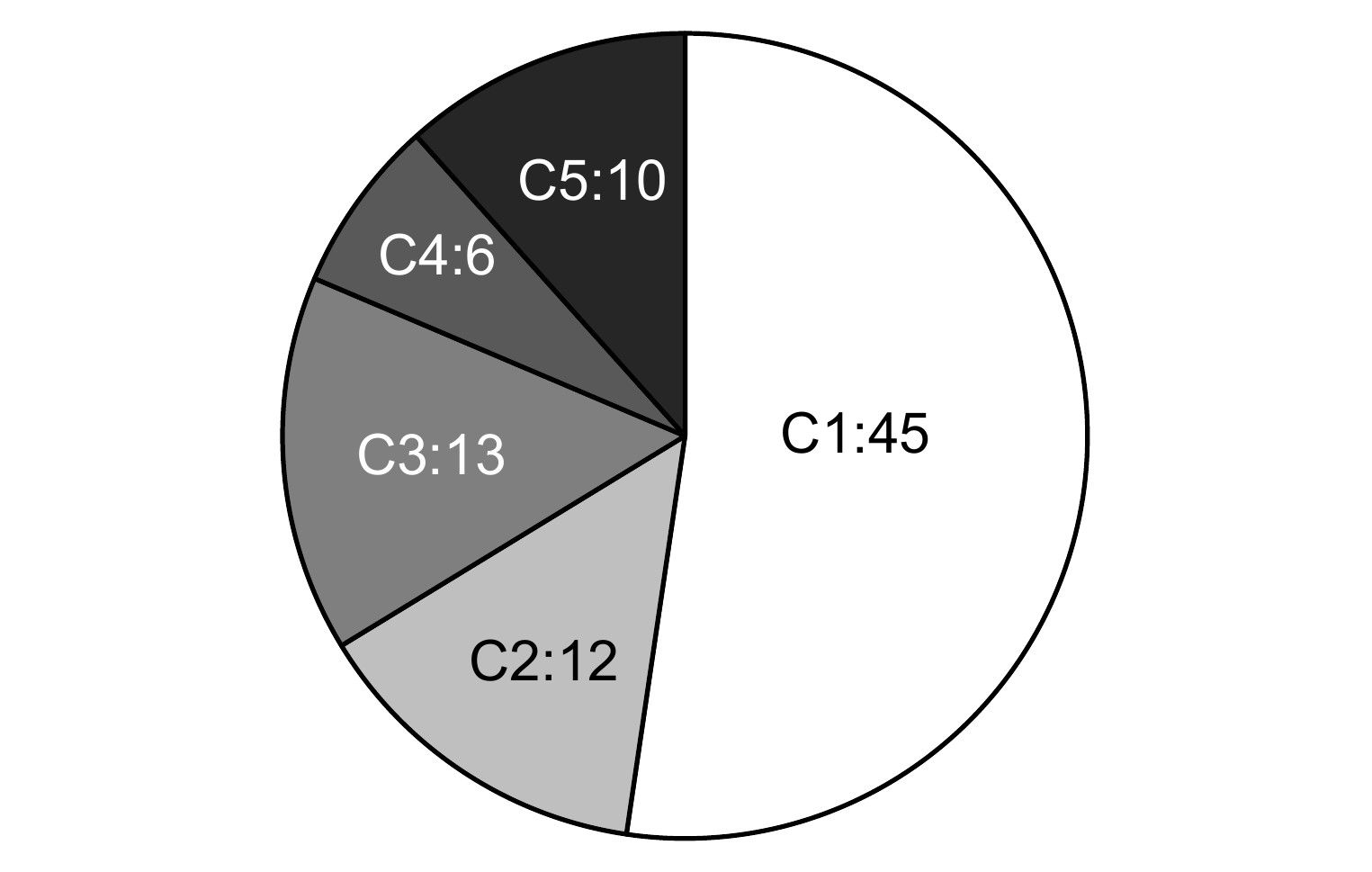}
\caption{Number of primary studies in each cluster}
\label{fig:clusters}
\end{figure}

%%%%%%%%%%%%%%%%%%%%%%%%%%%%%%%%%%%%%%%%%%%%%%%%%%%%%%%%%%%%%%%%%%
According to the similarities on the problem formulations and solutions, studied CSI approaches are grouped into the following five clusters. Fig. \ref{fig:clusters} illustrates the number of primary studies in each cluster. 
\begin{enumerate}
\item[\textbf{C1:}] Exploring logical scenarios without parameter trajectories (Section \ref{sec:Ins_no_mp})
\item[\textbf{C2:}] Exploring logical scenarios with parameter trajectories (Section \ref{sec:Ins_mp})
\item[\textbf{C3:}] Inductive reasoning methods (Section \ref{sec:IR})
\item[\textbf{C4:}] Deductive reasoning methods (Section \ref{sec:DR})
\item[\textbf{C5:}] Finding critical scenes for CV-based functions (Section \ref{sec:CV})
\end{enumerate}

\begin{figure}[!t]
\centering
\includegraphics[width=0.45\textwidth]{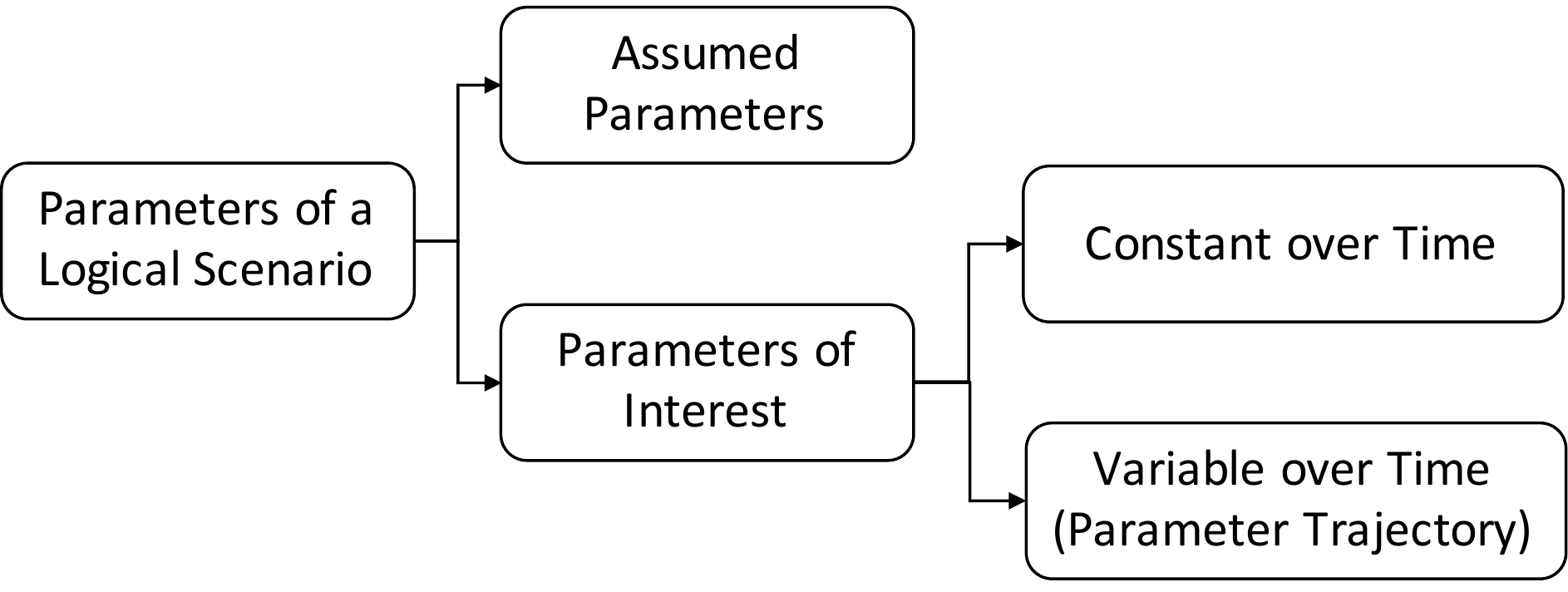}
\caption{Classification of parameters of a logical scenario}
\label{fig:logical_par}
\end{figure}

As depicted in Fig. \ref{fig:LoA}, a logical scenario can be instantiated into multiple concrete scenarios. The parameters of a logical scenario can be classified as illustrated in Fig.~\ref{fig:logical_par}. Assumed parameters have fixed values for all the instances (i.e. concrete scenarios), e.g. the number of vehicles in a scenario and the number of lanes. Parameters of interest construct the scenario space to be explored. These parameters include the ones that are constant over time (e.g. the weather condition or the position of a stationary obstacle on the road) and the ones that are variable over time (e.g. the speed of a surrounding vehicle or the perception error). If a parameter is variable over time, it can be represented as a parameter trajectory. Values of the parameters can be either categorical (e.g. weather, color and vehicle model) or numerical. Numerical values can be either continuous (e.g. speed, heading and sensor noise) or discrete (e.g. the number of other vehicles, the number of lanes and speed limit). 

The approaches to explore a logical scenario are elaborated in both \textbf{C1} and \textbf{C2}. The difference between these two clusters is that the logical scenarios in \textbf{C1} do not contain parameter trajectories. \textbf{C3} analyzes the methods to induce critical scenarios from different data sets. \textbf{C4} discusses systematic approaches to deduce critical functional scenarios. Most of the computer-vision (CV) based functions (e.g. object detection or classification) take a scene (i.e. a camera image) as input at each time step. The performance of such a function is mainly affected by the input scene rather than how the scene develops over time. Methods to find critical scenes for CV-based functions are summarized in \textbf{C5}.

The following subsections respectively introduce our analysis of these clusters. Except in \textbf{C4}, CSI approaches in each cluster are summarized with a gated tree, i.e. Fig. \ref{fig:instantiation_no_profile}, \ref{fig:instantiation_profile}, \ref{fig:inductive_methods} and \ref{fig:CV_solution}. Each node of these trees represents a component of a CSI solution. A parent node is connected with its children through either an "AND" gate or an "OR" gate. The "AND" gate implies that the parent note is constructed with all its children as necessary components. For example in Fig. \ref{fig:instantiation_no_profile}, all the methods in C1 (i.e. the root node) contain three components, namely a logical scenario model, an exploration method and a criticality assessment approach. If they are connected with an "OR" gate, the children will be the alternative solutions of the parent node. In other words, in a particular paper (e.g. \cite{Li2020}), the parent node (e.g. the Exploration Method node in Fig. \ref{fig:instantiation_no_profile}) will choose one of its children (e.g. Combinatorial Testing) as a solution. Not so many papers have been found in \textbf{C4}. Such a systematic figure is not given for \textbf{C4}, since a conclusion drawn from such few samples will not be representative. A systematic classification of CSI methods in \textbf{C4} will be part of the future work.

%%%%%%%%%%%%%%%%%%%%%%%%%%%%%%%%%%%%%%%%%%%%%%%%%%%%%%%%%%%%%%%%%%
\subsection{Exploring Logical Scenarios Without Parameter Trajectory}
\label{sec:Ins_no_mp}
% if openX mentioned in this cluster?
% SOI: black/white box
% remove researcher as subject
% check past tense when use people as the subject

\begin{figure}
\centering
\includegraphics[width=0.35\textwidth]{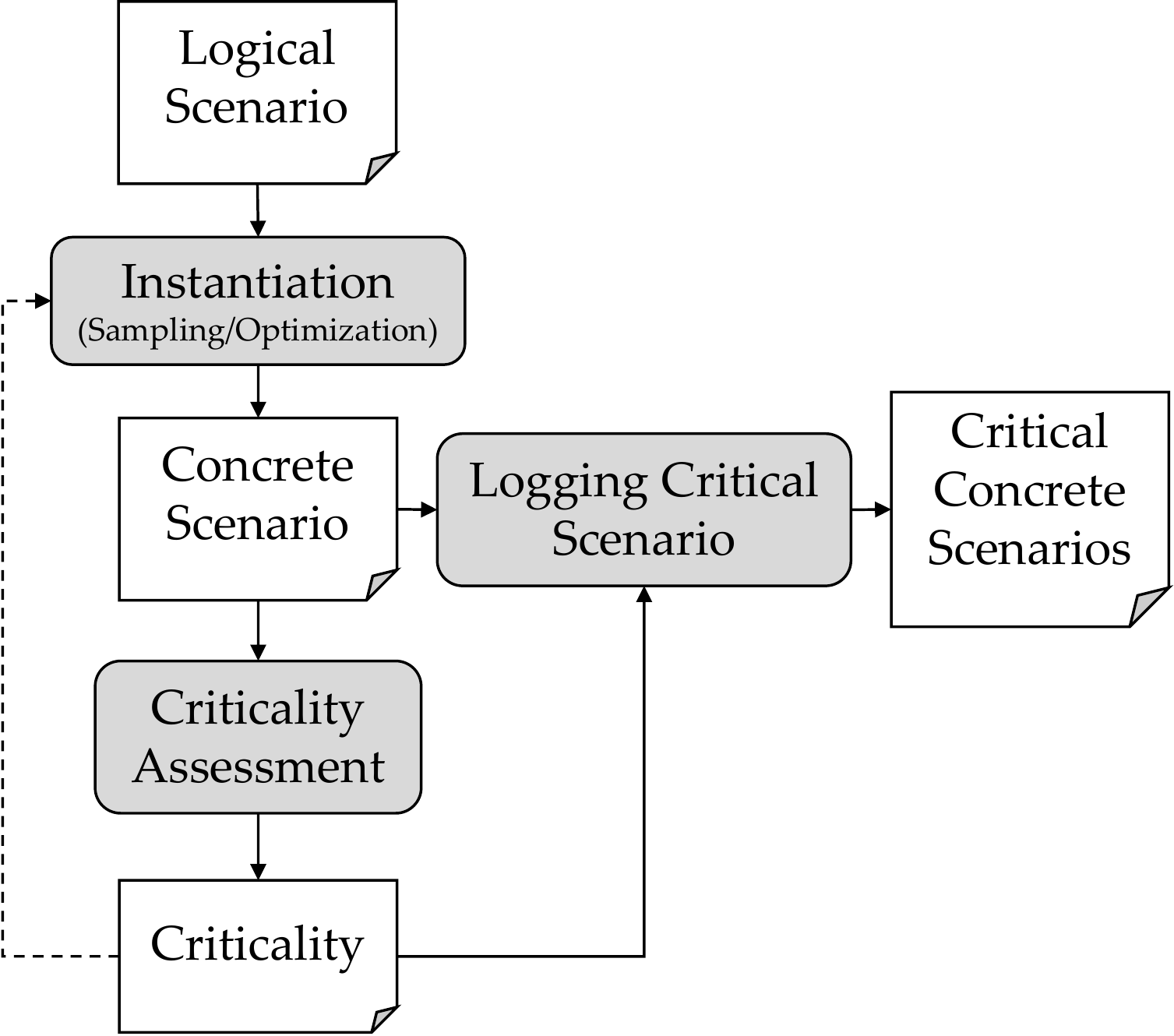}
\caption{Critical concrete scenario identification process} \label{fig:identification_process_critical_concrete_scenarios}
\end{figure}

\begin{figure*}
\centering
\includegraphics[width=0.88\textwidth]{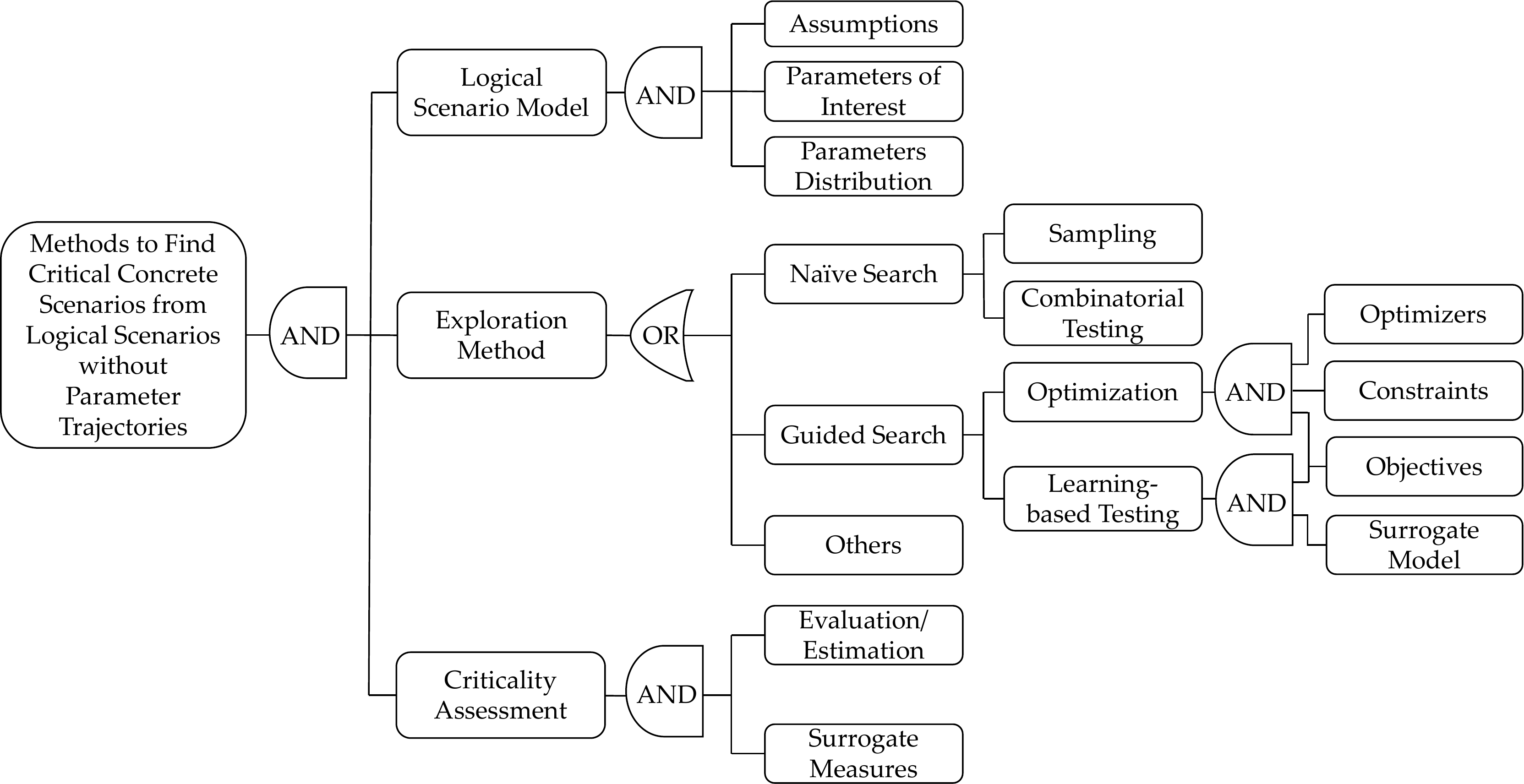}
\caption{Instantiation methods without parameter trajectories} \label{fig:instantiation_no_profile}
\end{figure*}

This section focuses on approaches to explore a logical scenario that does not contain a parameter trajectory. The exploration methods for logical scenarios with parameter trajectories are analyzed in Section \ref{sec:Ins_mp}. The exploration processes in this cluster are formulated as Design Space Exploration (DSE) or Search-based Testing (SBT) problems \cite{Gladisch2019}, whose flow chart is shown in Fig.~\ref{fig:identification_process_critical_concrete_scenarios}. Given a logical scenario, a set of concrete scenarios are generated with a parameter space exploration method. Among these generated concrete scenarios, critical ones are identified with a pre-defined criticality assessment method. Criticality assessment can be seen as a function that maps a point in the scenario space to a point in the scoring space. The scoring space is the quantitative evaluation of criticality for concrete scenarios. The evaluation is achieved through the surrogate measures since the criticality can hardly be measured directly. If the criticality is defined on multiple dimensions, each dimension in the scoring space refers to a criticality measure.
Based on Fig.~\ref{fig:identification_process_critical_concrete_scenarios} and the taxonomy in Fig. \ref{fig:Taxonomy}, all the CSI methods in this cluster are summarized in Fig.~\ref{fig:instantiation_no_profile}. The analysis of Fig.~\ref{fig:instantiation_no_profile} follows the taxonomy of \emph{Solution} category in Fig.~\ref{fig:Taxonomy}.

Regarding the analysis in Table \ref{tab:input_output}, most of the approaches in this cluster take logical scenarios as inputs and generate a set of critical concrete scenarios, which are further used to construct test cases or simulation cases. Exceptions include the approaches that focus on the criticality assessment of a given concrete scenario \cite{Wagner2018, Bithar2019}, and the refinement of a logical scenario \cite{Klischat2019, Althoff2018}. Criticality assessment is considered in this cluster because it is one of the essential steps to search for critical scenarios, as shown in Fig. \ref{fig:identification_process_critical_concrete_scenarios}. If a primary study only focuses on a criticality assessment approach, it usually implies that the given concrete scenario is generated from its previous step. Logical scenario refinement methods are included in this cluster only if they employ search-based methods. Instead of finding individual critical scenarios in the scenario space, they optimize the lower and upper bounds of the parameters of interest to derive the critical region. It can be treated as a prepossessing step of CSI methods to reduce the searching space.

The rest of this section explains Fig.~\ref{fig:instantiation_no_profile} based on all the primary studies in this cluster.

%%%%%%%%%%%%%%%%%%%%%%%%%%%%%%%%%%%%%%%%%%%%%%%%%%%%%%
\subsubsection{Scenario Model}\mbox{}
\label{sec:C1_scenario_model}

In this cluster, a logical scenario is modeled as a scenario space. Each dimension is a parameter of interest whose value can be either categorical or continuous. A logical scenario also specifies their value ranges or distributions. A concrete scenario is a vector in this scenario space with a fixed value for each dimension. As mentioned in Fig.~\ref{fig:SOTIF}, each parameter of interest correlates to one known scenario factor at one layer, as described in TABLE~\ref{tab:6_layers}. From the perspective of test (or simulation) case construction, each parameter specifies either:
\begin{itemize}
\item an \textbf{initial condition} including, e.g., the initial position, speed, or heading of the ego vehicle or other traffic participant; 
\item a \textbf{constant condition} including the weather condition (e.g., sunny, foggy, or raining) and the road topology;
\item or a \textbf{parameter of an assumed model} including, e.g., the size, field of view, or constant speed of a vehicle model
\end{itemize}

Parameter distributions can help to estimate the likelihood of exposure of a given concrete scenario. 9 out of 45 studies in this cluster consider realistic distributions of parameters (\cite{DeGelder2017a, akagi2019risk, Wagner2018, zhao2016accelerated, Feng2019b, Vehicles2020, Feng2020, Feng2020c, pop00022}) during the exploration. Other studies assume that the parameters follow a uniform distribution. The realistic parameter distributions are taken into account mainly for the following reasons:

\begin{itemize}
    \item \textbf{To estimate the failure rate of the SoI in all situations:} As discussed in Section \ref{sec:level_of_abstraction}, estimating the failure rate with importance sampling needs critical scenarios or critical region as input. In addition, it also needs to know the likelihood of exposure of each sampled concrete scenario, which can be approximated based on the parameter distributions.
    \item \textbf{To consider commonality as part of the criticality definition:} Compared to rare cases, those with higher probabilities to exposure in the real world are of significant interest. To this end, the commonality of a scenario is introduced into the definition of criticality \cite{DeGelder2017a, akagi2019risk, Wagner2018}, so that the exploration method can find common and hazardous scenarios. This commonality can be quantified with the help of parameter distributions in real traffic.
\end{itemize}

Parameter distributions are derived from an existing real-life driving database (i.e., the Naturalistic Driving Study (NDS) or Field Operational Test (FOT) data). The parameters can be assumed either mutually independent \cite{Bithar2019, 8490321} or dependent \cite{akagi2019risk}. Akagi et al. \cite{akagi2019risk} considered the parameter distribution and approximate it by a Gaussian Mixture Model (GMM), where parameter probability distributions represent the covariance of variables. Methods to approximate parameter distributions from a given data set are summarized in \cite{reiss2012approximate}. 

%%%%%%%%%%%%%%%%%%%%%%%%%%%%%%%%%%%%%%%%%%%%%%%%%%%%%%
\subsubsection{Exploration Methods}\mbox{}
\label{sec:C1_Exp_Mtd}

Fig.~\ref{fig:instantiation_no_profile} lists all the methods to explore the scenario space employed in the primary studies in this cluster. These exploration methods can be divided into two types: (1)~naive search (i.e., \emph{Sampling} and \emph{Combinatorial Testing}) and (2)~guided search (i.e., \emph{Optimization} and \emph{Learning-based Testing}). 

A naive approach for scenario exploration is to search randomly or systematically over the scenario space. In other words, samples are mutually independent. Therefore these approaches can be implemented in a parallel manner to reduce the exploration time. However, if critical scenarios are rare, these approaches can be inefficient as the probability of sampling a critical scenario is low. On the other hand, the guided search methods have the potential to be more efficient, since the searching direction at each iteration is adjusted according to the search result of the previous iteration, so as to converge the exploration to critical regions. Each exploration method under these two types is briefly introduced in the rest of this section.

\begin{figure}
\centering
\includegraphics[width=0.48\textwidth]{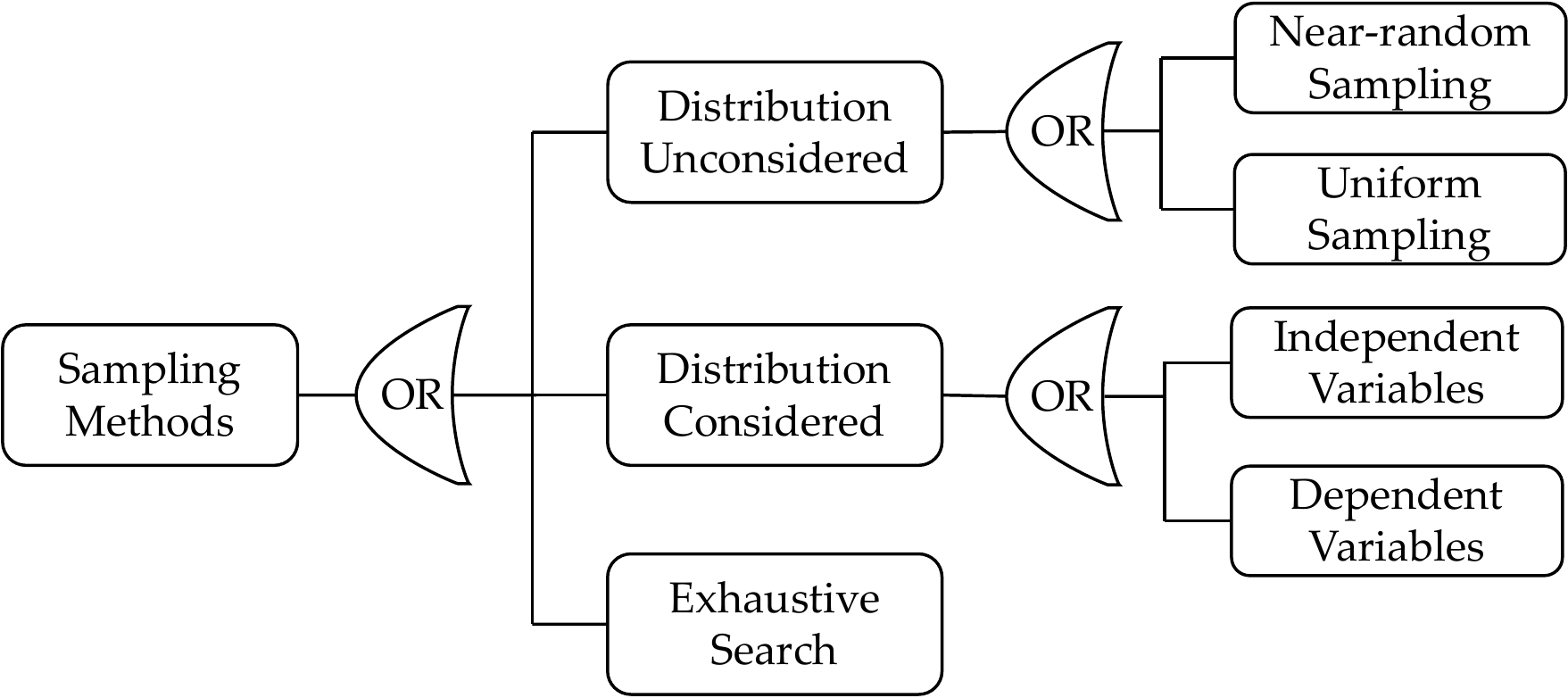}
\caption{Sampling method in scenario exploration} 
\label{fig:instantiation_sampling}
\end{figure}

\textbf{Sampling}: The sampling method instantiates a concrete scenario by randomly assigning each parameter's value in a logical scenario space. A predetermined number of samples are taken statistically based on probability distributions of parameters, and its sampling size is determined by the required coverage and computation time for simulation. The applied sampling methods are summarized in Fig.~\ref{fig:instantiation_sampling}. According to the parameter descriptions in logical scenarios, the sampling approach can be classified based on whether the parameter distributions are taken into account. We assume that a uniform distribution is adopted if the parameter distribution is not mentioned. Near-random sampling, such as Latin Hypercube sampling \cite{batsch2019performance}, can improve the coverage when the sampling size is small. It splits the multi-dimensional parameter space into even grids and selects samples in each grid with a given number. 

As mentioned in Section~\ref{sec:C1_scenario_model}, parameter distribution is considered for two purposes: failure rate estimation and commonality consideration. Since the failure rate estimation by importance sampling barely appears during the scenario space exploration phase, in this section we only consider parameter distribution used to model the commonality of the scenarios. Moreover, when considering the parameter distributions, relationships between parameters can be assumed either dependent or independent. Different assumptions require different sampling methods. Studies \cite{Bithar2019, 8490321} investigated the parameter distributions before sampling. In these studies, each dimension in sampled scenarios is viewed as independent and identically distributed (i.i.d.). The covariance among different parameters is not considered, and their values can be determined independently. Nevertheless, correlations among variables are also investigated in some other studies\cite{akagi2019risk, 8431291}. When considering the relationship among variables, more restrictions are imposed on the sampling. Instead of Monte-Carlo sampling, the Markov Chain Monte Carlo (MCMC) method \cite{akagi2019risk} can be applied efficiently with the knowledge of parameter distribution and covariance. The proposed approach in \cite{akagi2019risk} applies a risk index from naturalistic driving data to select risky traffic conditions efficiently.

If the scenario space is fully discrete and reasonably small, an exhaustive search is applicable to guarantee full coverage. As shown in Fig.~\ref{fig:instantiation_sampling}, it is considered as an extreme case of sampling. If the scenario database is available, the exhaustive search can be achieved through checking every test case derived from NDS/FOT data \cite{pop00220}. Otherwise, the exhaustive search can discretize continuous parameters and perform a full-scale grid search to examine every existing test case in the scenario space \cite{reiterer2019beyond,8431291,stumper2018towards}.
    
\textbf{Combinatorial testing}: Combinatorial Testing (CT) is a commonly used software testing method focusing on the identification of failures triggered only by particular combinations of inputs \cite{Kuhn2015}. The core of CT is to generate a minimum set of test cases (i.e. a covering array) that satisfy N-wise coverage\footnote{N-wise coverage states that all N-tuples of parameter values must be tested at least once, given an SoI model with the parameter list, values, and constraints to define parameter interactions \cite{choi2014design}}. Under the context of CSI for ADS, CT can be used to find the unknown combinations of influential factors that may fail the ADS or a particular AD function. An example is when a pedestrian crosses a road in front of the ego vehicle in the evening. The criticality is significantly influenced by the combination of different scenario parameters, such as the initial speed of the ego vehicle, visibility, and distance between the pedestrian and the vehicle. The covering array, as a matrix that stores the testing configurations, specifies test data where each row of the array can be regarded as a set of parameter values for a specific test\cite{Nie2011}. For example, if the required coverage is three-wise, the generated covering array should cover all combinations of values from arbitrary three parameters, where three specifies the number of parameters in combination. A good covering array can significantly reduce computational cost and improve test efficiency. However, finding the minimum covering array is an NP-hard problem \cite{Kuhn2015}. A test database comprises all test scenarios in the covering array, and it can cover all possible combinations of parameter values at a predefined degree. 

N-wise coverage can only be defined on discrete parameter space. However, in a real-life scenario, both continuous and discrete variables are prevalent in the parameter space of the logical scenario. A common method to handle continuous parameters is discretization, which is a process of quantizing continuous attributes by converting the continuous data into a finite set of intervals and assigning some specific data values to each interval \cite{liu2002discretization}. In some other studies, Tuncali et al.~\cite{Tuncali2018} circumvented the problem by a two-step approach. The first step performs CT on all the discrete parameters. In the second step, for each combination of the discrete parameters, they conducted optimization by simulated annealing on all the continuous parameters. The optimization is used as the falsification process to identify critical scenarios.

Instead of focusing on the minimum number of test cases in the covering array, covering array generation methods can also be customized. For example, in \cite{Gao2019, Xia2017, xia2018test, 8985542}, the generated covering array fulfills not only the N-wise coverage but also maximizes the overall complexity of all the scenarios.

\textbf{Optimization}: Studies in this class formulate the CSI approaches into optimization problems, which generically contain four parts, namely the design variables, the constraints, the objective functions, and the optimizer (i.e. the solver) \cite{kochenderfer2019algorithms}. The design variables have already been analyzed in Section \ref{sec:C1_scenario_model} and their ranges are restricted by the constraints. Therefore, as shown in Fig.~\ref{fig:instantiation_no_profile}, this section mainly analyzes the objective functions and the selected optimizers.

In the vast majority of studies in this class, the objective functions represent quantified criticality measures, which are discussed in Section~\ref{sec:C1_Critical_assessment_method} in detail. Other than criticality, the objective function can also include the similarity between scenarios (i.e. the distance in the scenario configuration space) to maximize the diversity of the identified critical scenarios \cite{Feng2019b,Vehicles2020}.

The choice of optimizer highly depends on the problem formulation, especially the transparency and complexity of the underlying models. A simulator is commonly employed when estimating criticality measures. Due to the high model complexity and unavailability of the interior structure of plant models, in many cases simulation and system analysis of the SoI can be computationally expensive. For this reason, the behavior of the SoI is treated as a black box. Thus, the corresponding search process can be regarded as a black-box optimization problem. For a black-box optimization problem, the input (i.e. scenario parameters) and output (i.e. simulation results) relationship can only be analyzed by exterior observation through simulations \cite{xiao2015optimization}.  To tackle this problem, various heuristic methods can be applied to approximate the fitting function and find the global minimum iteratively, including genetic algorithm \cite{Abdessalem2018, pop00010, cutrone2018framework}, Bayesian optimization \cite{gangopadhyay2019identification,Gladisch2019}, and simulated annealing \cite{pop00045}. In addition, the domain-specific heuristic technique, where the optimizer only suffices in a certain system, helps to identify multiple local minimums of feasible solutions by rule-based searching \cite{8490321}, heuristic simulation-based gradient descent \cite{Huang20173078} and zoom-in sampler \cite{Beglerovic2018}. Scenario searching is formulated as a two-step optimization problem in \cite{Feng2019b, Vehicles2020, Feng2020}. The first step of the optimization tries to find multiple local optimal solutions. In the second step, the neighborhood of the local optimal solutions is searched to find all the scenarios whose criticalities are within a given threshold.

\textbf{Learning-based testing}:
Learning-based testing methods \cite{meinke2011learning} aim to automatically generate a large number of high-quality test cases by combining a model checking algorithm with an efficient model inference algorithm, and integrating these two with the SoI in an iterative loop. It learns the properties of the SoI during optimization by training a surrogate model. The diversity can be optimized by maximizing the distances between samples in either the scenario space or the scoring space. Mullins et al. \cite{Mullins2018} used a surrogate model for training to improve the sampling coverage. It takes a set of samples as input and returns the estimated diversity (i.e. the mean distance) on the scoring space of the input samples. Nabhan et al. \cite{Nabhan2019a} used learning-based methods to try to maximize the distance between scenarios. It also considers non-safety-critical qualities (such as deceleration and jerk effects indicating passenger's comfort) during the scenario generation. 

%Compared to optimization, which only focuses on safety-critical indices, learning-based methods have the advantage to gain the scenario's property of non-functional qualities, such as estimated diversity among scenarios \cite{Mullins2018} and approximation of passenger's comfort \cite{Nabhan2019a}. 

%Learning-based methods have a strong dependency on the training data. With a set of NDS/FOT data which contains massive positive (normal driving) and negative (collision/malfunction) samples, support vector machine (SVM) for the criticality check by building a classifier in the parameter space offline with training data (e.g. \cite{stumper2018towards, ma2019driver}), which helps to estimate the criticality of underlying critical scenarios and improve the test procedure.
    
\textbf{Others}: This category refers to other methods to generate test cases and detect critical scenarios not included in our research, such as Tabu search, hill-climbing, and grammatical evolution. The review of common search algorithms in software testing can be found in \cite{mcminn2011search, cotta2008adaptive}.

%%%%%%%%%%%%%%%%%%%%%%%%%%%%%%%%%%%%%%%%%%%%%%%%%%%%%%
\subsubsection{Criticality Assessment Methods}\mbox{}
\label{sec:C1_Critical_assessment_method}

\begin{figure}[!t]
    \centering
    \includegraphics[width=0.48\textwidth]{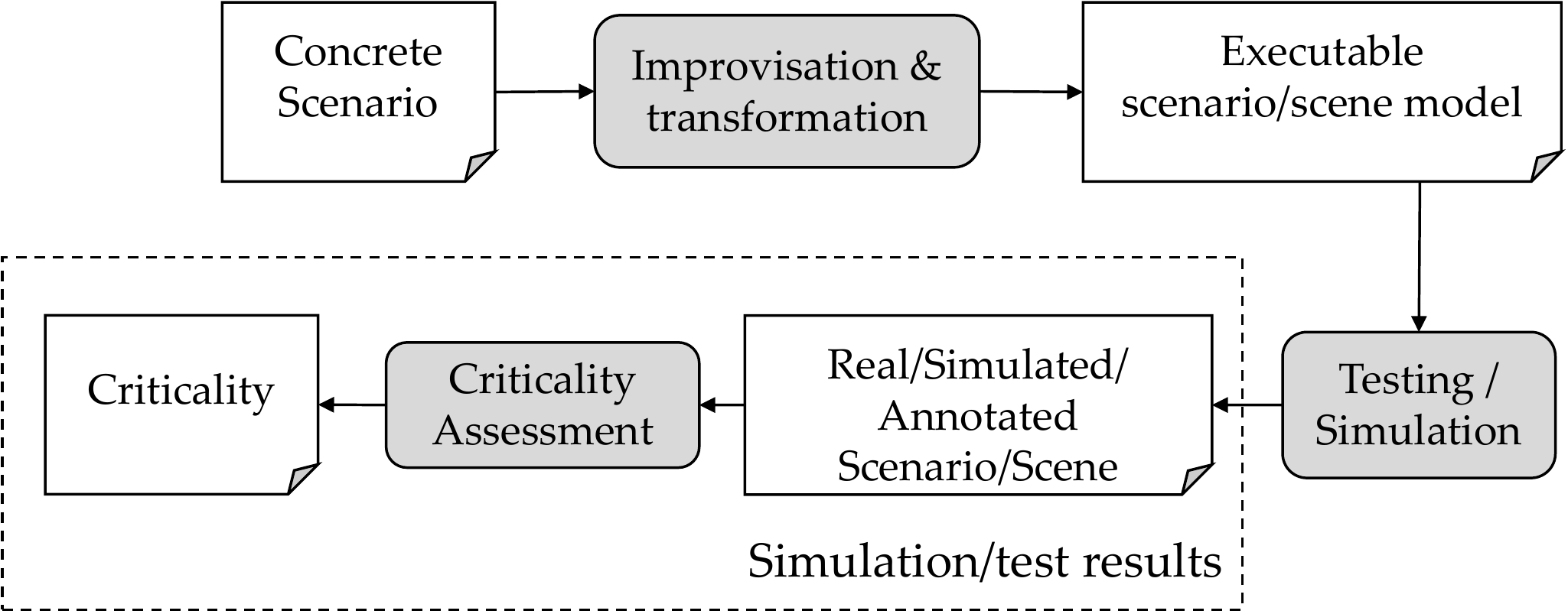}
    \caption{Test case generation from concrete scenarios and testing-based criticality assessment} \label{fig:test_case_generation}
\end{figure}

After attaining concrete scenarios from the CSI methods mentioned above, most of the primary studies adopt testing-based approaches to verify the criticality of derived scenarios, as shown in Fig.~\ref{fig:test_case_generation}. Different definitions of criticality are discussed in Section \ref{sec:criticality_def}. This section discusses how to assess the criticality of a concrete scenario under different criticality definitions.

In the criticality assessment phase, most of the studies utilize an X-in-the-loop simulation to estimate the criticality of a concrete scenario, where X represents the model, software, or hardware of the SoI as a black-box. Criticalities are assessed on the simulation results based on the predefined surrogate measures summarized in Table~\ref{tab:criticality_def}. Nevertheless, the criticality can also be assessed without X-in-the-loop simulations. Validation can be realized through real-world testing to analyze the performance of the proposed exploration method \cite{stumper2018towards, Vehicles2020}. 

As discussed before, criticality assessment is a function whose input is a concrete scenario, and whose output is a quantified criticality index. Most of the studied CSI methods in this cluster implement this function deductively with analytical approaches, while the others implement it inductively with machine learning approaches \cite{Mullins2018, Nabhan2019a}.

Inductive criticality assessment is generally applied in the data-driven approach. Instead of evaluating scenario criticality by the black-box simulation, it constructs an approximation model to emulate the real system behavior with the database. The database is composed of existing scenario data with labels, which indicates the criticality of the scenarios and is used in the training process of the approximation models. The labels of criticality measures can be attained directly from the training data in the scenario database. It can also be derived by simulation of concrete scenarios, where the simulation result is regarded as the ground truth.

Since criticality can hardly be measured directly, deductive criticality assessment is based on a surrogate measure to evaluate the criticality of a concrete scenario quantitatively. The surrogate measures can be either Boolean or numerical in the scoring space, depending on the types of exploration methods. For naive search methods, criticality measures are Boolean, since each sampled scenario needs to be evaluated as critical or non-critical. For guided search methods, since criticality is a part of the objective function, the criticality measure has to be quantified as numerical. Under each criticality definition, one or more surrogate measures were proposed in different primary studies, as summarized in Table~\ref{tab:criticality_def}. More measures of criticality can be found in \cite{Junietz2019}. Based on Table~\ref{tab:criticality_def}, the surrogate criticality measures used in this cluster are summarized as follows.

\textbf{KPI-based measures} are commonly used in this cluster due to the simplicity of implementation. It describes the proximity to a critical state through simulations driven by black-box testing. KPI-based methods calculate metrics by assessing a posterior measurement of the vehicle state. The metrics are used to evaluate the criticality in a scene, and the criticality of a scenario can be subsequently indicated by the worst scene. In the context of the safety-critical scenario measurement, KPI metrics refer to Time-to-X metrics (e.g., Time-to-Collision, Time-to-Brake, and Time-to-React), distance-based metrics (e.g., longitudinal and lateral distance), velocity-based metrics (e.g., relative speed), and acceleration-based metrics (e.g., required deceleration). A comprehensive analysis of KPI comparison can be found in \cite{Mahmud2017}. The metrics above are usually used as the measures of a particular function. However, compared to the implementation-specific category, the non-implementation-specific ones can be found only with the generic features of the SoI, for example, a highly abstract simulation model which is generic enough to represent different implementations of the SoI or a criticality measure that is only based on the environmental aspects. The critical scenarios are detected without information of the implemented function, which makes it possible to apply scenarios for system design \cite{Bithar2019} and early phase of verification \cite{stumper2018towards}. Also, \textbf{the driveable area} \cite{Klischat2019, Althoff2018} can be viewed as a special measure in this category, where the criticality is defined directly on a logical scenario by examining the range of relevant parameters. The SoI (e.g., general motion planning algorithms) will be more critical if the solution space is smaller with a less drivable area. Meanwhile, besides safety-critical applications, KPI-based approaches can also be applied in functional-critical studies. Unlike safety-critical measures, function-critical assessment emphasizes the performance of a particular module of the SoI (e.g., sensor, decision-making, and actuator), which does not necessarily propagate to an accident.

Besides the KPIs mentioned above, in order to increase the efficiency of critical scenario detection, \textbf{complexity} can be regarded as an auxiliary property of criticality in some studies \cite{Gao2019, Xia2017,xia2018test, 8985542}. Compared to traditional software testing, critical scenario generation focuses on finding triggering conditions instead of explicit software bugs. Different value selections correspond to different levels of complexity, which is treated as a priori knowledge before performing the test case generation. By our definition in Section \ref{sec:criticality_def}, they are viewed as non-implementation-specific and consequence-unaware, since the complexity is not exclusively oriented to any particular malfunction type.

Compared to KPI-based methods, \textbf{collision} is a more straightforward measure in the safety-critical assessment. It generates binary output according to whether a crash happens in a simulated scenario. However, some real-world collision scenarios may not manifest on simulation due to the low fidelity of the simulation model. In this cluster, collision analysis only appears in the implementation-specific category since collision avoidance has to be realized with certain types of ADAS/ADS.

A more complex and specific measure in implementation-specific class can be represented by \textbf{formal specifications} such as signal temporal logic (STL)~\cite{Tuncali2018, Tuncali2020}. This kind of method can be applied to either safety-critical or function-critical use cases. The safety-critical approach examines the scenarios that may lead to accidents, while the function-critical measure aims to find anomalies in subsystem-level function and test its robustness.

In addition, implementation-specific criticality can also be characterized by \textbf{performance boundary}, which can be divided into two types. The first type is the predefined boundary, where the exploration methods try to separate critical scenarios to non-critical ones through the boundary and find avoidable collisions \cite{batsch2019performance}. In the second type, the performance boundary is unknown at the beginning. Several performance modes are derived through clustering the scoring space. Scenarios around the boundaries of the performance modes are of great interest since slight changes of parameter values can contribute to the behavior change. It is assumed that faults tend to manifest in those scenarios \cite{Tuncali2018}. Moreover, we distinguish the study in \cite{Mullins2018} as the consequence-unaware type for the function-critical use case, since a scenario is regarded as critical if a small change of its configuration leads to significant changes in the SoI performance. By this definition, consequential malfunctioning behavior is not explicitly given. In the above-mentioned types of studies, the criticality assessment is realized by simulation. 

Apart from the safety-critical and function-critical consideration, \textbf{quality of service (QoS)} indices can also be applied as surrogate measures. In this cluster, the normal operations of the vehicle are assumed, and QoS is quantified to judge the performance of a system, such as by fuel consumption \cite{Mullins2018}, passengers' comfort \cite{Nabhan2019a}, and overall traffic quality \cite{Hallerbach2018b}.

%%%%%%%%%%%%%%%%%%%%%%%%%%%%%%%%%%%%%%%%%%%%%%%%%%%%%%
\subsubsection{Coverage}
\label{sec:Ins_no_mp_coverage}
\begin{figure}[!t]
\centering
\includegraphics[width=0.48\textwidth]{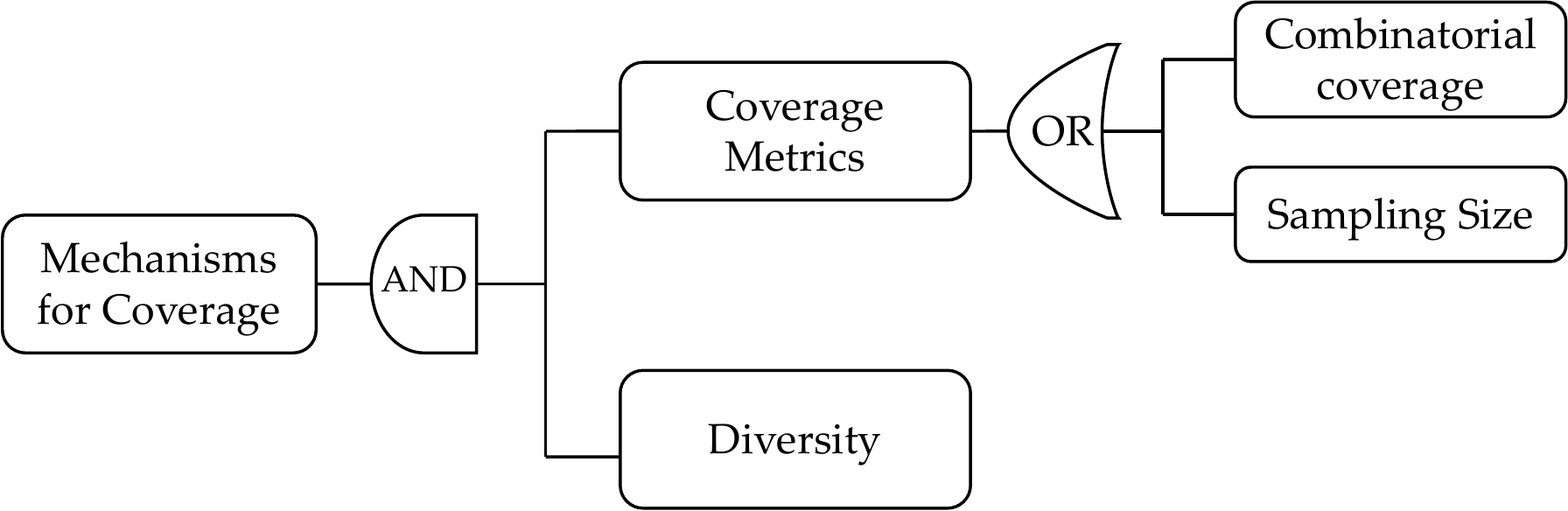}
\caption{Mechanisms for coverage of scenario space exploration methods} \label{fig:coverage_mechanism}
\end{figure}

For safety argumentation, coverage must be considered when exploring a logical scenario. However, not all the primary studies in this cluster explicitly discuss coverage. As shown in Fig.~\ref{fig:coverage_mechanism}, this section summarizes the consideration of coverage and the mechanisms to increase the diversity of the identified critical scenarios in this cluster.

% coverage metrics: 1) combinatorial coverage. 2) sampling size (have distribution, check variance value)
% diversity: 1) scenario space diversity. 2) criticality space diversity

Combinatorial coverage (i.e. N-wise coverage) \cite{ponn2019towards, Li2020, Tuncali2018, Tuncali2020, Gao2019} and sampling size \cite{batsch2019performance,stumper2018towards, Bithar2019} are the two metrics found in the primary studies in this cluster to measure the coverage of the exploration. These two coverage metrics are detailed in Section \ref{sec:C1_Exp_Mtd}.

%functional coverage method is a structural coverage metrics at the software architectural level as defined by ISO 26262 \cite{ISO26262}. Test automation tool, such as Vitaq \cite{Khastgir2017}, can be utilized to achieve functional coverage based on defined use case requirements. To this end, some other standard coverage metrics, such as statement coverage, branch coverage and call coverage can be fulfilled in a similar way.

Another definition related to coverage is the diversity of the identified critical scenarios. Similar scenarios have a high potential to reflect the same triggering condition. Diverse critical scenarios can help to identify different functional insufficiencies and hazardous events. The rest of this section discusses mechanisms to increase the diversity of the identified critical scenarios.

In the primary studies, diversity is quantitatively defined based on the distance in the scenario space. For sampling-based exploration methods, distances are generally estimated in the Euclidean space. Different measures, such as Voronoi interpolation \cite{Gladisch2019}, can be used at sampling to ensure no tests exist in close proximity. Moreover, various sampling methods, such as Latin Hypercube Sampling \cite{batsch2019performance}, can also improve the diversity of samples. It divides each parameter into intervals and ensures the scenario space is evenly covered.

On the other hand, for optimization or learning-based methods, diversity can be considered part of either the objective or the constraint. The diversity can be realized either by examining the distance among scenarios \cite{Mullins2018}, or by exploring multiple local minima \cite{wagner2019virtual, gangopadhyay2019identification}. For the latter approach, Wagner et al. \cite{wagner2019virtual} extracted key features from critical scenarios by principal component analysis (PCA) and added noise in a lower-dimensional component space to reconstruct more varied critical scenarios. Gangopadhyay et al. \cite{gangopadhyay2019identification} applied Bayesian Optimization to identify multiple minima regions from a non-convex function, where critical scenarios are situated in minima regions. Both methods mentioned above aim to find critical scenarios with diverse failure conditions.

%%%%%%%%%%%%%%%%%%%%%%%%%%%%%%%%%%%%%%%%%%%%%%%%%%%%%%
\subsubsection{Required Information}

When identifying critical concrete scenarios from a logical scenario, a CSI method needs a logical scenario as input and one or more surrogate measures to support criticality assessment. These have been discussed respectively in Sections \ref{sec:C1_scenario_model} and \ref{sec:C1_Critical_assessment_method}. Research in this cluster normally verifies the case study by simulation-based approaches. In the context of simulation, the completeness of a test case includes four parts, namely simulation objects (i.e. SoI), simulator and environment, criticality, and database. The simulation objects and criticality have been discussed in previous sections. Herein, we mainly focus on the simulator and database.

\textbf{Simulator}: The simulator provides the platform to implement and express SoI properties in vehicle dynamics and ADAS/AD function through modeling language. Many high-fidelity simulators (e.g., PreScan, CarMaker) also support virtual test driving to visualize simulation environments. We regard a simulator as the ad-hoc simulation platform if the simulator is developed in-house or not mentioned in the article. MATLAB, together with other programming language simulators, are viewed as low-fidelity simulation platforms as the system dynamics are simplified, or virtualization is omitted. The main target of the low-fidelity simulator is the realization of a system dynamics description and to enable the process of sampling, thus the model fidelity is not the focus.
 %We classify primary studies based on types of applied simulators and summarize in Fig.~\ref{}. 

\textbf{Database}: All primary studies referring to the database in this cluster employ it to analyze parameter distribution as prior knowledge. Based on our findings, we cluster the database into three types: naturalistic driving data~\cite{ma2019driver, pop00261, stumper2018towards, wagner2019virtual}, traffic flow data~\cite{akagi2019risk, pop00220} and collision accident data~\cite{8431291}. It is worth noting that NDS databases exclusive to autonomous driving (e.g., Safety Pool) are available but have not been found applied in the field of this cluster.
% EuroNcap

%Necessary information to generate scenarios:e.g. a set of influential factors(a logical scenario), database, 

%Necessary information to assess criticality: e.g. SUT as a black box, simulator

%prior knowledge, training data for model training/ criticality check

%%%% previous notes
%Simulators (academia, industry, collaboration)
%What a simulator can provide
% environment and sensing models (how the environment interact with the sensors)
% Vehicle dynamic models (how control signals can affect the behavior of the ego-vehicle)

% % notes0125

% consider layer5 for control:
%     Ponn2019a
%     Gladisch2019

% consider performance limitation of perception:
%     sensor range: Ponn2019a, Batsch2019a
%     sensor noise: Ponn2019a, Hallerbach2018b
    
%     Tuncali2018: detected (by simulation) perception error (color reason)

% ADAS as SOI that consider driver behavior
%     Ponn2019a: it has a driver model to generate driver behavior

%%%%%%%%%%%%%%%%%%%%%%%%%%%%%%%%%%%%%%%%%%%%%%%%%%%%%%%%%%%%%%%%%%
\subsection{Exploring Logical Scenarios with Parameter Trajectories}
\label{sec:Ins_mp}
In the previous section, the behavior of other actors (e.g., vehicles or pedestrians) is determined by several parameters of a predefined motion model (e.g., the velocity of a constant velocity model). In this way, possible behaviors of other actors are restricted, especially their reactions to the behaviors of the ego vehicle. A more arbitrary way to determine the behavior of another actor is to model it as a motion trajectory (position trajectory or acceleration trajectory). This can also be extended to any parameters whose development over time is of interest (e.g., the angle of the sun). In general, the variable over time parameters, also called parameter trajectories, are the values of the scenario that change during the simulation. Methods to explore a logical scenario including such parameter trajectories are discussed in this section. Such problems are also called stress testing in \cite{Koren2018,Du18,Koren2019} or adversarial testing in \cite{Qin2019}.

Replacing one parameter with a trajectory of parameters will significantly increase the exploration space. Therefore, the methods introduced in the previous section are not fully suitable for the problem in this cluster. The rest of this section discusses the methods in the primary studies for this type of problem.

\begin{figure}[!t]
\centering
\includegraphics[width=0.48\textwidth]{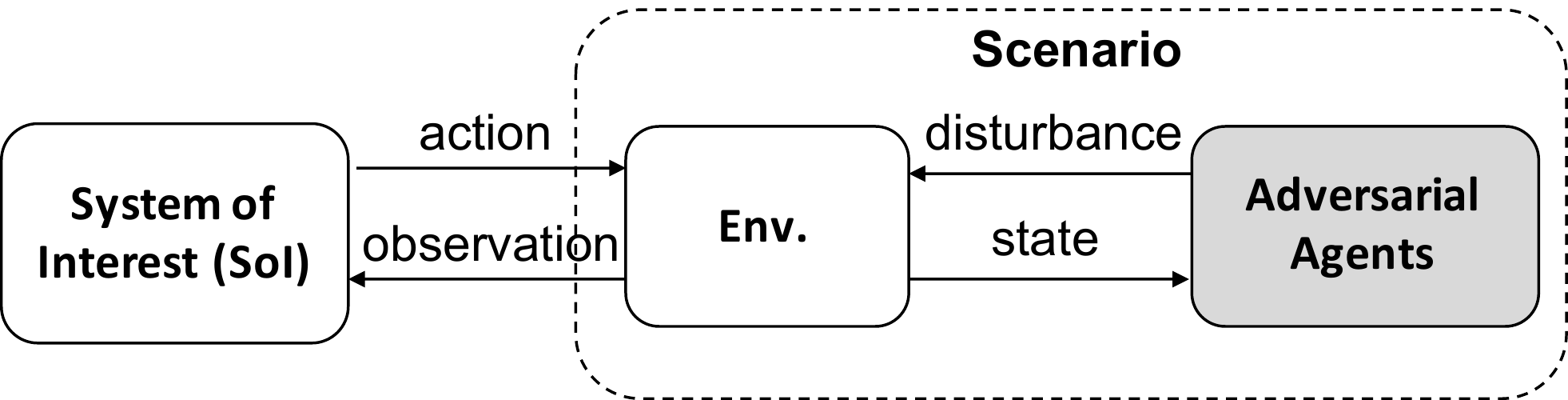}
\caption{Relation between the SoI, the environment and the adversarial agents \cite{Corso2020a}} \label{fig:scenario_model}
\end{figure}

In these studies, the SoI is challenged by a scenario in which the behavior (or the perceived behavior) of the other actors is set to make the system fail under a certain metric. Some methods treat the SoI as a back box e.g. \cite{Althoff2018, Klischat2019} while others may need to know the model of the system as part of the optimization process e.g. \cite{Tuncali2017}, or at least the order of the model \cite{7963024}. Fig.~\ref{fig:scenario_model}, based on \cite{Corso2020a}, shows the relationship between the SoI, the environment and the adversarial agents. The SoI generates actions based on the observations obtained from the environment, which is also influenced by the adversarial agents.  All the connections represented in this diagram, the actions and observations from the SoI, and the states and disturbances from the adversarial agents, are parameter trajectories.

\begin{figure*}[!t]
\centering
\includegraphics[width=0.88\textwidth]{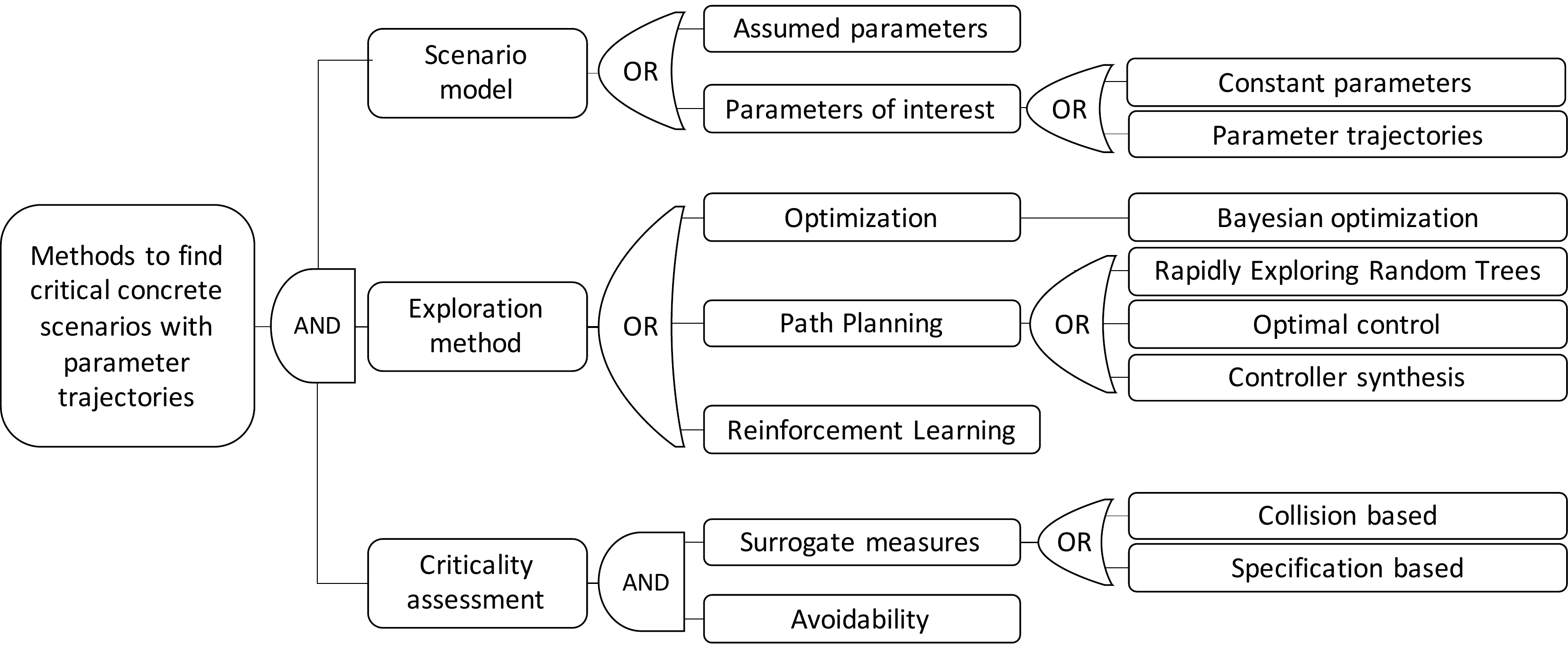}
\caption{Instantiation methods with parameter trajectories} \label{fig:instantiation_profile}
\end{figure*}

An overview of methods found in this cluster can be seen in Fig.~\ref{fig:instantiation_profile}. These methods require a scenario model, an exploration method, and a criticality assessment based on the taxonomy in Fig.~\ref{fig:Taxonomy}. The scenario model contains information about the constraints imposed to the scenario. The exploration method searches for a scenario that may lead the SoI to a hazardous event (as described in Fig.~\ref{fig:SOTIF}) based on the criticality assessment. This criticality is assessed based on a pre-defined surrogate measure.

The categories proposed in \cite{Corso2020a} have been used to classify the exploration methods of this section. Even though the scope of \cite{Corso2020a} is not within the automotive domain, and the methods analyzed only consider black-box safety validation, the proposed categories are still useful for the purpose of classifying the exploration methods of the present study. 

\subsubsection{Scenario model}

A scenario model includes a set of parameters. Some of them have a predefined value (assumed parameters) while others (parameters of interest) should be optimized to find a critical concrete scenario, as shown in Fig.~\ref{fig:instantiation_profile}. Examples of assumed parameters in the primary studies are the number of other actors (moving traffic participants) \cite{Chance2020}, the number of lanes \cite{Klischat2019}, and the position of the crosswalk \cite{Koren2018}.

Referring to Fig.~\ref{fig:logical_par}, parameters of interest in this cluster may include both constant (over time) parameters and parameter trajectories, or only parameter trajectories.
An example of constant parameters is the the number of agents in a particular scenario in \cite{Chance2020}, which is explored within a predefined range.

The exploration method aim to find the optimal parameter trajectories to challenge the SoI. These parameters generally include the motion profiles of other actors and the level of noise (e.g., \cite{Koren2018}). Parameter trajectories can also be used to model other factors such as weather conditions or traffic light patterns, as mentioned in \cite{Qin2019}.

Motion profiles can be restricted to longitudinal motion, as in \cite{Koschi2019}, or they can also include lateral positions. The lateral positions can be expressed with respect to the center of the lane, as in \cite{pop00009} or using absolute coordinates, as in \cite{Koren2018}. Besides positional restrictions, there can be other types of constraints, as in \cite{pop00009}, where the lateral velocity and the yaw rate were specified to be within certain values.

Including parameter trajectories significantly increases the searching space. In addition, the values of the same parameter at different time steps are not independent. Therefore, methods introduced in Section \ref{sec:Ins_no_mp} are not optimal for the problems in this cluster. Instead, most of the approaches in this cluster formulate the exploration as a sequential decision-making problem.

\subsubsection{Exploration method}

The exploration method aims to find a set of values and trajectories for the parameters of interest that force the SoI to fail under a certain metric. As mentioned before, for the classification of the methods, the categories proposed in \cite{Corso2020a} have been followed, where these exploration algorithms are divided into optimization, path planning, and reinforcement learning:

\textbf{Optimization:} 

These methods are similar to the ones introduced in the previous section. They assume that the values at different time steps of a trajectory are independent and include a cost function designed to guide the search for a trajectory that forces the SoI to fail. In \cite{Corso2020a}, simulated annealing, genetic algorithms and Bayesian optimization are included in this category.
\cite{Abeysirigoonawardena2019} is the only publication related to this category that has been identified. It uses Bayesian optimization to generate adversarial scenarios. Bayesian optimization is an algorithm proposed to optimize functions that are expensive to evaluate without knowledge about the structure of the function e.g. concavity or linearity \cite{frazier_tutorial_2018}. In \cite{Abeysirigoonawardena2019}, the optimization is not directly used to generate a trajectory. It is used to optimize the policy of the adversarial agents that decided their behavior based on their observations. No study has been found during our study that uses optimization techniques to directly generate trajectories. One of the reasons might be the complexity of the environment, which might make these methods impractical due to the exponential growth of the state space.

\textbf{Path Planning:}

Path planning algorithms aim to navigate the state space from a starting state to a goal state. In this case, the goal is to get from an initial state to one of the failure states using disturbances as control inputs \cite{Corso2020a}. In this category, the following algorithms have been identified:

\textbf{Rapidly exploring random trees (RRT):}
RRT is a search algorithm for path planning designed to handle high degrees of freedom \cite{lavalle_rapidly-exploring_1998}. This method is used in \cite{Koschi2019} to make an ACC system fail by generating motions of the leading vehicle. They use two methods to explore the tree: forward search and backward search. Forward search starts at a randomly generated safe state and generates the tree based on the actions of the leading vehicle aim to get the ego vehicle into an unsafe state. Backward search starts at a random unsafe state and searches backwards in time trying to reach a safe state. In both methods, they randomly generate the behavior of the leading vehicle and, based on that, they generate the response of the ego vehicle. In \cite{Tuncali2019}, the aim is to find almost-avoidable collisions or near-misses. To promote this kind of collision, RRT was used together with a custom cost function that estimated how avoidable a collision was based on the ratio of the collision surface, the collision speed, and the minimum time to collision. In both studies, each node of the tree includes the state of the system so that, when the tree grows, only a partial simulation is executed from an existing node of the tree to the new node, instead of running the simulation from the initial state.

\textbf{Optimal control:}
This is a control strategy that aims to optimize an objective function. It originated as an extension of the calculus of variations \cite{SARGENT2000361}. In \cite{7963024}, the process to search for critical scenarios is formulated as an optimal control problem, which finds the optimal sequence of control input (in the case study, it is the acceleration profile of the leading vehicle and the time instances of critical scenes) to optimize the objective function.

\textbf{Controller synthesis:}
Controller synthesis aims to generate of a controller given the complete model of a system and the environment that is guaranteed to achieve a specification \cite{Chuchu}.
In \cite{pop00009}, the authors use controller synthesis to generate scenarios where it is guaranteed to be possible to satisfy a set of formally defined specifications.

\textbf{Reinforcement learning:}

Reinforcement learning aims to program agents by giving a reward based on their actions without specifying how the task should be accomplished \cite{kaelbling_reinforcement_1996}. Actions might affect not only the current reward, but also the future ones \cite{sutton2018reinforcement}. The algorithms analyzed in this section model the problem as a Markov decision process (MDP). The methods found in this category are the following:

\textbf{Monte Carlo tree search:}
a heuristic search algorithm for decision making that takes random samples in the decision space and uses that information to build a search tree \cite{browne_survey_2012}. In \cite{Koren2018}, the authors compare its performance against deep-reinforcement learning and Trust Region Policy Optimization (TRPO) to solve the MDP used to model the scenario.

\textbf{Deep reinforcement learning:}
uses deep neural networks to approximate the value function, the policy or the model of reinforcement learning \cite{DeepReinforcementLearning}.
\cite{Koren2018} extended Adaptive Stress Testing (AST) originally proposed in \cite{AdaptiveLee} to test aircraft collision avoidance systems. AST formulates the scenario as an MDP and uses a Monte Carlo tree search to solve it. In \cite{Koren2018}, the authors proposed to extend AST by using deep reinforcement learning to solve the MDP and apply it to a set of autonomous vehicle scenarios. \cite{Corso2019} and \cite{Du18} improve the reward function previously proposed by using RSS \cite{RSS} to find scenarios where the ego vehicle performs improper actions and including a trajectory dissimilarity reward to find diverse failures. In \cite{Koren2019}, the authors extend the work performed in \cite{Koren2018} by replacing the original Multi-layer Perceptron Network with a Recurrent Neural Network. In \cite{Qin2019}, the authors use Deep Q-learning (a deep reinforcement learning method) to learn the behavior of the adversarial agents. The ego vehicle and the agents have Rulebooks (originally proposed in \cite{Rulebooks}) associated with them to constrain their behaviors.

\subsubsection{Criticality assessment}
\label{sec:C2_Critical_assessment_method}

The criticality assessment is used to define whether the SoI has failed to accomplish its requirements. This assessment can be categorized based on the surrogate measure used to determine the failure and based on whether the avoidability of the failure is analyzed or not.

The surrogate measures used in the literature of this study can be categorized as follows:

\textbf{Collision based:}
Only a collision is considered when analyzing the performance of the SoI.  In \cite{Koren2018}, any collision is considered critical by the algorithm, regardless of how the collision occurs. For this reason, in some of the results obtained, the pedestrian is considered responsible after studying the scenario. In \cite{Koschi2019}, the scenario model only allows for the generation of rear-end collisions caused by the ego vehicle, removing the need of having to identify the liable actor.

\textbf{Specification based:}
\cite{Corso2019} uses the rules from RSS \cite{RSS} to define the improper actions evaluated in the reward function. The reward function guides the search for scenarios in which the ego vehicle behaves inappropriately. In \cite{Qin2019}, the objective function is based on a Rulebook \cite{Rulebooks}. In \cite{7963024}, the behavior of the other actors is optimized to make the system violate its requirements.

Looking at how the avoidability of the failure is taken into account, the following classification can be made:

\textbf{Avoidability not considered:}
Most of the studies reviewed do not consider whether the failure could have been avoided or not. They simply compute that it was a failure.

\textbf{Avoidability considered:}
In \cite{Tuncali2019} the aim is to generate test cases in the boundary where the autonomous vehicle can no longer avoid a collision (almost-avoidable collisions or near-misses). To achieve that, they propose their version of RRT with a custom cost function that promotes collisions that are almost avoidable.
\cite{pop00009} proposes a method to find avoidable critical scenarios using controller synthesis. Avoidability is derived with a controlled invariant set (a subset of the safe states). States in this set are such that, as long as the disturbance is within a certain range, it is always possible to find a control input so that the next state is also in this set.

A complete classification of all the papers analyzed in this study based on the surrogate measure can be found in Table \ref{tab:criticality_def}.

\subsubsection{Coverage}
Coverage is not mentioned by most of the papers in this section. Some of them include a measure to promote better coverage, but none of them address it as the main focus. In \cite{Tuncali2019}, a novelty function is computed based on the Mahalanobis distance to achieve better coverage and avoid local minimum. \cite{Corso2019} includes a trajectory dissimilarity reward to promote the discovery of highly diverse failure scenarios.

\subsubsection{Required information}

All the studies analyzed in the section require a simulator. Some studies require access to the internal state of the simulator, as in \cite{Koren2018}, while others treat them as a black-box as in \cite{Koren2019}. In \cite{Koschi2019}, as mentioned before, a backward search is performed, where their method tries to get to a safe state from a future unsafe state. Due to the impossibility of computing the inverse of the ACC control law, the authors generate random inputs for the ACC vehicle to try to get the vehicle in the previous state. Then they simulate forward to ensure getting into an unsafe state again.

%%%%%%%%%%%%%%%%%%%%%%%%%%%%%%%%%%%%%%%%%%%%%%%%%%%%%%%%%%%%%%%%%%
\subsection{Inductive Reasoning}
\label{sec:IR}
\begin{figure*}
\centering
\includegraphics[width=0.8\textwidth]{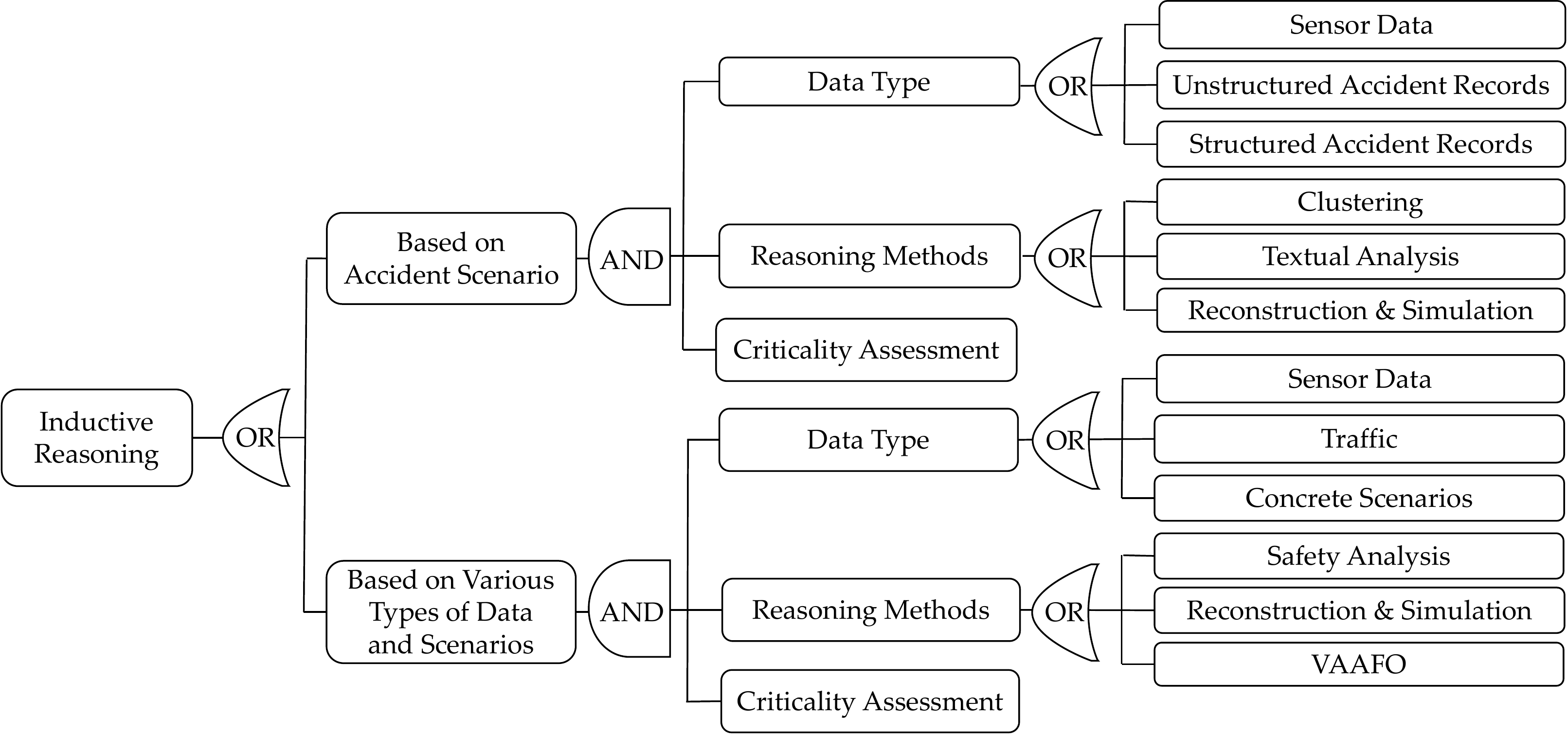}
\caption{Inductive reasoning methods to find critical scenarios} \label{fig:inductive_methods}
\end{figure*}

This section focuses on inductive reasoning methods. With these methods, the critical (functional or concrete) scenarios are induced from different data sources. The main data sources identified in the primary studies are summarized into two types: 1) based on accident scenarios only (Section \ref{sec:induc_based_on_accident_scenarios}), and 2) based on various types of data and scenarios (Section \ref{sec:induc_based_on_all_scenarios}). The former source relies on the accident database, including raw accident data, accident reports or records, while the latter source refers to the existing logical or concrete scenarios, natural driving data, traffic data or sensor data. 
CSI methods in this cluster are summarized as shown in Fig.~\ref{fig:inductive_methods}, which is also used as a basis for the structure of the following subsections (data type, reasoning methods and criticality assessment).
%Most of the accident databases or police reports are written in partially structured natural language.

\subsubsection{Based on accident scenarios}
\label{sec:induc_based_on_accident_scenarios}
These methods aim to find the common features for critical scenario identification from a variety of accidental scenarios. As shown in Fig.~\ref{fig:inductive_methods}, we analyzed each type from three perspectives: data type, reasoning methods and criticality assessment. %purpose and structure
% coverage content to de added

\textbf{Data type:}

The articles \cite{So2019,Gambi2019a,pop2019,pop2019,Hu2020,Kim20181,SUI2019,Stark2019a} used accident database (e.g., NHTSA\cite{NHTSA}, GIDAS\cite{GIDAS}, IGLAD\cite{IGLAD} to find the critical pre-crash scenarios. The used data type of the data source can be mainly grouped into sensor data, unstructured accident records (e.g., natural language document) and structured accident records (e.g., meta-data database). The sensor data refers to the recorded time series data in accidents (such as video data from the equipped camera sensor and GPS speed, yaw rate, acceleration, and target vehicle information from equipped radar sensor\cite{Kim20181}). The unstructured accident records are the police accident reports or protocols. In these reports, information for each scenario includes time, location, vehicle and driver details, before-crash maneuver, triggering events, crash descriptions of police officers, etc. These data are used for scenario classification and textual analysis. The structured accident records are meta data, characteristic data and categorized attribute data set from database\cite{Hu2020,SUI2019,Stark2019a}. These data types are mainly used for clustering analysis. 
%replace all others content in inductive method picture with the others identified content in this section

% The author \cite{So2019} analyzed 223,552 accident records from Korea National Police Agency. In paper \cite{Gambi2019a,pop2019,pop2019} they used the accident reports from NHTSA (National Highway Traffic Safety Administration) to transform it into simulation scenarios for the purpose of test case generation. In \cite{Hu2020} the author used 4255 accident data from IGLAD (Initiative for the Global Harmonization of Accident Data) to conduct the scenario mining. In\cite{Kim20181} sensor data were used in 363 accident scenarios from two sets of intersection accident data. Main data source for this study was from Naturalistic Driving Study (NDS), a database developed by a research project Second Strategic Highway Research Program (SHRP2). In\cite{SUI2019} 672 cases from China In-Depth Accident Study (CIDAS) data were taken in the final analysis. In In\cite{Stark2019a} the pre-crash matrix (PCM) data was analyzed from German In-Depth Accident Study (GIDAS) for the critical scenario identifications. 

\textbf{Reasoning methods:}

The main reasoning methods we identified are clustering, textual analysis, and reconstruction \& simulation. 
As a typical technique for statistical data analysis, clustering is widely used in many fields and domains. In the autonomous driving domain, clustering is also used for deriving and extracting typical or relevant driving scenarios from real driving data or databases which contain a large number of scenario instances. In this section, as the first observed reasoning method, clustering was applied to find typical pre-crash scenarios from accident scenarios databases\cite{Hu2020,SUI2019,Nitsche2017}. The derived scenarios are mostly functional scenarios, and they can be used to supplement the existing test suite or as new specific scenarios.
In these papers, clustering methods need to be applied to well-defined data. Therefore the first step of these methods is to interpret source data into a pre-defined feature vector (an n-dimensional vector of numerical features that represent some scenario variables or attributes). The selection of clustering methods depends on the data types of the features. The data type of each feature can be either numerical or categorical. If the feature vector contains categorical data, clustering methods based on euclidean distance cannot be directly adopted, such as the case in\cite{Hu2020}). Clustering methods for categorical data are introduced in \cite{bhagat2013review}. Features are selected based on different analyses, such as correlation analysis or relevance analysis \cite{SUI2019,Nitsche2017}(Most articles do not explain how these features are selected, more content regarding features could be found in section ontology \ref{sec:Ontology_Design}.) The data is subsequently prepared for clustering. Typical clustering methods include k-means and k-medoids\cite{Hu2020,SUI2019,Nitsche2017}. After clustering, scenario groups or clusters are formed based on the clustering results. One typical scenario is selected from one cluster with different methods. In the end, the derived pre-crash scenarios can be used as inspiration for the V\&V testing activities. In addition, they will be used to compare against the current test suite and to supplement it. 

The second observed method is textual analysis. Since accidents in accident databases are normally recorded with plain text or structured plain text, textural analysis or Natural Language Processing (NLP) techniques can help to extract features of accidental scenarios.
%The articles \cite{So2019,Gambi2019a,pop2019,pop2019} used textual analysis from textural accident database as reasoning methods.
Using NLP techniques, the accident description was parsed according to a pre-defined scenario ontology and generated typical critical scenarios based on a pre-defined sentence template\cite{Gambi2019a,pop2019,pop2019}. %The information is extracted through text-weight analysis or by searching for high-frequent wordings.
Information about the crash contributing factors (like weather, lighting, roads and the vehicles involved in the crash) were extracted from the reports. According to the parsed information, simulations were generated which can represent the accident scenarios depicted in the database. Trajectories of dynamic objects are modeled as a vector of way points. One vehicle in the simulation was then replaced with the ego-vehicle which was checked then to see if its behavior passes the oracle, which is to safely reach a goal point within a given time. %The goal point is behind the accident point to guarantee that the ego vehicle avoids the accident. 
The simulation's primary role in this case is to use the simulator to ensure the safety and performance of SoI in accident scenarios. The main steps are to analyze the incident database scenarios, reconstruct them, define the oracles, and finally execute the simulation with the scenarios.
%Reconstruction and simulation is also an important approach used as reasoning method to generate critical scenarios. 
Reconstruction in this case means that the data described in the database was reconstructed to the executable scenario in a simulator. Based on simulation results, critical scenarios are identified using a pre-defined set of KPIs to represent the criticality. The scenarios are derived from an accident database and the ADS is run to elicit the corresponding decisions from the system. In\cite{Kim20181}, the authors first select accident scenarios from intersection accident data according to two sensor systems (camera-based and radar-based). All the selected scenarios are classified into sixteen vehicle-to-vehicle pre-crash classes. Corresponding Safety Remaining Distances (SRDs) were calculated for each scenario for the two sensor systems. An SRD greater than zero suggests that it is possible to prevent an accident. Prevention rates for each system in each scenario class are calculated and subsequently compared with the compliance rate of each corresponding scenario to evaluate which sensor system is necessary for each scenario class. 
% add NLP and Stark GIDAS

\textbf{Criticality assessment:}
%\label{sec:C3_Critical_assessment_method}
As shown in table~\ref{tab:criticality_def}, since all the papers in this part took the data source based on accident scenarios only, the criticality definition of these articles is all related to safety-critical. The papers \cite{So2019,Hu2020,SUI2019} use an accident database and adopt clustering methods to find pre-crash scenarios. Therefore, their criticality definitions are all non-implementation-specific.  
%In \cite{Kim20181,Stark2019a,Gambi2019a,pop2019,pop2019} are all implementation specific. 
On the other hand, \cite{Kim20181} and \cite{Stark2019a} used reconstructed scenarios and then simulated them using collision as criticality to find critical scenarios.Therefore, they are all implementation-specific.  \cite{Gambi2019a,pop2019,pop2019} used NLP techniques and also collision as criticality simulation. Therefore they are also implementation-specific. 
% 44,46,65,101,103; 45,64,90,91,92
%more explanation of what features of implementation-specific
%Accident database --> safety critical. refer to table 5. (simulation: NLP and simulation, implementation specific) 
%In\cite{Gambi2019a,pop2019,pop2019} a scenario is critical if in the simulation, the ego vehicle (i.e., the system under test) cannot avoid the accident. In\cite{Hu2020,SUI2019} a critical scenario means a crash scenario. In In\cite{Kim20181} Safety- remaining distance (SRD) was used as a quantitative indicator. The author in\cite{Stark2019a} used the following KPIs for the criticality assessment: time gap (longitudinal front vehicle distance divides ego vehicle velocity) for ACC; TTC (time-to-collision) for AEB; TTLC (time-to-line-crossing) and DLC (distance-to-line crossing) for LKA. 

\subsubsection{Based on various types of data and scenarios
}
\label{sec:induc_based_on_all_scenarios}
Compared to the previous section, articles \cite{Qi2019,Yue9129787,VAAFO,Jenkins20183340} in this section do not use accident databases as the data source to find the critical pre-crash scenarios. Their used data type can be mainly grouped into sensor data, traffic data and others.

\textbf{Data type:}
The sensor data in this section refers to real sensor data from field operations. This data is mainly used for the virtual assessment of ADAS/AD functions to find critical scenarios. As shown in Fig.~\ref{fig:inductive_methods}, traffic data was also used here as a data type, refering to the public map data and traffic flow data. This data was used for microscopic traffic simulation (i.e., each vehicle and its dynamics are modeled individually, no detailed individual sensor models and function inside) parameterization to find critical scenarios. The other data types are concrete scenarios from available analyses or simulations. These data types are mainly used for the generation of new accident scenarios from prerecorded data. 
 
%refer to based on accident data section.
%The author in\cite{Qi2019} took a set of concrete scenarios from available analysis. In \cite{Yue9129787} public traffic data was used for the microscopic traffic simulation parameterization to find the critical scenario. In \cite{VAAFO} the author used sensor data from human driving data for the virtual assessment of automation in field operation. The data used in \cite{Jenkins20183340} was taken from a publicly available road traffic simulator. A set of concrete scenarios were used using the simulator for the pre-crash scenario generation. 

\textbf{Reasoning methods:}
%TODO: refer to previous section.
The main reasoning methods identified in this section are safety analysis, Virtual Assessment of Automation in Field Operation (VAAFO), simulation and others. 
Safety analysis was used as a method for scenario analysis and simplification for reducing the number of test scenarios for HAV test and evaluation\cite{Qi2019}. First, the concrete scenarios set are analyzed. Second, by analyzing concrete scenarios through the traversal of trajectories, trajectories that lead to collisions or test tasks uncompleted are obtained. By analyzing these trajectories, the SCPs (scenario characteristic parameters) of the corresponding scenario are obtained using functional decomposition \cite{Amersbach2019}, combined with fault tree analysis (FTA). By analyzing the overlap or relationship among the SCPs, the inclusion relationship among scenarios is obtained according to the SCPs included in different scenarios. By searching for the combination that contains the fewest scenarios but still covers all the SCPs, and using this set of scenarios to replace the original combination of test scenarios, the redundant evaluation scenarios were deleted.
Using simulation as the method, the authors in\cite{Yue9129787} used public traffic data to calibrate SUMO (simulation tool, Simulation of Urban MObility) to perform traffic simulation. Based on the simulation results, the data is post-processed to extract concrete crash scenarios. 

VAAFO \cite{VAAFO} is used as a method in parallel with human driving for critical scenario identification. In VAAFO (Virtual Assessment of Automation in Field Operation), the vehicle is driven by a human. It receives information from all sensors, but is not connected to the actuators  and instead operates in parallel while the human driver drives the vehicle. AD functions are running (simulating) within the perceived world. The trajectory planned by the AD function is compared with the human driving trajectory. The AD function may take a different decision than the human driver. This difference may consequently affect the behavior of other vehicles. The authors in \cite{VAAFO} state that accident scenarios made by human drivers will be recorded by the police. Potential critical scenarios are filtered further to identify real critical scenarios after correction of the world model and criticality metrics. The VAAFO method will identify scenarios that are risky for the AD but not for human drivers. 

As another method, the authors in\cite{Jenkins20183340} uses Long Short Term Memory (LSTM) networks to generate new collision data from prerecorded data. The data used to train the Recurrent Neural Networks (RNNs) comes from a simulation environment, but it could also make use of real accident data. In the example developed in the study, the data includes speed, the direction of vehicles and traffic light data. Once trained, the network can be used to generate new collision data starting from an initial seed that contains the initial speed, direction and traffic light state.

\textbf{Criticality assessment:}
The criticality definitions of the articles in this section are also shown in TABLE~\ref{tab:criticality_def}. The criticality definition in \cite{Qi2019} is based on safety analysis. This paper uses the term hazardous scenario, which refers to a scenario that may lead to harm, caused by the functional limitation or failures of the system. A scenario is critical if it is possible to have a collision in this scenario due to FuSa (Functional Safety) or a SOTIF problem with the ego vehicle. 
In \cite{VAAFO}, the criticality was defined via the assessment module, the simplest measure is whether a real collision has occurred or not, and the criticality definition also considered SOI. Therefore, it was also implementation-specific and related to a collision. 
%The definitions in papers \cite{Yue9129787} and \cite{Jenkins20183340} do not consider a specific SoI and the criticalities are based on collision or crash.
The definitions in papers \cite{Yue9129787} and \cite{Jenkins20183340} do not consider a specific SoI. \cite{Yue9129787} defined a scenario risk index and \cite{Jenkins20183340} used RNN for criticality definition based on the collected data in simulation, and both are therefore non-implementation-specific.
%criticality assessment according to methods classification, accident safety critical, TA clustering --> non implementation, Simulation --> implementation specific,

%In\cite{Qi2019} A scenario is critical if there exist a possible critical trajectory. A trajectory is critical if it leads to a collision or an unfinished task. In other words, a scenario is critical if it is possible to have a collision in this scenario due to FuSa or SOTIF problem of the ego vehicle. 
%In\cite{Yue9129787} the criticality was calculated using the Scenario Risk Index (SRI). The SRI is based on risk assessment. The magnitude of risk is equal to the probability of a risk event times the degree of loss. The loss calculation was based on the kinetic energy of collision. 
%In\cite{VAAFO} the paper says that for the assessment module, the simplest measure is whether a real collision has occurred or not. Further, multiple metrics for preventive or active safety can be implemented. Those can range from the safety in the current motion state (TTC, constant reserve time, etc.)  to motion prediction models (physics-based, maneuver-based, and interaction-aware). However, no particular metrics are given. 
%In the study\cite{Jenkins20183340} they only consider scenarios where there is a collision.
%They do not consider a specific system under test for this process. The aim of the study is to generate new accident scenarios from prerecorded data using a RNN (Recurrent Neural Networks).

\subsubsection{Mechanisms for Coverage}
Here we summarize and present our observations on how these articles evaluate the coverage results of the critical scenarios detected in their papers, all of which are based, of course, on the fact that the articles contain references to relevant content. Different methods also have different mechanisms for coverage definitions and evaluations.
For all articles that use clustering methods, their coverage is based on the completeness of the feature vectors (pre-defined numerical features that represent some scenario variables or attributes), i.e., if these feature vectors are complete, then their clustered typical scenarios can well reflect and cover the scenario space represented by the defined scenario variables or attribute.
The methods based on existing data (textual analysis and reconstruction\&simulation) do not explicitly discuss the coverage results of the critical scenarios they found.
%Since most articles in this section took data from different databases, the coverage therefore depends on the completeness of the database and the data coverage, while the coverage of clustering methods depends on the features completeness. 
%coverage corresponds to methods, different methods have different coverage, clustering: coverage --> feature vectors, typical scenario --> inspiration, check and supplement the current test suite/catalog,  

\subsubsection{Required Information}
%Summarize all type of database and with citation. Refer to table, name of database, simulator, FTA...Summary of all the info. 
All the required information in this section relates to sensor data, unstructured accident records, structured accident records, sentence template for NLP, simulator, FTA analysis, pre-crash scenario classes, traffic data and real driving sensor data. The involved databases are Korea National Police Agency Accident Database, NHTSA (National Highway Traffic Safety Administration), IGLAD\cite{CIDAS} (Initiative for the Global Harmonization of Accident Data), Second Strategic Highway Research Program (SHRP2), China In-Depth Accident Study (CIDAS) and German In-Depth Accident Study (GIDAS). 
%treat the method as black box, corresponds to figure inductive method, how the data will affect the results,  

%The author \cite{So2019} analyzed 223,552 accident records from Korea National Police Agency. In paper \cite{Gambi2019a,pop2019,pop2019} they used the accident reports from NHTSA (National Highway Traffic Safety Administration) to transform it into simulation scenarios for the purpose of test case generation. In \cite{Hu2020} the author used 4255 accident data from IGLAD (Initiative for the Global Harmonization of Accident Data) to conduct the scenario mining. In\cite{Kim20181} sensor data were used in 363 accident scenarios from two sets of intersection accident data. Main data source for this study was from Naturalistic Driving Study (NDS), a database developed by a research project Second Strategic Highway Research Program (SHRP2). In\cite{SUI2019} 672 cases from China In-Depth Accident Study (CIDAS) data were taken in the final analysis. In In\cite{Stark2019a} the pre-crash matrix (PCM) data was analyzed from German In-Depth Accident Study (GIDAS) for the critical scenario identifications. 

%%%%%%%%%%%%   Xinhai_Notes    %%%%%%%
%Definition: Methods to induce critical (functional or concrete) scenarios from a set of existing scenario data 

%Most of the accident databases or police reports are written in partially structured natural language.

%refer to ontology design section

%%%%%%%%%%%%%%%%%%%%%%%%%%%%%%%%%%%%%%%%%%%%%%%%%%%%%%%%%%%%%%%%%%
\subsection{Deductive Reasoning}
\label{sec:DR}
% refer to ontology design section
This section focuses on the methods that use a deductive reasoning approach based on different sources of knowledge to find critical functional or logical scenarios. Based on this scope, a total of 6 related articles were identified. 

Four of these papers \cite{Zhou2017a,weber2018,Huang2018,Xie2018} focus on finding pre-crash scenarios by systematically considering all the possibilities under a set of explicitly pre-defined assumptions. The identified pre-crash scenarios can be used as safety-critical operational situations as introduced in Fig. \ref{fig:SOTIF}. These four papers adopt a similar approach. As the first step, they take some available logical scenarios from function/system specification or safety analysis. Then, they define a structured road and initial conditions based on the function specification (e.g., 3 lane highway road) for the reasoning. Afterwards, the scenario is elaborated using the assumptions, e.g., by adding more vehicles and changing the vehicles' behaviors. The assumptions include: potential behaviors of vehicles on the road defined in the ODD, possible collision types, traffic rules, and function (SoI) features. All the identified pre-crash scenarios should be used for system verification and validation. 
In addition to finding pre-crash scenarios as critical operational situations, there is also an approach to find complete critical scenarios, which also include a triggering condition. The authors in \cite{Neurohr2020b} proposed a method to identify and to quantify the risk of critical scenarios for highly automated driving vehicles, considering both functional insufficiencies and failures. They proposed a new method to combine triggering conditions with the fault tree to deduce critical scenario from a given safety goal. The approach can be summarized into the following steps: The first step is to simulate the automated driving functions as components and interfaces. The second step is to identify potential hazards related to the AD system. Using a HAZOP-like method, they first identify generic vehicle-level hazards that are independent of the underlying implementation. Second, they use a HAZOP-like approach again to identify Functional Insufficiencies with hazardous effects. Each AD function is treated as a black box. Next, an “environmental fault tree” is used to identify the causes of functional inadequacies and the corresponding environmental conditions for triggering a hazardous scenario. 

% The author in\cite{Neurohr2020b} proposed a method to identify and quantify hazardous scenarios for highly automated driving vehicles with an emphasis on AD function limitations and failures. In this paper, a scenario is defined as the temporal development between several scenes in a sequence of scenes. Hazardous scenarios are identified with the following steps: 

% Step 1: model the automated driving functions as components and interfaces. 

% Step 2: identify hazards. They firstly identify generic vehicle-level hazards that are independent of the underlying implementation with a HAZOP (hazard and operability study)-like method. Secondly, they identify the Functional Insufficiencies with a HAZOP-like method. Here, each AD function is treated as a black box. 

% Step 3: identify the causes of the functional insufficiencies and the corresponding environmental conditions with “environmental fault tree” (a new method proposed in this paper to combine triggering conditions with a fault tree). 

% Step 4: quantify the risk of the identified hazardous scenarios (functional insufficiencies and environmental conditions). 

Ponn et al. \cite{Ponn} proposed a methodology for an intelligent selection of relevant scenarios for the certification of automated vehicles. They proposed a two-stage optimization framework to generate concrete scenarios. A detailed optimization method was not proposed. In the first optimization stage, the parameters of Layer one, two and five (refer to the 6-layered model discussed in Sec.~\ref{sec:scenario_def}) are first optimized by sensor analysis and consideration of driving behavior. In addition, the trajectory of the potential conflict partner (Layer L4) is determined. In the second stage, further objects are defined (to refine the logical scenario) by considering the complexity, and their trajectories are optimized. 
% The following factors are considered to increase the complexity of the logical scenario, which the detailed definition can be found in \cite{Schaub1996}: 

% \begin{itemize}
%     \item Number of elements 
%     \item Number of states per element 
%     \item Interdependency (The variables of a situation influence each other reciprocally)
%     \item Self-dynamics (Change of the system without direct intervention from the outside)
%     \item Intransparency (The observability of the system is incomplete, so there is a need for metrics to extract information from the system)
%     \item Multiple conflicting goals 
%     \item Openness of the target situation (Goals that are not clearly formulated)
%     \item Novelty (Structures of the system are not known and hypothesis formation is necessary for the exploration of known results)
% \end{itemize}

\textbf{Criticality assessment:}
\label{sec:C4_Critical_assessment_method}
The criticality definitions of the articles in this section are all shown in TABLE~\ref{tab:criticality_def}. All the articles in this section relate to collision and are non-implementation specific. No implementation of a specific system is needed for these approaches. Basically, a critical scenario should contain at least one accident threat or collision (i.e., one accident type may potentially happen).

\subsubsection{Mechanisms for Coverage}
The coverage in relation to the methods of this section depends on a set of explicitly predefined assumptions in the scenario complexity definition (e.g., potential behaviors of vehicles on the road defined in ODD, possible collision types, traffic rules, and function (SoI) features).
% 2 parts, one for pre-crash and another for the rest 2
%Assumption of scenario complexities...Pre-crash articles: assumption of traffic rules...

\subsubsection{Required Information}
The required information in this section also relates to the assumptions, such as potential behaviors of vehicles on the highway, possible collision types, traffic rules, set of base pre-crash scenarios, set of predefined basic maneuvers, set of generic hazardous scenarios, function features, and system specification. 
%based on defined scope of assumption in traffic rules...Assumptions are required info. 

%%%%%%%%%%%%%%%%%%%%%%%%%%%%%%%%%%%%%%%%%%%%%%%%%%%%%%%%%%%%%%%%%%
\subsection{Finding Critical Scenes for CV-based Functions}
\label{sec:CV}
\begin{figure*}
    \centering
    \includegraphics[width=0.90\textwidth]{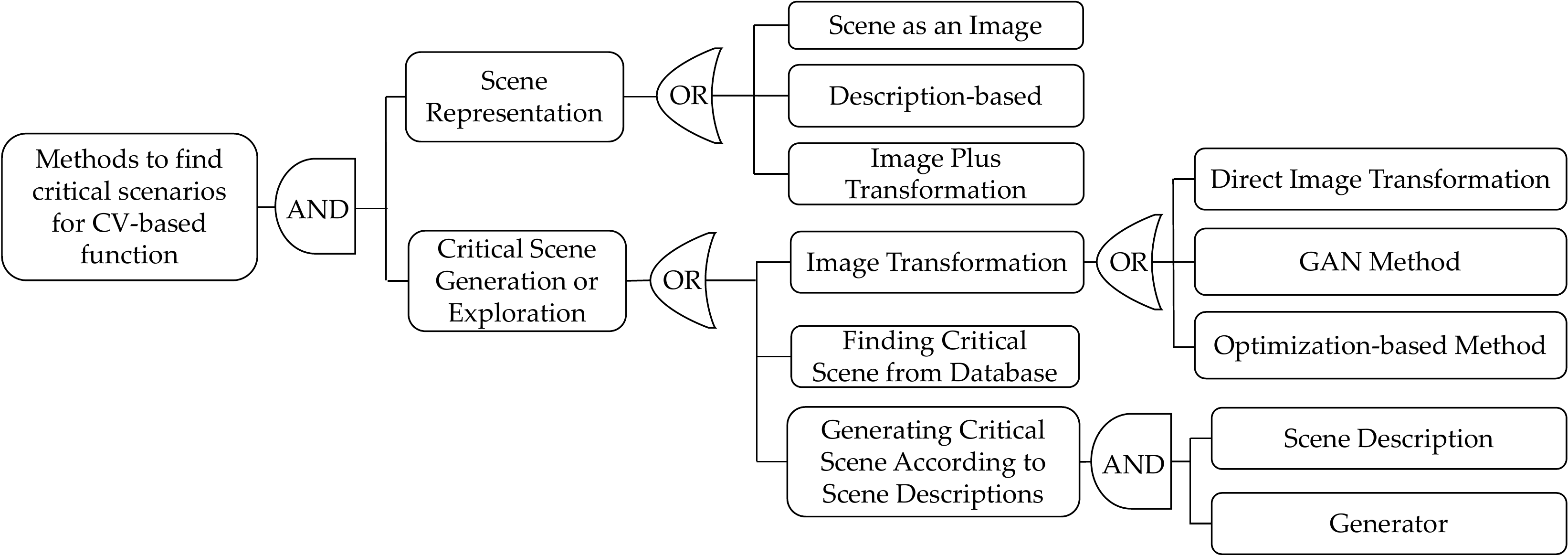}
    \caption{Methods to find critical scene for CV-based function} \label{fig:CV_solution}
\end{figure*}

This section focuses on the cluster of methods to find critical scenes to falsify a CV-based function. Based on the taxonomy in Fig. \ref{fig:Taxonomy}, methods in this cluster are summarized in Fig.~\ref{fig:CV_solution} and elaborated in the following subsections.

%%%%%%%%%%%%%%%%%%%%%%%%%%%%%%%%%
\subsubsection{Scene Representation}

In this cluster, scenes can be represented in the following three ways:

\textbf{Scene as an image:} To align with the terms defined in Fig. \ref{fig:LoA}, an executable scene is an image derived from a camera or stored in a database. Some methods in this cluster directly generate critical scenes as images \cite{CycleGANs, Scenic, DeepRoad, S.Yang.2017}. Some methods take images as inputs to explore for critical ones \cite{C.Zhang.2018, J.Wang.2018}.

\textbf{Description-based:} In this representation, a scene is described by a set of parameters \cite{C.Zhang.2018, J.Wang.2018, Scenic, Bolte.2019}. It can be used for both logical scenes and concrete scenes. A logical scene contains the ranges or the probability distributions of all the parameters. A concrete scene has a fixed value for each parameter. A tool is needed to transform a concrete scene into an image, based on, for instance, a video game (e.g., GTA V \cite{Scenic}) or a physics engine (e.g., Unreal engine \cite{S.Yang.2017}). 

\textbf{Image plus transformation:} This is used to represent a transformed image. Examples of image transformation include shearing, blurring, adding occlusions, and changing weather conditions \cite{DeepXplore}. With this representation, a concrete scene is a transformed image, while a logical scenario is represented as an original image together with a parameterized transformation (e.g., shearing with different angles) \cite{DeepXplore, DeepTest}.

Different scene representations are required by different methods to find or to synthesize critical scenes. These methods are introduced in the next sub-section.
%%%%%%%%%%%%%%%%%%%%%%%%%%%%%%%%%
\subsubsection{Critical Scene Generation or Exploration}

As shown in Fig. \ref{fig:CV_solution}, we identified the following three major types of methods to generate or explore a critical scene to falsify CV-based functions.

%%%%%%%%%%%%
\textbf{Image transformation:}
This type of methods assumes that a non-critical scene can be made more critical by changing or adding content on the image, for example, adding occlusions, intensifying the brightness, or changing weather into extreme conditions. Our literature search found the following image transformation approaches:
\begin{itemize}
    \item \textbf{Direct image transformation:} In this method, the image transformation is applied directly to the image without the help of ML algorithms or optimization algorithm. In general, direct image transformation (e.g., translation, scaling, shearing, rotation) \cite{DeepTest}is a simple technique. However, these transformations do not make the image more critical. They just generate different images, which in most of the cases are not realistic. Image transformation can also be performed by particular image processing tools. For example, Yang et al. \cite{S.Yang.2017} used Adobe Photoshop to add blur or rain effect through convolutional image transformation.  
    \item \textbf{GAN method:} A Generative Adversarial Network (GAN) can be trained to generate new images that are similar but specifically different from the dataset \cite{Zhaoqing.2019}. In the context of CSI, GAN can help to automatically generate a large amount of driving scenes with various weather conditions, which look realistic \cite{DeepRoad, CycleGANs}.
    \item \textbf{Optimization-based method:} Optimization algorithms can guide the exploration towards critical scenes, and also increase the coverage. For example, Pei et al. \cite{DeepTest,DeepXplore} propose an optimization-based method to find a set of combinations of direct image transformations to increase the neuron coverage (i.e., the proportion of the neurons activated by a set of test inputs). Increasing neuron coverage improves the quality of a test suite \cite{Harel.2020}. 
    %Image transformation can also be achieved by using the optimization algorithm. This method is similar with the GAN method where the image parameter and its transformation matrix are determined and calculated automatically by the optimization algorithm. In our literature study, we found that this technique is used for increasing neuron coverage for evaluating Ml for CV-based function \cite{DeepTest} \cite{DeepXplore}. By using this method, the erroneous behaviors of ML for CV-based function can be identified.
\end{itemize}

%%%%%%%%%%%%
\textbf{Finding critical scene from database:}
This type of methods traverses all the images in a database to find the non-implementation-specific critical ones \cite{Bolte.2019,J.Wang.2018,C.Zhang.2018}. Images in the databases are firstly manually encoded into predefined parameterized scene descriptions. Thereafter, the core of this type of methods is the approach to assess the criticality of each scene according to its scene description. The criticality can be assessed based on the complexity of the scene using a pre-trained machine learning (ML) model. Discussion on criticality assessment methods in this cluster is given in Section \ref{sec:C5_Critical_assessment_method}. 
%The method mostly used for this purpose is ML-based. In particular, ML is used either to find or predict critical scene in the database \cite{J.Wang.2018,C.Zhang.2018, Bolte.2019}.

%%%%%%%%%%%%
\textbf{Generating critical scene according to scene descriptions:}
This type of methods generates scenes according to a parameterized scene description with predefined semantics and syntax. The scene description may include environmental information such as weather conditions, light conditions and time. It may also include the information of road users such as the relative position, heading, dimension, car model, color, etc. 
According to these scene descriptions, images are generated with the help of a third-party tool, such as GTA V \cite{Scenic} and the Unreal Engine \cite{S.Yang.2017}. The generated scenes can support data augmentation (see examples in \cite{Scenic}), the verification of CV-based functions, and the identification of influential factors. Similar to the image transformation method, this method can also minimize the effort and budget compared to physically collecting these data. For instance, it can generate a potentially critical scene by adding specified occlusions, different weather conditions, or poor illumination. 

%%%%%%%%%%%%%%%%%%%%%%%%%%%%%%%%%
\subsubsection{Criticality Assessment}
\label{sec:C5_Critical_assessment_method}

In this cluster, all the identified critical scenes are function-critical. They may cause unintended results of CV-based functions. Most papers associate critical scenes with challenging environment situations such as bad weather, occlusions, and various light conditions. The criticality can be assessed either directly on an image \cite{Scenic, S.Yang.2017, CycleGANs, Dreossi.2019, Bolte.2019, DeepXplore, DeepTest, Bolte.2019}, or on a scene description \cite{J.Wang.2018, C.Zhang.2018}. Surrogate measures to assess the criticality of scenes are summarized in TABLE \ref{tab:criticality_def}. 

For the methods that directly assess criticality on images, most of them feed the image under assessment to the SoI (i.e., the function under verification) as input to assess its performance. Therefore, the assessed criticality is implementation-specific. In TABLE \ref{tab:criticality_def}, surrogate measures for these methods include failures and differential behaviors, which are explained at the end of this sub-section. In this literature review, only one paper \cite{Bolte.2019} was found to assess non-implementation-specific criticality on images. The corresponding surrogate measure for this method \cite{Bolte.2019} in TABLE \ref{tab:criticality_def} is predictability, which is also explained at the end of this sub-section. It is categorized as non-implementation-specific since the criticality assessment process does not involve the SoI.

As discussed previously, a scene description is a list of parameters, each of which supports the description of one influential factor. Complexity in TABLE \ref{tab:criticality_def} is the surrogate measure to access the criticality of a scene based on all the parameter values in its description. Complexity intends to reflect how likely the performance of CV-based functions in general will be affected by this scene. Therefore, complexity is categorized as non-implementation-specific. Detailed explanation of complexity is elaborated at the end of this subsection.

%To assess criticality on a scene description, a complexity score is given to each parameter value. The score reflects how likely this parameter will affect the performance of a CV-based function. For example, foggy weather will have a higher complexity score than sunny weather. The sum of the complexity scores of all the parameters denotes the complexity of the concrete scene.  a criticality is associated with the complexities of all the influential factors is considered as listed in Table \ref{tab:criticality_def}. 

%As for the remaining method which is complexity, we consider it as a method that assess criticality based on a scene description. This method is categorized as non implementation specific in TABLE \ref{tab:criticality_def}. The detail explanation of this method can be found at the end of this sub-section.

%In our literature study, we identified that a pre-trained ML model can automatically assess the criticality of a scene description. A grading or ranking approach is commonly used for this purpose. Semantic descriptor is needed in the grading approach to describe the scene \cite{J.Wang.2018} \cite{C.Zhang.2018}. The image description may cover information such as characteristics of road scene, the topology information of traffic elements, and driving conditions.

The detail explanation of each surrogate measure in TABLE \ref{tab:criticality_def} for this cluster is as follows: 

%%%%%%%%%%%%
\begin{itemize}
    \item \textbf{Failures:} It can help to find scenes that may lead the CV-based function to a specific failure mode (e.g., mis-detection or mis-classification) \cite{Scenic, S.Yang.2017, CycleGANs, Dreossi.2019}. This measure is used to find consequence-aware implementation-specific critical scenes. 
    
    \item \textbf{Differential behaviors:} This category is used to assess the Deep Learning (DL) algorithms by measuring the number of differential behaviors (i.e., different outputs from multiple similar DL systems with the same input)\cite{DeepXplore} \cite{DeepTest}. In safety-critical systems such as automated driving systems, differential behavior can lead to disastrous consequences (e.g., collision, crash) \cite{DeepXplore}.
    
    \item \textbf{Complexity:} This category measures how challenging the scene is for the SoI. In the primary studies, a complex scene can be specified according to project experience \cite{Scenic,DeepRoad}. The complexity of a scene can also be evaluated by a trained machine learning model. \cite{J.Wang.2018,C.Zhang.2018}
    
    %the criticality of scene based on a complexity scene metric. The complexity scene metric can be obtained by giving a grade or score to the scene according to road semantic complexity (i.e., scene data are encoded semantically to analyze road types, scene types, and challenging conditions of the scene) and traffic element complexity (i.e., the scene data are encoded to the matrix of traffic elements to identify the distance and viewpoint of vehicle in the scene)\cite{J.Wang.2018}. Another approach is by classifying the scene into several levels based on its complexity \cite{C.Zhang.2018}. As for the evaluation of scene complexity, ML approaches such as support vector machine (SVM) \cite{C.Zhang.2018} or support vector regression (SVR) \cite{J.Wang.2018} are considered. 
    
    \item \textbf{Predictability:} For one frame in a video stream, its criticality can be assessed based on how difficult it can be predicted according to its previous frames. Bolte et al. \cite{Bolte.2019} utilized an image prediction algorithm to predict each frame and compared the predicted frame with the real frame. The big difference implies low predictability. In their method, images with relevant objects (e.g., vehicles and pedestrians) and low predictability are considered critical.
\end{itemize}

%%%%%%%%%%%%%%%%%%%%%%%%%%%%%%%%%
\subsubsection{Mechanisms for Coverage}
Coverage is rarely discussed in the primary studies in this cluster. The only explicitly mentioned coverage is the neuron coverage (i.e., the proportion of neurons activated by the whole test set) \cite{DeepXplore}, which assesses the test adequacy of neural networks. \cite{Harel.2020} This coverage is implementation-specific, and can not reflect the completeness of the identified scenes. A coverage defined based on the ODD can be a potential future research topic.

%%%%%%%%%%%%%%%%%%%%%%%%%%%%%%%%%
\subsubsection{Required Information}
We have identified general required information when generating or finding critical scene for CV-based function as follows:

%%%%%%%%%%%%
\textbf{Scene generator:} The scene generator is used for description-based or image transformation-based method to generate a new scene. The scene generator can be semantic-descriptor tools, game editor (Unreal) \cite{S.Yang.2017}, photo editor (Adobe) \cite{DeepXplore}, etc.   

%%%%%%%%%%%%
\textbf{Database:} 
Most of the methods in this cluster required database. A critical scene can be synthesized according to an existing scene in a database \cite{DeepRoad, DeepTest, DeepXplore, CycleGANs}. A database can also be used to train the machine learning model to access criticality of a new scene \cite{J.Wang.2018,C.Zhang.2018}.
%The datasets can be divided to two categories, namely image or video datasets and environment datasets.

%%%%%%%%%%%%
\textbf{Influential factors:}
The influential factors are required when specifying a critical scene description \cite{Scenic, S.Yang.2017, Dreossi.2019}.

%%%%%%%%%%%%%%%%%%%%%%%%%%%%%%%%%%%%%%%%%%%%%%%%%%%%%%%%%%%%%%%%%%
%\subsection{Required Information}

%Necessary information to generate scenarios:

%Necessary information to assess criticality:

% \begin{figure}
% \centering
% \includegraphics[width=0.9\textwidth]{Domain_Knowledge}
% \caption{The categories of required domain knowledge} \label{fig:Domain_knowledge}
% \end{figure}

% \begin{table}[htbp]
% \caption{How domain knowledge is used in different steps}
% \label{tab:Domain_knowledge}
% \centering
% \begin{tabular}{|C{1.5cm}|L{3.5cm}|L{3.5cm}|L{3.5cm}|}
% \hline
%  $\downarrow$ & 
%  \multicolumn{1}{c|}{\textbf{Real data}} & \multicolumn{1}{c|}{\textbf{Quantitative knowledge}} & \multicolumn{1}{c|}{\textbf{Qualitative knowledge}}\\ 
% \hline
% \textbf{Reasoning} 
% & % real data 4 reasoning
% behavior of other vehicles

% to validate the proposed scenario catalogue \cite{Zhou2017a}
% & % quantitative models 4 reasoning

% & % Qualitative knowledge 4 reasoning
% to list factors that need to be considered for scenario modeling \cite{koopman2019}
% to categorize critical scenarios based on collision types \cite{Zhou2017a,weber2018}

% \\
% \hline
% \textbf{Formalization} 
% & % real data 4 formalization

% & 

% &

% \\
% \hline
% \textbf{Instances} & 

% & 

% &

% \\
% \hline
% \textbf{Testing/Sim.} & 

% & 

% &

% \\
% \hline
% \end{tabular}
% \end{table}

\section{The Evaluation Category}
\label{sec:Validation}
%Based on different cluster, which validation methods are used in the reviewed articles. 

This section mainly focuses on the evaluation category defined in Section~\ref{sec:Tax} to answer the question "How are the validity of the approach and the identified critical scenarios assessed?" for each cluster defined in Section~\ref{sec:Solution}.

\textbf{C1: Exploring logical scenarios without state trajectories}\\
Almost all the primary studies in this cluster perform case studies as the evaluation method to validate their proposed methods and the identified critical scenarios. Most of these case studies are realized by simulation. %Besides simulation, there are also papers, which use real experiment data for the validation. 
%In \cite{Feng2019b, Vehicles2020, Feng2020} the authors compared the failure rate estimated by their method with the failure rate measured from real testing. 
The validation in these papers shows that critical scenarios can be found using the proposed methods based on the defined criticality metrics. However, in these papers, there are no validations on the criticality of the scenarios themselves. In other words, these articles did not tackle whether the identified critical scenario in simulation is also critical in the real world, or the gap between real world and simulation. Regarding coverage evaluation, as shown in Fig.~\ref{fig:coverage_mechanism}, different mechanisms like N-wise or sampling size were proposed in this cluster, but there was no validation of the coverage results in real-world interpretation. 

%%% Criticality check method used real data for validation. %%%

%coverage meaning for real world scenario?
%Criticality means for real world scenario. 
%There are no validations on the criticality of the scenarios on the performance boundary. 
%A case study design, as a technology empirical evaluation research methodology and a way to generalize from observed case study outcomes,

\textbf{C2: Exploring logical scenarios with state trajectories}\\
The evaluation methods in this category are almost all case studies and very similar to the ones of category C1.

\textbf{C3: Inductive reasoning methods}\\
Since the papers in this category used either existing databases or available scenarios as data sources, the most used evaluation method is comparing the identified scenarios with the scenarios in the regulations. In the papers that generate functional or logical pre-crash scenarios by using clustering from real data, they directly compare these functional scenarios with the test scenario of the regulations or testing organizations, e.g., Euro NCAP (\cite{So2019,SUI2019,Hu2020}). For other papers that generated concrete scenarios, the simulation results, like simulated damage or KPIs using the proposed methods, were compared directly with the recorded scenario or the test results. There was no validation of the coverage results of the identified critical scenarios. 
%Because the majority of the articles in this section used data from different databases, the coverage is determined by the completeness of the database and the data coverage. 
%The validation work can be summarized into 2 perspectives: 1)	How many simulations can be successfully generated from accident descriptions? 2) How well the generated scenarios represent the description?  

\textbf{C4: Deductive reasoning methods}\\
Not all the papers in this section performed a validation of the proposed methods. One interesting validation method used in this cluster, however, is presented in \cite{Zhou2017a}. The authors used different knowledge, like possible collision types and traffic rules, to generate the pre-crash functional scenario and further evaluated the generated scenario catalog with respect to the naturalistic driving database and test regulations (Euro Ncap, UNECE, etc.). The coverage in this section was mostly based on the defined assumptions of the scenario definition, and there was no validation of the coverage. 
%In \cite{Zhou2017,Huang2018} the validation of the identified test case catalog is evaluated by a naturalistic driving database and test regulations of level 1 automated vehicles. 
%further explain the zhou2017 validation methods, how to validate functional scenario . therefore these methods don't cover the entire parameterhttps://www.overleaf.com/project space hence there is no guarantee that all errors of the SoI could be detected.

\textbf{C5: Finding critical scenes for CV-based functions}\\ 
%Most paper performed case study in this cluster to evaluate the methods. 
Almost all the articles in this category used a case study as the evaluation method. Similar to the papers in C1 and C2, there are no validations made to determine whether the identified critical scenario in simulation is also critical in the real world for the computer vision system. For this cluster, as mentioned in \ref{sec:CV}, the coverage depends mainly on the used dataset. No validation of the coverage results has been found in this cluster.
%%% Impossible to validate coverage

\section{Relation to Other Relevant Directions}
\label{sec:Other_Dir}
In this section, we briefly present the other relevant directions discovered during the literature review but are out of scope of the survey. These relevant directions have important implications and relationships to the main idea of the CSI method. In the following section, we first give a brief introduction to these methods, then explain how they relate to our article, and finally list the differences.

%supporting or complimentary

%%%%%%%%%%%%%%%%%%%%%%%%%%%%%%%%%%%%%%%%%%%%%%
\subsection{Online risk assessment}

Online risk assessment refers to the methods which evaluate and predict in real time if a situation could be dangerous and result in harm. As summarized in \cite{Lefevre2014}, these methods analyze the predicted trajectories of other vehicles and the potential colliding point. The methods presented in section~\ref{sec:C1_Critical_assessment_method} are mainly offline methods, whereas the methods here are for real-time vehicle application. 
%Online in vehicle algorithm,
%measure CSI can refer 
%in 6.1 the papers are offline assessment, 

\textbf{Relation:}
For both online and offline approaches, the identified influential factors as well as the critical scenario based on unexpected behaviors or events can be used. Second, the end-to-end scene risk prediction used in online risk assessment can also be used offline to determine a scene's criticality. Furthermore, risk indicators such as Time-To-Collision and Time-To-React can be used for both methods. %However, they can hardly be transformed into test cases.

\textbf{Differences:}
The online methods can only utilize the real street data whereas the CSI methods can utilize ground-truth environmental information. Offline methods, on the other hand, are not so easily transferred to online methods because the offline approach typically requires  exhaustive searches, and computing all the potential trajectories of the vehicles is computationally costly hens not real-time capable.
% data source 

%%%%%%%%%%%%%%%%%%%%%%%%%%%%%%%%%%%%%%%%%%%%%%
\subsection{Scenario-based Function Assessment}

Function assessment is one of the main reasons for using scenarios in the development, verification and validation of ADSs. It is also one of the primary reasons for finding critical scenarios as discussed throughout this paper. Here we can consider both broad descriptions of scenarios, e.g., \cite{Bach2016,roesener2016scenario},
%\footnote{In addition to not mentioning how the approaches help in identifying critical scenarios, these two references also fall outside the time period included in the review.}
used for functional assessment, as well as different ways to achieve coverage of the scenario space in general, e.g., through Monte-Carlo (MC) simulations. Furthermore, \cite{pop00261} suggests a method using a heuristic search and a fitness function to ensure that the scenarios specified are actually tested. It is also possible to do functional assessment using MC simulations of agent and simulation models \cite{wang2017prospective}. In \cite{wang2017prospective} these models are parameterized with different data sources to reflect actual driving scenarios.

\textbf{Relation:}
As mentioned, function assessment is a common vein across many of the papers considered in this review, and one of the primary purposes for using scenarios in general is to achieve some kind of assessment of the function. This paper primarily focuses on the assessment of critical scenarios whereas \cite{Bach2016,roesener2016scenario,pop00261,wang2017prospective} consider scenarios in a broader sense of AD function assessment. Having a ''driver's license test'' for ADSs would perhaps also be considered a function assessment. This aspect is however not considered in this review nor further in this section.

\textbf{Differences:}
The primary difference to the approaches of this section is the lack of explicit discussion about how to find or identify critical scenarios. In \cite{Bach2016,roesener2016scenario} scenarios are used as important descriptors to achieve function assessment, but there are no discussions about how to identify critical ones. Ensuring that specified scenarios are indeed tested, as described in \cite{pop00261}, is surely relevant, but again outside of the scope of this review as \cite{pop00261} disclose no details with respect to finding critical scenarios. One could however note that the described methods could be used to ensure that critical as well as non-critical scenarios that are specified are executed. 
Monte-Carlo simulation, in general, is not covered in this review as MC only provides a brute force methodology for covering scenario space. However, importance sampling, as suggested in \cite{Zhao2015b}, is covered since it can be used to find the scenarios that are more critical to the system. The authors of \cite{Zhao2015b} have also considered using subset simulation for accelerated testing of ADSs \cite{zhang2018accelerated}. Similar to importance sampling, subset simulation guides the MC search for relevant scenarios using a KPI. If this KPI is chosen to capture a safety-critical aspect, this approach would also yield critical scenarios. 

%%%%%%%%%%%%%%%%%%%%%%%%%%%%%%%%%%%%%%%%%%%%%%
\subsection{Scenario-based System Design}
Scenarios have also been suggested for supporting system design. Authors in \cite{stark2019towards} outline a method to identify critical spots of scenario space (white spots) through simulation of an ADAS feature. These white spot areas are subsequently used to design the ADAS system such that it fulfills the requirements derived from the white spots identified in previous performance assessment(s). The paper \cite{gyllenhammar2020towards} in a related vein, suggests the inverse, namely to define the ODD of the ADS using a world model of the intended use case for the ADS, where this world model is proposed to be constructed based on scenarios and populated using real traffic data. Paper \cite{lee2020identifying} on the other hand suggests the opposite again, that is, to use simulation assessment as a means to determine the ODD of the ADS by removing road segments where the ADS perform poorly (c.f. white spots of \cite{stark2019towards}).

\textbf{Relation:}
In all three cases above, scenarios are used (implicitly) to support the proposed process. Noteworthy is that both \cite{lee2020identifying,stark2019towards} consider the parts of scenario space (in their respective definitions) that result in critical situations for the system being assessed. 

\textbf{Differences:}
Even though scenarios are considered, neither of the approaches discuss how to explicitly find the critical scenarios (white spots of \cite{stark2019towards}), which is the primary focus of this review. Thus, deploying the methods described in this review would increase the efficiency of \cite{lee2020identifying,stark2019towards} in terms of needed resources for simulations.

%%%%%%%%%%%%%%%%%%%%%%%%%%%%%%%%%%%%%%%%%%%%%%
\subsection{Fault Injection}
Fault Injection (FI) is a method that accelerates the occurrence of faults to test and evaluate faults in systems. It is an important part of the test process in the automotive standard ISO 26262~\cite{ISO26262} and is distributed over the development stages. The author in ~\cite{DriveFI} applied the fault injection method to find and mine situations and faults that can be dangerous to the vehicle's safety.

\textbf{Relation:}
The target of article~\cite{DriveFI} is to find safety-critical situation and faults by using causal and counter-factual reasoning about the behavior of the ADS under a fault for safety analysis. FI method can therefore also be used to find critical scenario and situation. 

\textbf{Differences:}
The primary goal of FI is to evaluate the faults in system and analyze the system behavior based on the injected faults. The basic purpose of FI differs from that of the CSI method, but it is an interesting and well-established method for testing the system in autonomous driving applications. 
%literature fault injection needed

%%%%%%%%%%%%%%%%%%%%%%%%%%%%%%%%%%%%%%%%%%%%%%
\subsection{Ontology Design and Influential Factor Analysis}
\label{sec:Ontology_Design}
The articles and methods here focus on finding and analyzing the influencing factor, and utilizing a domain ontology to capture and represent the environment of the system under test. As presented in Fig~\ref{fig:SOTIF}, a scenario condition can be modeled as a  combination  of  several scenario  factors. The main steps of the ontology design method in the section are similar to the method in the reasoning section regarding both deductive and inductive reasoning. They took input or knowledge from different sources, such as system specification, sensor error type, system knowledge or analysis like FTA, or FMEA analysis~\cite{cao2018}. 

\textbf{Relation:}
The articles in this section are not included in the main content. The reason is that they focus mainly on the use of ontology design to find triggering conditions or influencing factors, but these identified influencing factors can be further used in CSI method to find the critical scenarios. 

\textbf{Differences:}
The main difference between this section's approaches is that no further case study or application were performed to identify critical scenarios. 

% also divide into deductive and inductive groups
% What are the focuses of these papers? INfluence factors, features
% what are the connections to our survey paper --> refer to SOTIF figure
% Different to our scope
% different methods of these papers. 

%Triggering  conditions  andhazard  operational  condition  are  two  major  componentsof  a  safety-critical  scenario.  Annex  B2  of  ISO/PAS  21448further assumes that a scenario condition can be modeled as a  combination  of  several  scenario  factors  (e.g.,  heavy  rain,glare,  slippery  road  surface,  etc).  Under  this  assumption,an  unknown  critical  scenario  can  be  attributed  either  anunknown  scenario  factor  or  an  unknown  combination  ofknown scenario factors. Not  all  scenarios  containing  one  or  more  triggeringconditions  as  critical  scenarios.  The  triggered  hazardousbehavior may not propagate to a vehicle-level hazard own-ing to the resilience of the system. In addition, the hazardpropagated from the triggering conditions may not lead toan accident without a hazardous operation condition.
% Include these? 
% \cite{queiroz2019geoscenario}
% \cite{bagschik2018ontology}

%%%%%%%%%%%%%%%%%%%%%%%%%%%%%%%%%%%%%%%%%%%%%%
%\subsection{Scenario Space reduction}
%Space reduction approaches by reducing non-critical scenarios are %included in this literature review. 

%\cite{Bach2018} by reducing repeating scenarios  

%Functional decomposition \cite{Amersbach2019}

%%%%%%%%%%%%%%%%%%%%%%%%%%%%%%%%%%%%%%%%%%%%%%
\subsection{Formal Methods}

%\textbf{Relation:}
%Formal methods are used to define and verify unambiguous specifications of computer systems. These methods rely on a sufficiently complete abstract model of the real world and formal specifications that should resemble the informal requirements that should be imposed on the system. \cite{58215}

%\textbf{Differences:}
%This literature review only covers the papers which use formal methods to find counter examples as critical scenarios.

%It is necessary to represent the relations between the behaviors of the agents and environmental conditions with hybrid automata, which need to be sophisticated enough to reflect reality and simple enough to guarantee computational feasibility.

Formal methods are used to define and verify unambiguous specifications of computer systems. These methods rely on a sufficiently complete abstract model of the real world and formal specifications that should resemble the informal requirements that should be imposed on the system. \cite{58215}

\textbf{Relation:}
This literature review covers the papers which use formal methods to find counter examples as critical scenarios.

\textbf{Differences:}
It is necessary to represent the relations between the behaviors of the agents and environmental conditions with hybrid automata, which need to be sophisticated enough to reflect reality and simple enough to guarantee computational feasibility.

%%%%%%%%%%%%%%%%%%%%%%%%%%%%%%%%%%%%%%%%%%%%%%
\subsection{Unknown Unknowns Detection in Computer Vision}
Unknown unknowns in computer vision (CV) functions (e.g., object classification) refer to the case where the employed predictive model (e.g., an deep neural network) assign incorrect labels to instances with high confidence \cite{unknown_unknowns}. Explainable AI \cite{Dosilovic.2018} approaches can be adopted to detect and avoid unknown unknowns. 

\textbf{Relation:}
These unknown unknowns are typically caused by the mismatch between training data and the cases encountered at test time. Since the CV functions are almost blind to such errors, they can be considered as implementation-specific critical scenes.

\textbf{Differences:}
According to the authors' understanding, these approaches should be considered in Section \ref{sec:CV}. However, no automated driving applications of these approaches were found during our literature search.

\subsection{Data Augmentation}
Data augmentation is used for machine learning based functions to expand the size of the training set by generating new training data.

\textbf{Relation:}
Section \ref{sec:CV} covers methods to generate new scenes according to a predefined scene description. These generated scenes can also be used for data augmentation.

\textbf{Differences:}
Traditional data augmentation methods focus on how to generate a big amount of different scenes as training data. Methods introduced in Section \ref{sec:CV} focus on generating new and also potentially critical scenes.

\section{Discussion}
\label{sec:Diss}
% \subsection{Open Issues and Challenges}

This section discusses how the results of the literature review (as presented in Sections \ref{sec:Tax}, \ref{sec:Problem}, \ref{sec:Solution} and \ref{sec:Validation}) answer the three research questions defined in Section \ref{sec:planning}, repeated below.
\begin{enumerate}
%\item[\textbf{RQ1:}] How can the state-of-the-art CSI methods for ADS and ADAS be classified?
\item[\textbf{RQ1:}] What would be a taxonomy that allows to systematically categorize and compare state-of-the art CSI methods for ADS and ADAS?
\item[\textbf{RQ2:}] What is the current status of CSI methods research with respect to this taxonomy?
\item[\textbf{RQ3:}] What are the remaining problems and challenges for further investigation?
\end{enumerate}
We have addressed RQ1 by analyzing relevant industrial standards, as discussed in Section \ref{sec:Background}, and the primary studies collected with the methodology introduced in Section \ref{sec:methodology}. The taxonomy was derived through iterations and validated by applying it to the collected studies. 
The proposed taxonomy, as presented in Section \ref{sec:Tax}, can sufficiently categorize all the primary studies. To support further research and engineering in areas related to critical scenarios and their identification methods, this taxonomy can help researchers to provide a contextual understanding - the big picture of such methods-, and provide guidance regarding CSI method design and adoption.

%RQ1 is answered in Section \ref{sec:Tax} based on relevant industrial standards as discussed in Section \ref{sec:Background}, and the collected primary studies with the methodology introduced in Section \ref{sec:methodology}. The proposed taxonomy can sufficiently categorize all the primary studies. In the future research, this taxonomy can help researchers to better understand the big picture of such methods, and to address their own research on the big picture.

RQ2 is answered in Sections \ref{sec:Problem}, \ref{sec:Solution} and \ref{sec:Validation} by analyzing all the primary studies according to the taxonomy. The answer to this question is also used as the basis to answer RQ3.

RQ3 is answered by internal discussions among all the authors; "remaining problems and challenges" are presented 
% dispersedly 
in Sections \ref{sec:Problem}, \ref{sec:Solution} and \ref{sec:Validation}. Some important points are summarized and discussed in the rest of this section, including suggestions for future work.

%With the ultimate goal being safety of automated driving and ADAS, the purpose of CSI is to address the underlining sources of harm, as depicted in Fig.~\ref{fig:SOTIF}, early in the development life-cycle. 
%As discussed in Section \ref{sec:Validation}, there are two dimensions to evaluate a CSI method, namely coverage and efficiency. Coverage reflects the completeness of the identified critical scenarios, while efficiency denotes the speed to conduct the iterative process in Fig. \ref{fig:ODD_CS}. 
%Most of the primary studies consider how to fasten the CSI process, however, the consideration on coverage is still not sufficient. 

%Each point includes: summary of collected information, open issues

\subsection{Coverage}
\label{sec:dis_coverage}
%Considering software testing work on coverage, coverage could be defined in at least 3 ways (ADD SUITABLE REFERENCE):
%(i) with respect to the scenario space - "parameters defined" - i.e. a white- (or gray-) box view; (ii) w.r.t. the inputs of the SUT, and (iii) w.r.t. the requirements on the SUT - although this would %be difficult for an ADS. One approach could for (iii) to start out considering the ODD. 
%The papers we surveyed had a primary emphasis on the first approach. Further work on coverage metrics is needed. 
%Summary: Are interpolatability a problem? Both inside critical set of scenarios as well as resulting in "hidden critical scenarios" within the non-critical regions of scenario space? 

%Referring to Fig.~\ref{fig:source_unknown_CS}, unknown critical scenarios may arise due to a number of different factors. When considering coverage it is important to consider (at least) all these three: triggering conditions, operational situations and the combination of the two. 

Within the context of this paper, coverage can be defined in 3 ways (referred to as coverage types in the following): 1) the coverage of the exploration with respect to the given scenario space; 2) the coverage of all the critical scenarios in the given scenario space (i.e., the proportion of the identified critical scenarios among all the critical scenarios within the given scenario space); and 3) the coverage of all the critical functional insufficiencies and their triggering conditions under a given functional scenario or an ODD. The type 1 coverage evaluates the exploration of the scenario space. The type 2 coverage evaluates the effectiveness of a \textit{logical $\rightarrow$ concrete} CSI method. As shown in Fig. \ref{fig:ODD_CS}, identified critical scenarios are used to support the identification of critical functional insufficiencies so as to improve the safety of the intended functionalities. Therefore, the type 3 coverage is essential for safety analysis.
These three types of coverage are discussed in the rest of this section.

The type 1 coverage is valid for the CSI methods finding critical concrete scenarios from a given logical scenario. Most of these methods are covered in clusters C1, C2 and C5. Due to the huge size of the scenario space, reaching full coverage is practically impossible. To quantify the level of coverage, coverage metrics can be defined. Adopted coverage metrics include sampling size (for sample-based exploration methods) and combinatorial coverage (for combinatorial testing methods). It is difficult to directly measure the type 2 and type 3 coverage. The type 2 coverage can be indirectly reflected by the coverage metrics defined for the type 1 coverage. As discussed before, the type 3 coverage is of the most importance. However, the relationships between these three types of coverage have not been sufficiently discussed in the primary studies. Understanding these relationships can help to determine a sufficient coverage level (in terms of a particular coverage metric) on the scenario space. It can also help to analyze the completeness of the safety analysis. These relationships can be discussed from the perspective of how critical scenarios may be missed.
%In the primary studies, there are methods trying to increase the number of identified critical scenarios with a given coverage level. Details are discussed in Section \ref{sec:Ins_no_mp_coverage}.

\textbf{Relation between type 1 and type 2 coverage:} With potential limitations of an adopted exploration method, it may not be possible to identify all the critical concrete scenarios in a given scenario space (i.e., the given logical scenario). According to Fig. \ref{fig:identification_process_critical_concrete_scenarios}, these limitations can stem from the coverage level employed by the instantiation process or the criticality assessment method. Since full coverage is not reached, some critical scenarios may not be covered by the specified coverage level. Even though a critical scenario is reached during the exploration, it may still be assessed as noncritical due to the limitation of the criticality assessment method. If a simulator is used for criticality assessment, its fidelity (i.e. the influential factors in the simulator can represent those factors in real world) will affect the accuracy of the criticality assessment. Increasing the fidelity of simulators is one way to solve this problem. However, a high-fidelity simulator normally entails high computational resources, and thereby increases power and time consumption. Another way to solve this problem is to design surrogate criticality measures that can tolerate some simulation error (i.e. the differences between simulation and real results). For example, instead of collision, Time To Collision (TTC) can be used to find more potentially critical scenarios. However, as analyzed in Section \ref{sec:C5_Critical_assessment_method} and shown in Table \ref{tab:criticality_def}, no such surrogate measure was found in the primary studies to find critical scenes for the CV methods. On the other hand, some critical scenarios may also be filtered out by the limitation of the adopted surrogate criticality measure. For example, if longitudinal TTC is used as the surrogate measure, critical scenarios caused by lateral collision will be missed. Therefore an appropriate approach to determine the criticality measure can be another future research direction. In addition, another way to increase the fidelity of the simulator is to bring the vehicle into the loop of the simulation, such as the methods introduced in \cite{VAAFO} and \cite{Feng2020}.

\textbf{Relation between type 2 and type 3 coverage:} Even if all the critical concrete scenarios within the given logical scenario are identified, one can still not claim a full coverage of the type 3 coverage under the corresponding functional scenario. Due to the potential misalignment between the logical scenario and the functional scenario, there might be critical scenarios that are not covered by the logical scenario.
The reason for the misalignment is the assumptions made when formalizing the functional scenario (referring to Fig. \ref{fig:LoA}). For example, some influential factors might be missing, or the models of some influential factors might be too simplified (e.g., using a constant speed model to represent the behavior of another vehicle). Section \ref{sec:Ontology_Design} briefly discusses how to find and formulate influential factors. Even though it is not the focus of this survey paper, it is an important topic for the safety analysis in terms of SOTIF. A literature review on influential factor identification could be a valuable future work. 

Instead of being identified from critical scenarios, functional insufficiencies can also be derived by analyzing the AD functions with safety analysis methods, such as the ones introduced in \cite{CV-HAZOP} and \cite{Neurohr2020b}. As inspired by traditional safety engineering \cite{Leveson2010}, the coverage of the derived functional insufficiencies can be increased by conducting both top-down (e.g., fault tree analysis) and bottom-up (e.g., Failure Modes and Effects Analysis (FMEA)) safety analysis methods, where the top refers to the vehicle level safety goals, and the bottom refers to the malfunctioning behaviors of the components together with the corresponding triggering conditions. To this end, a potential valuable future research direction could be to propose a methodology that combines these safety analysis approaches and the CSI approaces introduced in this paper.

%In addition, when generating the covering array for combinatorial testing approaches, the relations between the influential factors are not considered. For example, some factors can be mutually exclusive (e.g. summer day v.s. icy road), while some other factors might be heavily related (e.g. rainy day v.s. wet road surface). In the generated covering array, if there are test cases that do not satisfy one or more of the relations, simply removing those test cases may affect the combinatorial coverage. 
%Therefore, another future research direction can be the covering array generation methods that efficiently consider the relations between influential factors.

In addition, as discussed in Section \ref{sec:Ins_no_mp_coverage}, some primary studies assume that close (in terms of the distance on the scenario space) critical scenarios are likely to reflect the same functional insufficiency with the same triggering condition. To this end, increasing the diversity of the identified critical scenarios can increase the coverage of the identified functional insufficiencies. However, explicit coverage metrics for the identified functional insufficiencies and triggering conditions have not been found in the primary studies. 

%For inductive CSI methods finding critical functional scenarios (i.e. cluster C3), 

When systematically deducing safety-critical operational situations (methods in cluster C4), the coverage is defined by the assumptions made about the possible behaviors of the involved traffic participants (including the ego vehicle) and the possible road topology. In the primary studies, these assumptions are made on layers L1 and L4 of the 6-layer scenario description model in Table \ref{tab:6_layers}, since they all consider highway scenarios. For urban scenarios, the layer L2 in the 6-layer model (e.g., traffic lights) should also be considered. In addition, as discussed in Section \ref{sec:Validation}, coverage of the deduced safety-critical operational situations can be evaluated against an accident database.

%To increase the coverage when deducing critical scenarios, 
%In a sense, such a top-down vs. a bottom-up approach could be related to traditional safety engineering methods.
%FMEA would correspond to the bottom up analysis (based on components or functions, and scenarios), whereas FTA and similar system level methods would  correspond to analyzing a system in a more top-down fashion (e.g. somewhat similar to going from the ODD to critical scenarios). 

Corresponding to Fig \ref{fig:SOTIF}, to achieve a complete safety analysis, coverage needs to be explicitly defined for hazards, malfunctioning behaviors, triggering conditions and safety-critical operational situations. As a precondition, this entails an explicit coverage definition for all the influential factors. These factors are classified into layers as explained in Table \ref{tab:6_layers}. A more detailed classification of these factors in each layer is necessary to facilitate the systematic analysis of both safety-critical operational situations and triggering conditions. Therefore, this would represent another relevant future research topic. 

%One other important aspect to consider is the coverage of the real world use case in relation to the used scenario model or scenario space - i.e. to what extent these have been validated. Do the specified scenarios space span all the possible/relevant situations in the real world for the considered use case of the ADS?  
%One way to achieve this is to specify the ADS itself using a scenario model, as for example suggested by \cite{gyllenhammar2020towards} where the authors proposed to use the ''world model'' of the use case to quantify the operating conditions of the ODD that would allow the ADS to operate in the use case. By such a construction the corresponding scenario space would indeed relatively well correspond to the use case intended, at least if the data used for quantification allows for induction to the real operational situation of the ADS.

%\subsection{Efficiency}
%As shown in \ref{fig:ODD_CS}, the goal of CSI is to identify the underling functional insufficiencies and triggering conditions. Therefore, in general, there are three ways to increase the efficiency: 1) to reduce the of the whole CSI process, 2) to increase the percentage of true critical scenarios among all the identified scenarios, and 3) to reduce the number of the identified critical scenarios that reflect the same functional insufficiency with the same triggering condition. As disussed in Section \ref{sec:Solution}, most of the CSI methods in the primary studies focus on the first two ways. In most of the cases, these two ways are in conflict with increasing coverage. For the third way, 

\subsection{The ALARP Principle}

Identifying all the functional insufficiencies with all the triggering conditions is practically impossible. The iteration in Fig. \ref{fig:ODD_CS} should be governed by the ALARP principle (i.e. to guarantee that the risk of harm is "As Low As Reasonably Practicable"), see e.g.,  \cite{Leveson2010}. The core concept behind "reasonably practicable" is to control the risk of harm to an expected level. Therefore it is necessary to know how to control risk; and how to determine the expected level.

ISO 26262 \cite{ISO26262} answers these two questions for the risk caused by systematic faults and random hardware faults. In ISO 26262, the risk of harm is determined by the severity and the probability of the harm. The ALARP principle in ISO 26262 is to control the residual risk (the risk remaining after the deployment of all the safety measures) to a reasonable level. 

For the harm caused by functional insufficiency, risk can be defined based on the severity of the harm and the probability of exposure of the safety-critical scenario. In some of the primary studies, this probability of exposure is estimated according to the historical traffic data of human-driven vehicles. However, automated driving vehicles may exhibit different driving behaviors. Therefore, human-driving data may not precisely reflect the situations with automated driving vehicles. To this end, an effective approach to estimate the risk of harm related to functional insufficiencies represents a future research direction. In particular, given the novelty and unknowns of automated driving, this emphasizes the important role of continuous data gathering to support methodology improvements. An ability to better estimate the risk of harm would  can support the determination of the stop condition of the iterative process illustrated in Fig. \ref{fig:ODD_CS}.

%The iteration in Fig. \ref{fig:ODD_CS} is a way to control the risk caused by functional insufficiencies, where CSI methods are an essential part. How to determine the expected level to control the risk stem from functional insufficiencies is not discussed in this paper. This level can help to determine the coverage level discussed in Section \ref{sec:dis_coverage}. 

%\subsection{Future work beyond the scope of our literature survey}

\subsection{Scenario Space Explosion}

Due to the openness of the driving environment, a vast amount of influential factors compose an ODD to explore. The exploration effort can be reduced by "divide and conquer" approaches. To our understanding, 
%As far as we know, 
the scenario space can be partitioned from two perspectives.  

As suggested in \cite{gyllenhammar2020towards}, the first perspective is to divide an ODD into applicable use cases. Each use case is a sub-space of the ODD represented as one functional scenario. In this way, critical scenarios are identified for each use case, instead of the whole ODD. The challenge of this method is to guarantee that the union set of all the use cases is equivalent to the whole ODD.

Different SoIs provide a second perspective for partitioning a scenario space (i.e. a logical ODD). As discussed in \cite{Amersbach2019}, triggering conditions for different automated driving functions (i.e. SoIs) contain different influential factors. For example, the angle of the sun may affect some perception functions, but it will not affect the vehicle dynamic control functions. Therefore, the number of influential factors for a particular AD function is much lower than that for the whole ADS. To this end, the exploration effort of a CSI method can be reduced by finding triggering conditions for each individual AD function. The identified triggering conditions for all the AD functions can be systematically combined to support the final safety analysis on the vehicle level. However, the propagation of a functional insufficiency is not deterministic due to the resilience of the downstream AD functions \cite{DriveFI}. For example, if a vehicle is not detected in some individual frames by the computer vision function, this can be resolved by the following object tracing and sensor fusion algorithms for most cases. Therefore, to have a complete safety analysis, it is necessary to explicitly analyze the resilience of the system (i.e. to identify the scenarios where unintended behaviors from upstream functions cannot be resolved), especially to analyze the resilience of the object tracking functions and the sensor fusion functions. According to this literature review, only one paper \cite{Tuncali2020} was found to indirectly touch this topic. Investigating resilience as just described, would thus provide a useful research direction. Moreover, a systemic view would in addition be required to ensure that the partitioning into the SoIs - and the ensuing composition of "evidence" is complete, for example with respect to common cause failures. As discussed in this section this requires new methods that combine different approaches to critical scenarios identification, coverage and validation. 

\subsection{Other Sources of Harm}
Fig. \ref{fig:SOTIF} shows the sources of harm that are considered in this paper. Besides these, there are also other aspects that can threaten the safety of an ADS, such as the ones listed below. For a complete safety analysis, all the sources of harm should be considered together, since a combination of multiple sources of harm may lead to a new hazardous event. 

\textbf{Potential lack of Communication:}
As shown in \cite{Shetty2021}, some functional insufficiencies cannot be resolved without V2V (Vehicle to Vehicle) and V2I (Vehicle to Infrastructure) communication. Examples of such functional insufficiencies include occlusions, traffic violation by other road users, and the uncertainty of behavior prediction. 

\textbf{Cyberattack:}
An ADS can be hacked, especially when V2X (vehicle to everything) communication is adopted. Standard ISO/SAE 21434 \cite{ISO-SAE-21434} discusses how to mitigate the risk regarding cybersecurity.

\textbf{Misuse:} Misuse is considered in ISO/PAS 21448 \cite{SOTIF} but not emphsized in this paper. It can be considered as part of a scenario on layer L0 in Table \ref{tab:6_layers}.

\textbf{Problems outside the embedded system:} ISO 26262 \cite{ISO26262} and ISO/PAS 21448 \cite{SOTIF} only focus the the safety issues caused by the embedded system. Other vehicle failures, such as a flat tire, a broken suspension or a battery fire, can also lead to significant harms.

\section{Conclusion}
\label{sec:Conc}
For assuring safety of autonomous and automated driving, it is essential to be able to efficiently and effectively derive critical scenarios, i.e. situations that cause potential risks of harm (safety risks). Such critical scenarios need explicit consideration in ADS design and V\&V efforts. Moreover, the use of state of the art techniques for CSI promises to reduce the V\&V space by focusing on critical scenarios.

In this paper, we contribute to the challenge of safety and quality assurance of ADS and ADAS. In particular, we presented the results obtained from a systematic literature review in the area of CSI methods. Moreover, we introduced a taxonomy focusing on practitioners for supporting them in selecting the right CSI method for identifying critical scenarios according to their project-specific needs. The introduced survey covers assumptions, levels of abstraction, metrics stating criticality, methods for the generation of critical scenarios and coverage measures.

In addition, we discussed research gaps obtained after analyzing the primary studies, for example the coverage metrics, the criticality measures, the identification and classification of influential factors, and the lack of a methodological framework to combine different CSI methods and also to connect the CSI methods with other safety analysis processes. Hence this survey may also be considered helpful in providing guidance regarding future research directions.

\appendices
%\section{Proof of the First Zonklar Equation}
%Appendix one text goes here.

% you can choose not to have a title for an appendix
% if you want by leaving the argument blank
%\section{}
%Appendix two text goes here.

% use section* for acknowledgment
\ifCLASSOPTIONcompsoc
  % The Computer Society usually uses the plural form
  \section*{Acknowledgments}
\else
  % regular IEEE prefers the singular form
  \section*{Acknowledgment}
\fi
The authors of this paper have received support from a number of initiatives including: TECoSA - the Vinnova Competence Center for Trustworthy Edge Computing Systems and Applications at KTH Royal Institute of Technology and the  ECSEL projects InSecTT (grant agreement No 876038 - \url{www.insectt.eu}) and ArchitectECA2030 (grant agreement No 877539 - \url{https://autoc3rt.automotive.oth-aw.de/}). The ECSEL projects have received funding from the ECSEL Joint Undertaking which receives support from the European Union’s Horizon 2020 research and innovation programme and Austria, Sweden, Spain, Italy, France, Portugal, Ireland, Finland, Slovenia, Poland, Netherlands, Turkey. Finally, support is also acknowledged from the CD-Laboratory for Quality Assurance Methodologies for Autonomous Cyber-Physical Systems (QAMCAS) ( \url{https://qamcas.ist.tugraz.at/}) with financial support by the Austrian Federal Ministry for Digital and Economic
Affairs, the National Foundation for Research, Technology and Development and the Christian Doppler Research Association.

\section*{Disclaimer}
The document reflects only the view of the authors. The European Commission and the companies (i.e. Sigma Technology Consulting, Scania, AVL List GmbH, TU Graz and Zenseact) are not responsible for any use that may be made of the information it contains.

%The authors would like to thank...
%\cite{CV-HAZOP}

% Can use something like this to put references on a page
% by themselves when using endfloat and the captionsoff option.
\ifCLASSOPTIONcaptionsoff
  \newpage
\fi

% trigger a \newpage just before the given reference
% number - used to balance the columns on the last page
% adjust value as needed - may need to be readjusted if
% the document is modified later
%\IEEEtriggeratref{8}
% The "triggered" command can be changed if desired:
%\IEEEtriggercmd{\enlargethispage{-5in}}

% references section

% can use a bibliography generated by BibTeX as a .bbl file
% BibTeX documentation can be easily obtained at:
% http://mirror.ctan.org/biblio/bibtex/contrib/doc/
% The IEEEtran BibTeX style support page is at:
% http://www.michaelshell.org/tex/ieeetran/bibtex/
\bibliographystyle{IEEEtran}
% argument is your BibTeX string definitions and bibliography database(s)
\bibliography{main.bib}
\end{document}